\documentclass[twocolumn,superscriptaddress,
amsmath,amssymb,aps,pra]{revtex4-1}

\usepackage{graphicx}
\usepackage{physics}
\usepackage{comment}
\usepackage{color}
\usepackage{dcolumn}
\usepackage[normalem]{ulem}

\usepackage[colorlinks,urlcolor=blue,citecolor=blue,linkcolor=blue]{hyperref}

\newcommand{\vect}[1]{{\mathbf #1}}
\renewcommand{\k}{{\bf k}}
\newcommand{\p}{{\bf p}}

\newcommand{\q}{{\bf q}}
\newcommand{\Q}{{\bf Q}}

\newcommand{\0}{{\bf 0}}

\newcommand{\beq}{\begin{equation}}
\newcommand{\eeq}{\end{equation}}

\newcommand{\area}{\mathcal{A}}
\newcommand{\sch}{Schr{\"o}dinger }

\DeclareMathOperator*{\Simiq}{\simeq}
\newcommand{\Frac}[2]{\displaystyle\frac{#1}{#2}}

\begin{document}

\title{Multiple polaron quasiparticles with 
dipolar fermions in a bilayer geometry}

\author{Antonio Tiene}
\affiliation{Departamento de F\'isica Te\'orica de la Materia
  Condensada \& Condensed Matter Physics Center (IFIMAC), Universidad
  Aut\'onoma de Madrid, Madrid 28049, Spain}

\author{Andrés Tamargo Bracho}

\affiliation{Departamento de F\'isica Te\'orica de la Materia
  Condensada \& Condensed Matter Physics Center (IFIMAC), Universidad
  Aut\'onoma de Madrid, Madrid 28049, Spain}

\author{Meera M. Parish}
\affiliation{School of Physics and Astronomy, Monash University, Victoria 3800, Australia}
\affiliation{ARC Centre of Excellence in Future Low-Energy Electronics Technologies, Monash University, Victoria 3800, Australia}

\author{Jesper Levinsen}
\affiliation{School of Physics and Astronomy, Monash University, Victoria 3800, Australia}
\affiliation{ARC Centre of Excellence in Future Low-Energy Electronics Technologies, Monash University, Victoria 3800, Australia}

\author{Francesca Maria Marchetti}
\affiliation{Departamento de F\'isica Te\'orica de la Materia
  Condensada \& Condensed Matter Physics Center (IFIMAC), Universidad
  Aut\'onoma de Madrid, Madrid 28049, Spain}

\date{\today}

\begin{abstract}
We study the Fermi polaron problem with dipolar fermions  
in a bilayer geometry, where a single dipolar particle in one layer interacts with a Fermi sea of dipolar fermions in the other layer. By evaluating the polaron spectrum, we obtain the appearance of a series of attractive branches when the distance between the layers diminishes. We relate these to the appearance of a series of bound two-dipole states when the interlayer dipolar interaction strength increases. 
By inspecting the orbital angular momentum component of the polaron branches, we observe an interchange of orbital character when system parameters such as the gas density or the interlayer distance are varied. 
Further, we study the possibility that the lowest energy two-body bound state spontaneously acquires a finite center of mass momentum when the density of fermions exceeds a critical value, and we determine the dominating orbital angular momenta that characterize the pairing. 
Finally, we propose to use the tunneling rate from and into an auxiliary layer as an experimental probe of the impurity spectral function. 
\end{abstract}

\maketitle

\section{Introduction}
\label{sec:intro}
Over the last two decades, significant progress has been made in manipulating ultracold gases of dipolar atoms and molecules. The surge in experimental activity in this field is motivated by the expectation that the anisotropic and long-range nature of dipole-dipole interactions can lead to exotic states of matter~\cite{Lahaye_RPP2009,Baranov-Zoller_ChemReviews2012,Chomaz_RPP2023}.
Significant progress has already been made with dipolar gases of highly magnetic atoms such as Cr~\cite{Griesmaier-Pfau_PRL2005,Stuhler_PRL2005}, Dy~\cite{Lu_PRL2011,Lu-Lev_PRL2012}, and Er~\cite{Aikawa_PRL2012,Aikawa-Ferlaino_PRL2014}, where the achievement of quantum degeneracy has allowed the investigation of droplets, supersolids and other quantum phenomena~\cite{Bottcher_2021}. 
However, in order to access the regime of strong dipole-dipole interactions, one requires other cold-atom platforms such as Rydberg atoms~\cite{Loew-Pfau_JPB2012} or heteronuclear molecules~\cite{Moses_NatPhys2017,Shaffer_NatComm2018}.

Fermionic polar molecules in layered geometries are particularly promising for 
realizing quantum phases with strong and tunable dipolar interactions. 
By confining dipolar molecules to two-dimensional (2D) layers, inelastic losses are suppressed, while the sign and strength of the dipolar interactions can be precisely controlled~\cite{Valtolina_Nature2020}. 
Degenerate 2D Fermi gases have already been achieved  
with KRb~\cite{DeMarco-Jun_Science2019,Tobias-Ye_PRL2020} and NaK~\cite{Schindewolf-Bloch_luo_Nature2022,Duda-Bloch-Lou_NP2023}, while  other molecules, including LiCs~\cite{Deiglmayr_PRL2008,Repp_PRA2013} and NaLi~\cite{Park-Ketterle_Nature2023}, are also being explored.
Moreover, recent experiments with ultracold KRb molecules have demonstrated the possibility to image and control multiple layers individually~\cite{Tobias-Ye_Science2022}, thus expanding the range of scenarios that can be explored with 2D dipolar gases.

From a theoretical standpoint, 
dipole-dipole interactions are expected to generate ordered phases of fermions in layered geometries. Of particular interest is the 
configuration where all dipole moments are aligned perpendicularly to the confining planes, such that the system has rotational symmetry. 
In the case of a single layer, intralayer superfluid $p$-wave pairing can be driven by dressing polar molecules with a microwave field~\cite{Cooper-Shlyapnikov_PRL2009,Levinsen-Cooper_PRA2011}. Furthermore, density-wave instabilities with dipolar fermions  
have been 
proposed~\cite{Bruun-Taylor_PRL2008,Yamaguchi-Miyakawa_PRA2010,Sun-DasSarma_PRB2010}, including the spontaneous appearance of a stripe phase~\cite{Parish-Marchetti_PRL2012} and Wigner crystallization~\cite{Matveeva2012} at sufficiently high densities or strong interactions.   
In the case of bilayers, in addition to density-wave instabilities~\cite{Marchetti-Parish_PRB2013}, new interlayer bound pairs can arise due to 
the attractive part of the dipolar interaction~\cite{Klawunn-Duhme-Santos_PRA2010,Potter-Demler_PRL2010,Pikovski-Santos_PRL2010,Klawunn-Santos_PRA2010}. 
Such pairing and associated interlayer superfluidity has been studied in the case of both balanced~\cite{Bruun-Taylor_PRL2008,Pikovski-Santos_PRL2010,Mazloom-Abedinpour_PRB2018} and imbalanced populations~\cite{Mazloom-Abedinpour_PRB2017}. The latter includes the possibility of realizing a Fulde-Ferrell-Larkin-Ovchinnikov (FFLO) modulated pairing phase~\cite{Lee-Shlyapnikov_PRA2017}. Most notably, the stability of the FFLO state can be enhanced by the long-range character of the interlayer dipolar interaction, where different partial waves contribute to the pairing order parameter.

\begin{figure}
    \centering
    \includegraphics[width=0.7\columnwidth]{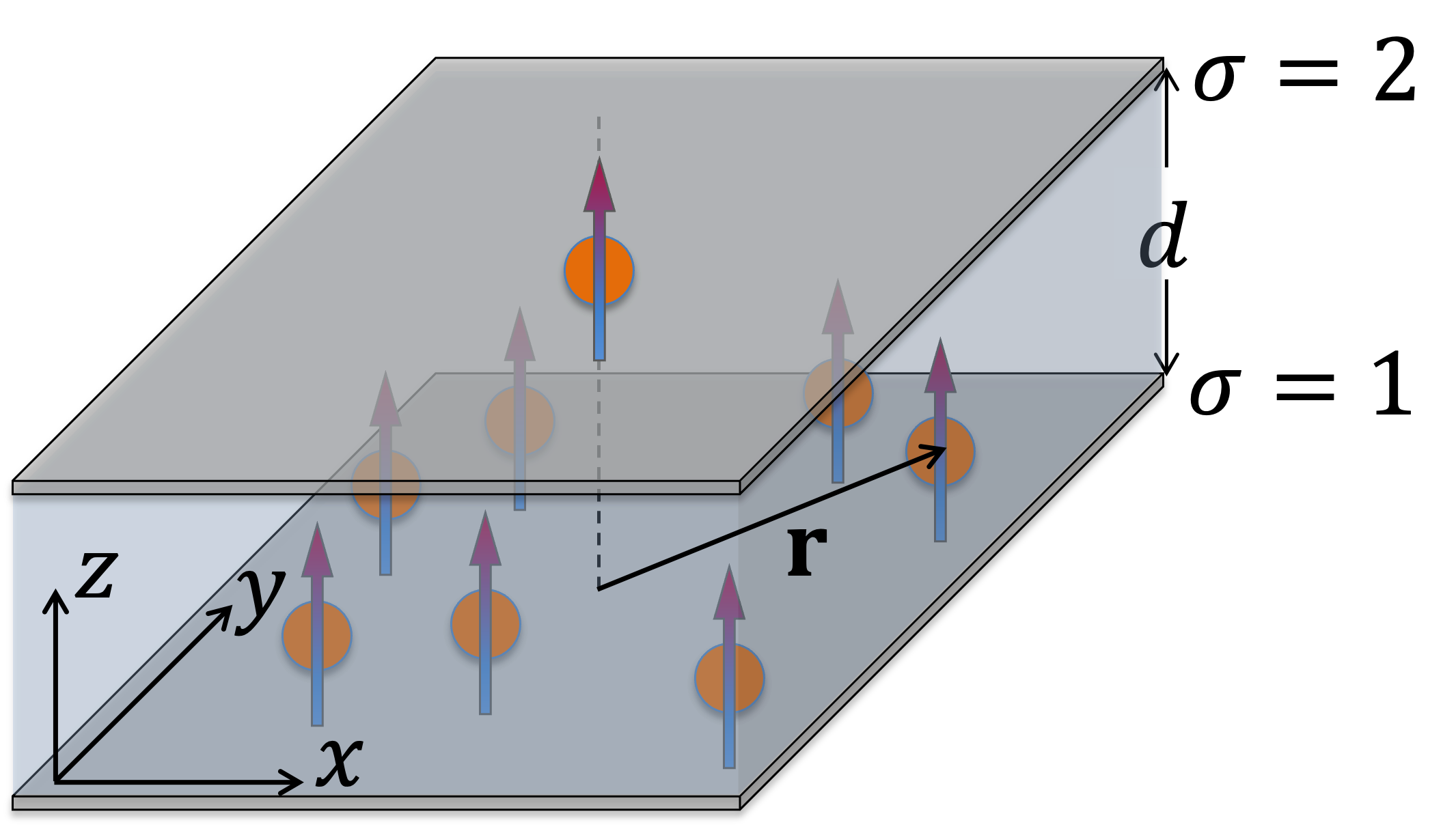}
    \caption{Schematic representation of the bilayer geometry considered: a Fermi gas of dipoles (such as polar molecules) is confined in the bottom layer $\sigma=1$, while a single dipole with the same perpendicular alignment is confined in the layer $\sigma =2$, generating an impurity problem where two inter-layer dipoles attract each other at short distances $r\lesssim d$ and repel each other at large distances $r\gtrsim d$ (see Fig.~\ref{fig:potentials}). Tunneling between the two layers is prevented by a barrier (blue filled region between the two layers). 
    }
    \label{fig:bilayer}
\end{figure}

In this work, we consider the limit of extreme population imbalance for 
dipolar fermions in a bilayer, 
as illustrated in Fig.~\ref{fig:bilayer}. Specifically, we have an impurity problem, where a single dipolar particle in one layer interacts with a Fermi sea of identical dipolar fermions in a different layer. This so-called ``Fermi-polaron'' problem has previously been considered in Refs.~\cite{Klawunn-Recati_PRA2013,Matveeva-Giorgini_PRL2013} for the case of dipoles in a bilayer geometry~\footnote{The ``repulsive Fermi polaron'' problem with dipolar fermions in a single layer geometry has been considered in Ref.~\cite{Bombin-Boronat_PRA2019}.}, and the general problem of an impurity in a 2D Fermi gas has been extensively studied in other 2D platforms such as ultracold atoms~\cite{Zollner2011,Parish_PRA11,Schmidt_PRA2012,Ngampruetikorn2012,Parish-Levinsen_PRA2013,Tajima_review2021} and doped semiconductors~\cite{Sidler_2016, Efimkin_PRB_2021,Tiene_PRB2022,Huang_PRX23}. Our scenario can be readily realized with polar molecules; however, note that our setup is quite general and could in principle apply to other dipoles such as  
Rydberg atoms.  
We consider two possible solutions of this problem. In the first, 
we generalize the interlayer bound state between two dipoles  to include the effects of an inert Fermi sea and how it blocks the occupation below the Fermi momentum. In the second, we consider the possibility of ``polaron'' quasiparticles, where the impurity is dressed by particle-hole excitations of the Fermi sea.
Throughout, we compare our results with those previously obtained within a $T$-matrix formalism which focused on the limit of weak dipolar interactions~\cite{Klawunn-Recati_PRA2013}.

By employing a variational Ansatz, we evaluate the polaron spectrum  to reveal that it is characterized by a series of attractive polaron branches, where the number of branches increases when the distance between the two layers decreases or  
the dipole moment increases.   
We associate the appearance of these polaron branches to the series of two-body bound states that, similarly to the 2D hydrogen atom~\cite{Yang_PRA1991}, is characterized by a specific orbital angular momentum component and a principal quantum number. In the limit of vanishingly small density of fermions, the attractive polaron energies recover those of the dipole-dipole bound states.
By evaluating the orbital angular momentum component of each polaron branch, we observe that the partial wave character of the branches evolve and interchange when we either increase the Fermi density or, at a fixed density, we increase the bilayer distance.

In contrast to the well-studied case of contact impurity-medium interactions~\cite{Massignan_2014,Scazza_2022}, a distinctive feature of finite-range dipole-dipole interactions is that the energy of the lowest energy polaron branch can either redshift, i.e., lower its energy, with increasing Fermi density, or blueshift. We find that this depends on the precise value of the dipolar strength, a quantity related to the specific value of the dipole moment and the layer separation. We explain this qualitative different behavior in terms of the contribution of hole scattering in the polaron formation, which we find is particularly important for the dipolar potential. 

Further, we consider the possibility that the lowest dipole-dipole bound state spontaneously acquires a finite center of mass momentum when the density of the Fermi sea increases.
Because of the long-range nature of the dipole-dipole interaction, this finite-momentum bound state mixes different orbital angular momentum components. We show that, while for small densities, $s$-wave pairing dominates close to the transition, for larger densities, the bound state acquires $p$- and $d$-wave components. These results agree with those found at finite but large imbalance in Ref.~\cite{Lee-Shlyapnikov_PRA2017}.

The paper is organized as follows. In Sec.~\ref{sec:model} we introduce the model of identical dipolar fermions in a bilayer geometry, and we discuss the relevant length and energy scales including their typical values in current experiments for either strongly magnetic atoms or heteronuclear molecules. 
Section~\ref{sec:molecule} describes the properties of the two-body interlayer bound states, generalized to the case where a Fermi gas in one of the layers acts to block the occupation of states below the Fermi sea.
In Sec.~\ref{sec:polaron} we describe the spectral properties of the Fermi polaron, while 
in Sec.~\ref{sec:observable} we show that the polaron spectral function can be probed analogously to radiofrequency spectroscopy by introducing an auxiliary layer to the system. Conclusions and perspectives are gathered  in Sec.~\ref{sec:concl}.

\section{Model}
\label{sec:model}
We investigate the configuration schematically represented in Fig.~\ref{fig:bilayer}. A Fermi sea of dipoles (e.g., polar molecules) is confined in one layer, with index $\sigma=1$, where  the dipole moments of the molecules are aligned perpendicularly to the plane by an external field. Additionally, a single dipolar molecule with the same perpendicular alignment occupies layer $\sigma = 2$.  
The Hamiltonian describing the systems is (we set $\hbar = 1$ and the system area $\area = 1$):
\begin{equation}
    \hat{H} = \sum_{\k, \sigma}
    \epsilon_\k \hat{c}_{\k,\sigma}^\dag \hat{c}_{\k,\sigma}^{}+ 
    \sum_{\k\k'\q} V_{\q} \hat{c}_{\k-\q,1}^{\dag} \hat{c}_{\k' + \q,2}^\dag  \hat{c}_{\k',2}^{} \hat{c}_{\k,1}^{}\; ,
\label{eq:hamil}
\end{equation}
where $\hat{c}_{\k,\sigma}^{\dag}$ ($\hat{c}_{\k,\sigma}^{}$) is the creation (annihilation) operator of a fermionic dipole with momentum $\k$ in layer $\sigma$. Dipoles in different layers have the same mass $m$ and their kinetic energy is $\epsilon_\k = k^2/2m$. 

In writing Eq.~\eqref{eq:hamil} we have implicitly assumed that the intralayer correlations are weak, such that the gas in layer $1$ is in a Fermi liquid phase where it can be treated as approximately non-interacting. This neglects the possibility of density instabilities which can occur in a (perpendicularly alligned) 2D dipolar Fermi gas with large dipole moments or high densities~\cite{Parish-Marchetti_PRL2012}.

On the other hand, the second term in the Hamiltonian~\eqref{eq:hamil} describes the dipolar interaction between a molecule in layer $1$ and one in layer $2$. The interlayer dipolar potentials in real  
and momentum space 
are, respectively, given by~\cite{Pikovski-Santos_PRL2010,Li-Das-Sarma_PRB2010,Klawunn-Santos_PRA2010}:
\begin{subequations}
\begin{align}
\label{eq:interd-real}
    V(\vect{r}) &= D^2\frac{r^2-2d^2}{(r^2+d^2)^{5/2}}\; ,\\
    V_\q &= -2\pi D^2 q e^{-qd} \; ,
\end{align}
\label{eq:dipolar}%
\end{subequations}
where $\vect{r}$ is the planar separation. Here $d$ is the layer separation and $D^2$ is the dipolar interaction strength. 
Out of these variables it is profitable to introduce a dimensionless dipolar strength
\begin{align}
    U_0 &= \frac{m D^2}{d} = \frac{D^2}{E_0 d^3}
 \; ,
\label{eq:intU0}
\end{align}
where $E_0\equiv 1/md^2$ is the energy scale associated with the layer separation. We see that $U_0$ increases by either increasing the dipolar interaction strength or by moving the two layers closer to each other. 

\begin{figure}
    \centering
    \includegraphics[width=1.0\columnwidth]{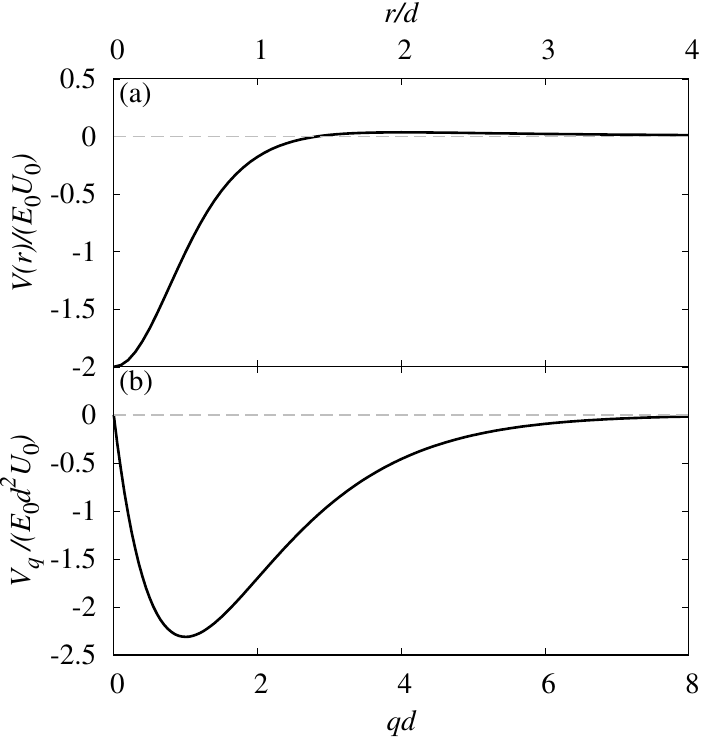}
    \caption{Interlayer dipolar interaction potential, Eq.~\eqref{eq:dipolar}, in real (a) and momentum (b) space.}
    \label{fig:potentials}
\end{figure}

The interlayer dipolar potential is plotted in Fig.~\ref{fig:potentials} in real and momentum space. As expected, in real space the potential is attractive at short distances $r\lesssim d$, where the dipoles are effectively arranged head-to-tail, while it is repulsive at large distances $r\gtrsim d$, where the dipoles are arranged side-by-side. Note that the interlayer potential has a vanishing zero-momentum contribution, i.e.,
\begin{equation*}
    V_{\q=\0}=\int d\vect{r}\, V(\vect{r}) = 0 \; .
\end{equation*}
This implies that the two-dipole bound state becomes very shallow when $U_0\to 0$~\cite{Klawunn-Santos_PRA2010}, as discussed further in Sec.~\ref{sec:molecule}.

The dipolar interaction strength $D^2$, which has the dimensions of energy times volume, is related to either the permanent magnetic dipole moment $\mu_m$ of a magnetic atom or to the dipole moment $D_e$ of an atom or a molecule induced by an electric field~\cite{Lahaye_RPP2009}:
\begin{equation*}
    D^2 = \begin{cases}
        \Frac{\mu_0}{4\pi} \mu_m^2 &\text{magnetic}\\[0.7em]
        \Frac{1}{4\pi \epsilon_0} D_e^2 &\text{electric}
    \end{cases}\; ,
\end{equation*}
where $\mu_0$ is the vacuum permeability and $\epsilon_0$ the vacuum permittivity. Usually heteronuclear molecules with induced electric dipole moments display much stronger dipole-dipole interactions than atoms with a permanent magnetic moment. In order to quantify the dipolar interaction, it is convenient to introduce the dipolar length $a_{dd}$~\cite{Lahaye_RPP2009,Chomaz_RPP2023}~\footnote{Note that Refs.~\cite{Lahaye_RPP2009,Chomaz_RPP2023} use a different definition of $C_{dd}$ but the same definition of $a_{dd}$: 
    $C_{dd}^{\text{Lehaye}} = 4\pi C_{dd}^{\text{Chomaz}} = 4\pi D^2 = 12\pi \frac{a_{dd}}{m}$. 
}:
\begin{equation}
    D^2 = \Frac{3 a_{dd}}{m} \; .
\end{equation}
For heteronuclear molecules, $a_{dd}$ can be up to 3 orders of magnitude larger than for magnetic atoms. Typical values of $a_{dd}$ for the highly magnetic atom $^{164}$Dy and for the heteronuclear molecules KRb and LiCs are given in Table~\ref{tab:add_values}. Furthermore, in current experiments, typically $d \simeq 500$~nm;  
however very recently a new super-resolution technique
which localizes and arranges dipolar molecules on a sub-$50$~nm scale has been implemented~\cite{Du-Ketterle_arxiv2023}. The corresponding values of $U_0$ that can be achieved for both highly magnetic atoms and heteronuclear molecules are listed in Table~\ref{tab:add_values}. Throughout this work we consider the dipoles to be structureless fermions such that we do not make a distinction between magnetic atoms or heteronuclear molecules.

\begin{table}[]
\begin{center}
\begin{tabular}{|c|c|c|c|c|}
    \hline
    & $a_{dd} (a_0)$ & $U_0 (d=50~\text{nm})$ & $U_0 (d=500~\text{nm})$ & $U_F$\\ \hline 
    $^{164}$Dy & 130.7 & 0.4 & 0.04 & 0.07\\ \hline
    KRb & $2\times 10^3$ & 6.3 & 0.6 & 1.1 
    \\ \hline
    LiCs & $2\times 10^5$ & 634.8 & 63.5 & 110  
    \\
    \hline
\end{tabular}
\end{center}
\caption{Typical values of  the dipolar length $a_{dd}$ in units of the Bohr radius $a_0 = 0.0529$~nm for the highly magnetic atom $^{164}$Dy~\cite{Chomaz_RPP2023}, and for the heteronuclear molecules KRb~\cite{Lahaye_RPP2009,Valtolina_Nature2020} and LiCs~\cite{Carr_NJP2009}. The corresponding value of the dimensionless dipolar interaction strength $U_0$~\eqref{eq:intU0} is given in the second and third column for two different bilayer distances $d$. In the fourth column we have $U_F$~\eqref{eq:new-units} (which is independent of $d$) for a typical Fermi gas density $n\simeq 10^8$~cm$^{-2}$. }
\label{tab:add_values}
\end{table}

The density $n$ of dipoles in layer $1$ is related to the Fermi momentum $k_F$ by
\begin{equation}
    k_F = \sqrt{4\pi n}\; . 
\end{equation}
From this, we can define a many-body dimensionless parameter $U_F$ analogous to the dipolar strength $U_0$:
\begin{align}
    U_F &= U_0 k_F d  = U_0  \sqrt{\Frac{2E_F}{E_0}} = m D^2 k_F \; .
\label{eq:new-units}
\end{align}
This dimensionless interaction strength characterizes the extent of many-body correlations in the system. As discussed above, we neglect the intralayer interaction and assume an ideal Fermi gas in layer $1$. Strictly speaking, this requires us to consider sufficiently 
small values of $U_F$ such that there are no ordered phases. Specifically, in the case of perpendicular orientation of the dipole moments, it has been found that the translational symmetry is broken for $U_F \gtrsim 6$ towards the appearances of a stripe phase~\cite{Parish-Marchetti_PRL2012}, while Wigner crystallization can occur for $U_F \gtrsim 25$~\cite{Matveeva2012}. Nevertheless, to comprehensively characterize bound-state and polaron properties in the presence of a Fermi gas, we must extend our investigation to higher density values. Therefore, to preserve the assumption of an ideal Fermi gas in layer $1$, we will implicitly assume that there is a small but finite temperature that induces the melting of the strongly correlated phases without strongly affecting the impurity physics.

Typically, the density of the 2D Fermi gas in experiments is $n\simeq  10^8$~cm$^{-2}$. 
This leads to  values of $U_F$ listed in the fourth column of Table~\ref{tab:add_values}.  
We furthermore list typical values for the dimensionless parameters $E_F/E_0$ and $1/k_Fd$ at different bilayer separation $d$ in Table~\ref{tab:kF-values}.

\begin{table}[]
\begin{center}
    \begin{tabular}{|c|c|c|}
        \hline
         & $d=50$~nm & $d=500$~nm \\ \hline 
         $E_F/E_0$ & 0.015 &  1.6\\ \hline
         $1/k_Fd$ & 5.7 &  0.56\\         
         \hline
    \end{tabular}
\end{center}
\caption{Values of the dimensionless density parameters $E_F/E_0$ and $1/k_Fd$ that can be accessed in experiments on dipolar Fermi gases with a bilayer distance $d$ and a typical density of the Fermi gas $n\simeq  10^8$~cm$^{-2}$.}
    \label{tab:kF-values}
\end{table}
%

\section{Dimer states}
\label{sec:molecule}
Due to the attractive part of the dipolar interaction, two dipoles can form an interlayer bound state. In this section, we discuss 
the properties of this two-body bound (dimer) state, generalized to the case where the Fermi gas in layer $1$ is inert and acts to block the occupation below 
the Fermi momentum $k_F$. We thus consider a general two-body state with a center-of-mass momentum $\Q$ described by:
\begin{equation}
    |M_2^{(\Q)}\rangle = \sum_{\k>k_F} \eta_{\k}^{(\Q)} \hat{c}_{\Q-\k,2}^{\dag} \hat{c}_{\k,1}^{\dag} |FS \rangle\; ,
\end{equation}
where the sum over the relative momenta $\k$ is restricted by Pauli blocking, $k>k_F$, while $|FS \rangle \equiv \prod_{\q<k_F} \hat{c}_{\q,1}^{\dag} |0\rangle$ describes the Fermi sea in layer 1, and $\eta_{\k}^{(\Q)}$ is the two-body wavefunction.
The energies $E$ can then be found by solving the corresponding \sch equation:
\begin{equation} 
    E \eta_{\k}^{(\Q)}  =\left(\epsilon_{\k} + \epsilon_{\Q - \k}\right)\eta_{\k}^{(\Q)}  +\sum_{\k'>k_F}V _{|\k-\k'|} \eta_{\k'}^{(\Q)}\; ,
\label{eq:M2schrod}
\end{equation}
which can be readily solved by numerical diagonalization. 

Before discussing the solution of Eq.~\eqref{eq:M2schrod}, it is useful to classify the dimer states according to their orbital angular momentum component. First, if the impurity momentum $Q=0$, the system is rotationally symmetric, and angular momentum is a good quantum number. Furthermore, when $E_F=0$, the center of mass and relative motion decouple, and therefore the energy at finite $Q$ is simply related to the energy at $Q=0$ via $E^{(Q)}=E^{(Q=0)}+Q^2/4m$, allowing us to take advantage of the rotational symmetry at $Q=0$. The presence of the Fermi sea complicates matters because then the center of mass motion no longer decouples. Thus, for a dimer state where both $\Q$ and $E_F$ are finite, the system is no longer rotationally invariant and orbital angular momentum 
is not conserved. 

To proceed,  
we expand the dimer wavefunction as a Fourier series over the orbital angular momentum basis $e^{i\ell\varphi}$:
\begin{equation}
    \eta_{\k}^{(\Q)} = \eta_{k\varphi}^{(Q)} = \sum_{\ell\in \mathbb{Z}} e^{i\ell\varphi} \tilde{\eta}_{k\ell}^{(Q)}\; ,
\end{equation}
where $\varphi$ is the angle between $\k$ and $\Q$. 
The dimer \sch equation~\eqref{eq:M2schrod} now reads 
\begin{multline}
    E \tilde{\eta}_{k\ell}^{(Q)}  = \left(2\epsilon_{\k} + \epsilon_{\Q}\right)\tilde{\eta}_{k\ell}^{(Q)} - \Frac{kQ}{2m} \left(\tilde{\eta}_{k,\ell-1}^{(Q)} + \tilde{\eta}_{k,\ell+1}^{(Q)}\right) \\ 
    + \int_{k_F}^\infty\Frac{dk'\, k'}{2\pi} \tilde{V} (k,k', \ell) \tilde{\eta}_{k'\ell}^{(Q)}\; .
\label{eq:finiteQ-eq-emme}
\end{multline}
Here, we have taken the continuum limit, $\sum_{\k'} \to \int \frac{d\k'}{(2\pi)^2}$ and decomposed 
the interlayer potential in the orbital angular momentum basis
%
%
%
%
by using the fact that the potential is diagonal in angular momentum, i.e.,
\begin{equation}
    \int_{0}^{2\pi} \Frac{d\varphi}{2\pi} \Frac{d\varphi'}{2\pi} e^{-i\ell\varphi} V_{|\k-\k'|}   e^{i\ell'\varphi'}\\ = \delta_{\ell \ell'} \tilde{V} (k,k',\ell)\; ,
\label{eq:diagonal}    
\end{equation}
where $\k=(k,\varphi)$ and $\k'=(k',\varphi')$.
Note that the potential $\tilde{V} (k,k',\ell)$ is real.

As expected, Eq.~\eqref{eq:finiteQ-eq-emme} becomes diagonal in $\ell$ when $Q=0$ since, in this limit, the orbital angular momentum is a good quantum number. Note also that, because the potential $\tilde{V} (k,k',\ell)$ is symmetric under the exchange $\ell \mapsto -\ell$, the eigenvectors for $\ell\ne 0$ can either be symmetric or antisymmetric  solutions:
\begin{equation}\             \tilde{\eta}_{k\ell}^{(\pm,Q)} = \Frac{\tilde{\eta}_{k\ell}^{(Q)} \pm \tilde{\eta}_{k,-\ell}^{(Q)}}{2}\; ,
\label{eq:symm-a}
\end{equation}
where $\tilde{\eta}_{k0}^{(+,Q)} = \tilde{\eta}_{k0}^{(Q)}$, while $\tilde{\eta}_{k0}^{(-,Q)}=0$.
In terms of these, the \sch equation reads ($\ell \geq 0$):
\begin{multline}
    E \tilde{\eta}_{k\ell}^{(\pm,Q)}  = \left(2\epsilon_{\k} + \epsilon_{\Q}\right)\tilde{\eta}_{k\ell}^{(\pm,Q)} - \Frac{kQ}{2m} \left(\tilde{\eta}_{k|\ell-1|}^{(\pm,Q)} + \tilde{\eta}_{k\ell+1}^{(\pm,Q)}\right) \\ 
    + \int_{k_F}^{\infty}\Frac{dk' k'}{2\pi} \tilde{V} (k,k', \ell) \tilde{\eta}_{k'\ell}^{(\pm,Q)}\; .
\label{eq:finiteQ-eq-emme_pm}
\end{multline}
One can see that the \sch equation~\eqref{eq:finiteQ-eq-emme} now becomes block diagonal using the symmetric and antisymmetric dimer wavefunctions, i.e., the equations for $\tilde{\eta}_{k0}^{(Q)}$ and $\tilde{\eta}_{k\ell}^{(+,Q)}$ decouple from those for $\tilde{\eta}_{k\ell}^{(-,Q)}$, the former $l=0$ states being in general the lowest energy solutions~\footnote{Note that, as mentioned previously, the \sch equation becomes diagonal in $\ell$ when either $Q=0$ or $E_F=0$, such that symmetric and antisymmetric solutions become degenerate.}.  

In the following two subsections, we first analyze the two-body limit, i.e., the limit $E_F\to 0$ where there is a single particle in each of layers $1$ and $2$, and we study the appearance of additional bound dimer states 
when the interlayer dipolar strength $U_0$ increases. Then, in Sec.~\ref{sec:finite}, we consider the effect of a finite density of fermions in layer $1$ and how this can lead to the dimer spontaneously acquiring a finite center of mass momentum.

\begin{figure}[t]
    \centering
    \includegraphics[width=1.0\columnwidth]{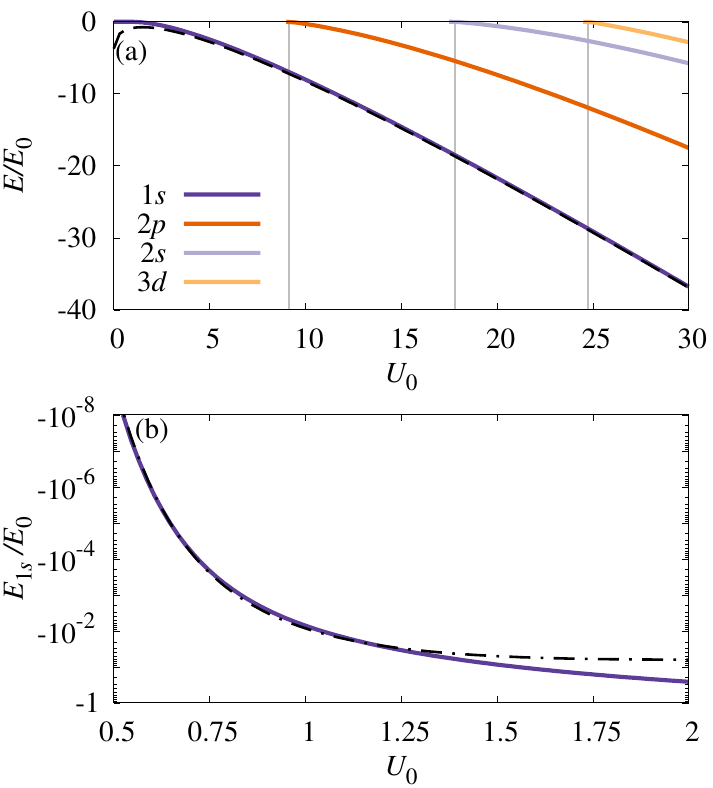}
    \caption{Energies of the bound-state dimers 
    at $E_F=0$ and $Q=0$ as a function of the dipolar interaction strength $U_0$. The energies are labelled with $n\ell$, where $\ell$ is the orbital angular momentum ($\ell=0$ is $s$-wave, $\ell=1$ is $p$-wave, and $\ell=2$ is $d$-wave) and $n$ is the eigenvalue index, where $n \ge \ell+1$.   
    (a) 
    Vertical (gray) lines are the binding thresholds for the $2p$, $2s$, and $3d$ states (see Table~\ref{tab:mol_threshold}). The black (dashed) line is the analytical expression~\eqref{eq:approx3} for $E_{1s}$ valid for $U_0 \gg 1$.
    (b) $U_0$ dependence of $E_{1s}$ for small values of $U_0$ and comparison with the analytical expression~\eqref{eq:approx2} valid for $U_0 \ll 1$ (black [dot-dashed]).
    }
    \label{fig:molecular_en}
\end{figure}

\begin{table}[t]
\centering
\begin{tabular}{|c|c|c|c|c|c|c|c|c|}
    \hline
    & $2p$ & $2s$ & $3d$ & $3p$ & $4f$ & $3s$ & $4d$ & $\dots$ \\ \hline
    $U_0$ & $9.2$ & $17.8$ & $24.7$ & $35$& $48$ & $52$ & $59.5$ & $\dots$ \\
    \hline
\end{tabular}
\caption{Threshold values of $U_0$ for the binding of excited dimer states.}
\label{tab:mol_threshold}
\end{table}
%
\subsection{Vacuum dimer}
\label{sec:molecule_Q0}
As explained above, in the absence of a Fermi sea in layer 1, the center-of-mass and relative motion decouple. 
We therefore solve the two-body problem at $Q=0$.
The dimer states can be labelled by the orbital angular momentum $\ell$ and the eigenvalue index $n$ (where increasing values of $n$ indicate larger energies eigenstates),  which we assume to be $n \ge \ell+1$, in analogy with the 2D hydrogenic atom~\cite{Yang_PRA1991}. 

We plot in Fig.~\ref{fig:molecular_en} the energies of the dimer states for increasing values of $U_0$, i.e., for either increasing values of the dipole moments $D^2$ or smaller bilayer distances $d$.
As expected, the $1s$ state is always bound for $U_0\ne 0$, even if it becomes 
very shallow when $U_0 \ll 1$, i.e., the binding energy goes exponentially to zero. This has been already analyzed by Ref.~\cite{Klawunn-Santos_PRA2010} and traced back to the fact that the interlayer dipolar potential has a vanishing zero-momentum contribution, in which case one cannot use Landau's formula for the energy of the bound state $E_{1s} \propto - 
\exp[4\pi/\int d\vect{r}\,V(r)]$.
Instead, Ref.~\cite{Klawunn-Santos_PRA2010} found an approximation for the $1s$ 
energy when $U_0 \ll 1$ by employing the Jost function formalism, leading to:
\begin{equation}
    {E}_{1s} \Simiq_{U_0\to 0} -E_0 e^{-\frac{8}{U_0^2} \left[ 1 - U_0 + \frac{U_0^2}{4} \left(\frac{5}{2} + \gamma-\ln2  
    \right) \right]}\; ,
\label{eq:approx2}
\end{equation}
where $\gamma$ is the Euler-Mascheroni constant.
For large values of $U_0$, variational calculations~\cite{Yudson_PRB1997} show that
\begin{equation}
    {E}_{1s} \Simiq_{U_0 \gg 1} -E_0\left(2U_0 - 4\sqrt{\Frac{3U_0}{2}} + \Frac{15}{4}\right)\; .
\label{eq:approx3}
\end{equation}
We see that these perturbative expressions match well with our numerical results in the two limits, as shown in  Fig.~\ref{fig:molecular_en}.

In addition to the $1s$ bound state, we find that the interlayer dipolar potential can bind an increasing number of dimer states 
with increasing $U_0$. The corresponding thresholds are indicated as grey vertical lines in Fig.~\ref{fig:molecular_en} and are listed in Table~\ref{tab:mol_threshold}.
In order of increasing $U_0$, the sequence of additional dimer states that eventually bind is $2p$, $2s$, $3d$, $3p$, $4f$, $3s$, $4d$, $\dots$.
Note that this order can change when we introduce the effects of Pauli blocking at finite $E_F$, even though the orbital angular momentum remains a good quantum number for $Q=0$.
\begin{figure}[]
    \centering
    \includegraphics[width=1.0\columnwidth]{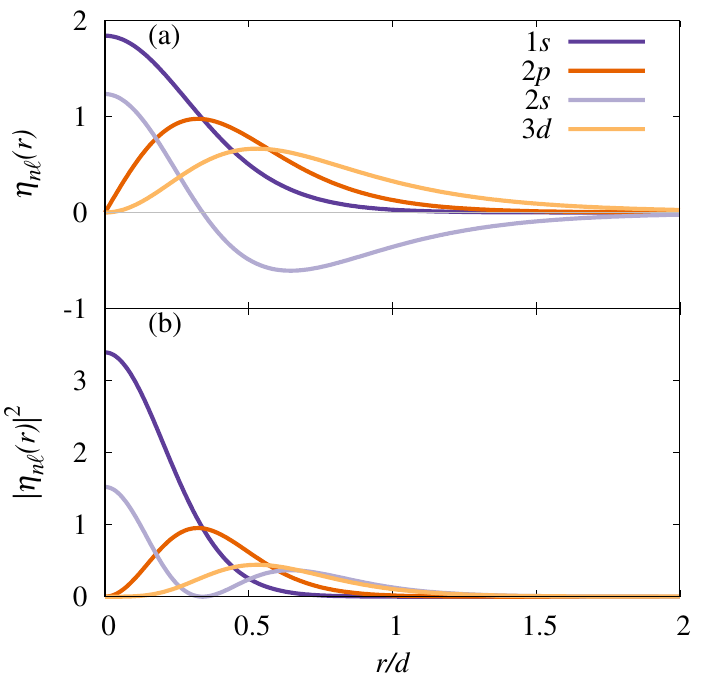}
    \caption{Real-space dimer wavefunctions in the case of $E_F=0$, Eq.~\eqref{eq:moleta}. We show (a) $\eta_{n\ell} (r)$ and (b) $|\eta_{n\ell} (r)|^2$ for $U_0=30$.
    }
    \label{fig:wave-f}
\end{figure}

For each bound state with energy $E_{n\ell}<0$, we evaluate in Fig.~\ref{fig:wave-f} the corresponding dimer eigenfunctions in 
real space,
\begin{equation}\label{eq:moleta}
    \eta_{n\ell} (r)  = i^\ell \int \Frac{k\,dk}{2\pi} J_{\ell} (kr) \tilde{\eta}_{k\ell}^{(Q=0,n)} \; ,
\end{equation}
where $J_{\ell} (x)$ is the Bessel function of the first kind. We find that, at small distances $r\lesssim d$, the dimer wavefunctions can be well approximated as
\begin{subequations}
\begin{align}
    \eta_{1s} (r) &\simeq \lambda_{1s} e^{-\alpha_{1s} r^2}\; ,\\
    \eta_{2s} (r) &\simeq \lambda_{2s} e^{-\alpha_{2s} r^2} (1-\beta_{2s} r)\; ,\\
    \eta_{2p} (r) &\simeq \lambda_{2p} r e^{-\alpha_{2p} r^2}\; ,\\
    \eta_{3d} (r) &\simeq \lambda_{3d} r^2 e^{-\alpha_{3d} r^2}\; .
\end{align}
\end{subequations}
This is similar to the hydrogenic atom in 2D~\cite{Yang_PRA1991}, with the difference being that the dipolar potential leads to a stronger confinement to shorter distances than the Coulomb potential, i.e., the wavefunctions are concentrated at $r<d$.

Apart from the two-body bound states, we might wonder about the scattering properties of the interlayer dipolar potential~\eqref{eq:dipolar}. In particular, one can show~\cite{Klawunn-Santos_PRA2010} that there is only a very restricted parameter regime of scattering energies $E\lesssim E_0$ and values of $U_0\sim 1$ where the scattering properties of the dipolar potential  recover those of a short-range contact-like attractive potential. 
This, as also discussed in Ref.~\cite{Klawunn-Santos_PRA2010}, is due to the fact that for $U_0<1$ the potential leads to a very shallow bound $1s$ state, while for $U_0>1$ there are additional states becoming bound. We  
discuss these aspects in App.~\ref{app:phase-sh}, where we explicitly evaluate the scattering phase shift within the variable-phase method.

\begin{figure}
    \centering
    \includegraphics[width=1.0\columnwidth]{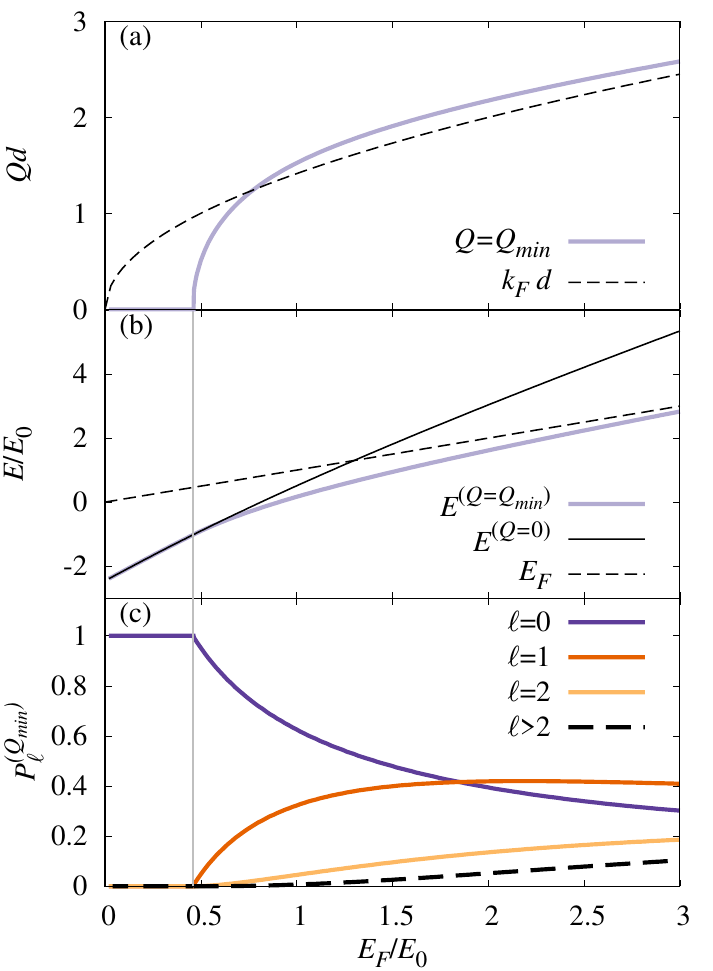}
    \caption{Spontaneous appearance of a finite dimer center-of-mass momentum when increasing the density of fermions in layer $1$ for $U_0=5$.
    (a) Values of the center-of-mass momentum $Q_{min}$ corresponding to the lowest dimer energy. 
    (b) Dimer ground state energy for $Q=Q_{min}$ (purple solid line) and $Q=0$ (black solid line). The dashed line is the energy of the normal state, $E_F$. 
    The (gray) vertical line indicates the threshold for $Q_{min} \ne 0$.   
    (c) Probability~\eqref{eq:mol_probability} that the dimer lowest eigenstate at $Q_{min}$ contains an orbital angular momentum component $\ell$.
    }
    \label{fig:En-and-Qmin-vs-EF}
\end{figure}
%
\subsection{Finite $E_F$ and finite-$Q$ dimers}
\label{sec:finite}
Similarly to the case of other attractive interaction potentials, the dimer ground state spontaneously acquires a finite center-of-mass momentum $Q_{min}$ above a threshold density of fermions~\cite{Parish_EPL11,Parish-Levinsen_PRA2013,Cotlet_PRB20,Tiene_PRR20}. 
Such a dimer state 
can be regarded as the extreme imbalance limit of the FFLO phase in spin-imbalanced  superconductors~\cite{FuldeFerrell_PR64, Larkin_64}. Over 
the last few decades there has been significant interest in studying this inhomogeneous superfluid phase in a variety of physical systems -- see, e.g., recent reviews~\cite{Casalbuoni_RMP04, Radzihovsky_RPP10}.
The possibility of generating such a phase has already been studied in Ref.~\cite{Lee-Shlyapnikov_PRA2017} for
fermionic dipolar molecules in a bilayer geometry with a finite imbalance of layer
densities. Here, it was found that, when the imbalance exceeds a critical value, the system undergoes a transition from
a uniform interlayer superfluid phase to the FFLO phase and that this phase is  enhanced by the long-range character of the interlayer dipolar interaction. Indeed, it has previously been shown that unscreened Coulomb interactions significantly stabilizes the FFLO phase~\cite{Parish_EPL11}.

In this section, we consider this problem from the perspective of the extremely 
imbalanced limit, with a single particle in layer $2$. We show in Fig.~\ref{fig:En-and-Qmin-vs-EF} the spontaneous formation of a dimer state with finite center-of-mass momentum 
for the  specific case of $U_0=5$. In this case, we observe that the $1s$ dimer state has a minimum at $Q=0$ for $E_F \le 0.46 E_0$. For larger densities of the Fermi sea in layer $1$, the lowest energy solution is for $Q=Q_{min} \ne 0$, with energy $E^{(Q_{min})} < E^{(Q=0)}$, in which case the orbital angular momentum ceases to be a good quantum number. For the range of $E_F$ studied in this work ($E_F \lesssim 6 E_0$), we do not observe any unbinding of the finite $Q$ dimer state, 
i.e., we find that the dimer energy remains below the energy $E_F$ of the normal state $|N\rangle = \hat{c}_{\0,2}^{\dag} \hat{c}_{\k_F,1}^{\dag} |FS \rangle$.

As also commented in Ref.~\cite{Lee-Shlyapnikov_PRA2017}, the long-range nature of the dipole-dipole interaction can mix different angular momentum components of the paired state and thus enhance the FFLO regime.
We evaluate  the probability that the lowest energy dimer eigenstate contains an orbital angular momentum component $\ell$ as
\begin{equation}
    P_{\ell}^{(Q_{min})} = \int_{k_F}^{\infty} \Frac{k\,dk}{2\pi} |\tilde{\eta}_{k\ell}^{(n=1,+,Q_{min})}|^2\; .
    \label{eq:mol_probability}
\end{equation}
We plot $P_{\ell}^{(Q_{min})}$ as a function of $E_F$ in Fig.~\ref{fig:En-and-Qmin-vs-EF}(c). When $Q=0$, the lowest energy state is $1s$ because $\ell$ is a good quantum number. However, for $E_F> 0.46E_0$ where the lowest dimer state develops a minimum at $Q_{min} \ne 0$, the lowest eigenvalue is characterized by several orbital angular momentum components $\ell$. As the transition from $Q=0$ to finite $Q_{min}$ is second order, the $s$ component ($\ell=0$) dominates close to the transtion. For larger values of $E_F$, the dimer state acquires first a $p$-wave component  ($\ell=1$) and, later, a smaller contribution from  the $\ell=2$, as well as $\ell>2$ components. 

In the following section, we illustrate how the properties of the ground and excited states of the two-body  bound dimers  affect the polaron properties, including its spectral function.

\section{Polaron}
\label{sec:polaron}
We now analyze the spectral properties of the polaron formed by the  
dipolar impurity in layer $2$ which is dressed by particle-hole excitations of the Fermi sea of dipoles in layer $1$. To this end, we employ a variational ansatz~\cite{Chevy_PRA2006} describing a polaron with zero center-of-mass momentum $Q=0$ as the superposition between the bare impurity weighted by the variational parameter $\phi_0^{}$ and a single particle-hole excitation, described by $\phi_{\k\q}^{}$:
\begin{equation}
     |P_3\rangle = 
     \left(\phi_0^{} \hat{c}_{\0,2}^{\dag} + \sum_{\k \q} \phi_{\k\q}^{} \hat{c}^\dag_{\q-\k,2} \hat{c}^\dag_{ \k,1} \hat{c}_{\q,1}^{} \right) |FS \rangle\; .
\label{eq:P3}
\end{equation}
Here, we use a notation where $\k$ is the momentum of the particle states ($k>k_F$) and $\q$ of the hole states ($q<k_F$). The polaron state is  normalized so that $1 = |\phi_0^{} |^2 + \sum_{\k\q} |\phi_{ \k\q}^{}|^2$. 

The variational ansatz in Eq.~\eqref{eq:P3} has previously been successfully employed to describe the impurity problem in different contexts, including ultracold atoms~\cite{Zollner2011,Parish_PRA11,Schmidt_PRA2012,Ngampruetikorn2012,Parish-Levinsen_PRA2013} and doped semiconductors~\cite{Sidler_2016,Efimkin_PRB_2021,Tiene_PRB2022,Huang_PRX23}.
In the case of contact interactions, truncating the dressing of the Fermi sea to a single particle-hole excitation has been demonstrated to be an excellent approximation, with an almost exact cancellation of higher order contributions~\cite{Combescot2008}. 

Previous work on the dipolar case~\cite{Klawunn-Recati_PRA2013} employed 
a $T$-matrix approach to evaluate the lowest attractive polaron branch energy, but the entire spectrum of excitations and the important role played by excited dimer states have not previously been considered. Furthermore, we find that there are important qualitative differences between the variational ansatz~\eqref{eq:P3} and the $T$-matrix approach. This is unlike the case of a pure contact interaction, where the variational ansatz in Eq.~\eqref{eq:P3}  is completely equivalent to the 
$T$-matrix approach within a ladder approximation~\cite{Combescot2007}. This highlights the important role played by the longer-range parts of the dipolar interaction potential, despite the dipole-dipole interaction formally corresponding to a short-range interaction 
in the sense that one can define an asymptotic region and an associated 2D scattering length~\cite{baranov2011}.   
We discuss in App.~\ref{app:Tmatrix} the precise relation between the ansatz~\eqref{eq:P3} and the $T$-matrix approach employed in Ref.~\cite{Klawunn-Recati_PRA2013}.

By minimizing the expectation value $\langle P_3 | (\hat{H} - E)| P_3 \rangle$ with respect to the variational parameters $\phi_0^{}$ and $\phi_{ \k\q}^{}$, the polaron spectral properties can be found by solving the eigenvalue problem
\begin{subequations}
\begin{align} 
\label{eq:EigProb_phi0}
    E \phi_0^{} &= \sum_{\k, \q} V_{|\k-\q|}\phi_{ \k\q}^{} \\
\nonumber
    E \phi_{ \k\q}^{} &=  E_{\k\q} \phi_{ \k\q}^{}  + V_{|\k-\q|}\phi_0^{}\\
    &+ \sum_{\k'} V_{|\k-\k'|} \phi_{ \k'\q}^{} - \sum_{\q'} V_{|\q-\q'|} \phi_{ \k\q'}^{} 
    \; ,
    \label{eq:EigProb_phikq}
\end{align}
\label{eq:eigen}%
\end{subequations}
where $E_{\k\q} = \epsilon_{\k} - \epsilon_{\q} + \epsilon_{\q-\k}$ and where we have omitted the terms that are zero because $V_\0=0$. 
The polaron spectral function $A(\omega)$ can be evaluated as usual from the impurity Green's function $G(\omega)$~\cite{Mahan2000book}:
\begin{subequations}
\begin{align}
\label{eq:imp_Gf}
    G(\omega) &= \sum_n \Frac{|\phi^{(n)}_0|^2}{\omega - E^{(n)}+i\eta}\\
    A(\omega) &= -\Frac{1}{\pi} \Im G(\omega)\; .
\label{eq:imp_sf}%
\end{align}
\label{eq:pol_spectral}
\end{subequations}

Similarly to the dimer problem, it is profitable to project the eigenvalue problem~\eqref{eq:eigen} onto an eigenbasis of the orbital angular momentum. This simplifies the numerical solution considerably, because we will see that the polaron branches are characterized by a small number of
partial waves $\ell$. Thus, we consider the following Fourier series of the dimer-hole wavefunctions:
\begin{equation*}
        \phi_{kq\varphi}^{} = \sum_{\ell\in \mathbb{Z}} e^{i \ell \varphi} \tilde{\phi}_{kq\ell}^{}
\; .
\end{equation*}
The eigenvalue equations in~\eqref{eq:eigen} now read:
\begin{subequations}
    \begin{align}
        E \phi_0^{} &= \int \frac{k\,dk}{2\pi} \frac{q\,dq}{2\pi}\sum_{\ell\in \mathbb{Z}} \tilde{V} (k,q,\ell) \tilde{\phi}_{kq\ell}^{} \\
\nonumber
    E \tilde{\phi}_{kq\ell}^{} &=  2\epsilon_\k \tilde{\phi}_{kq\ell}^{} -\Frac{kq}{2m} \left(\tilde{\phi}_{kq\ell-1}^{} + \tilde{\phi}_{kq\ell+1}^{}\right)\\
\nonumber
    &+ \tilde{V} (k,q,\ell) \phi_0^{}
%
    + \int \frac{ k'dk'}{2\pi} \tilde{V} (k,k',\ell) \tilde{\phi}_{k'q\ell}^{}   \\ 
    &- \int \frac{q'dq' }{2\pi} \tilde{V} (q,q',\ell) \tilde{\phi}_{kq'\ell}^{} \; ,
    \end{align}  
\end{subequations}
where we have employed Eq.~\eqref{eq:diagonal}. 
As for the dimer wavefunction, 
it is convenient to introduce symmetric and antisymmetric solutions for the exchange $\ell \mapsto -\ell$:
\begin{equation}
    \tilde{\phi}_{kq\ell}^{(\pm)} = \Frac{\tilde{\phi}_{kq\ell}^{} \pm \tilde{\phi}_{kq,-\ell}^{}}{2}\; ,
\label{eq:pol-m-symm}    
\end{equation}
where $\tilde{\phi}_{kq0}^{(+)} = \tilde{\phi}_{kq0}^{}$ and  $\tilde{\phi}_{kq0}^{(-)} =0$. Restricting to $\ell\ge 0$, the eigenvalue equations become
\begin{subequations}
\begin{align}  
    E \phi_0^{} &=  \int \frac{k\,dk}{2\pi}\frac{q\,dq}{2\pi}  \sum_{\ell \geq 0} (2-\delta_{\ell,0}) \tilde{V}(k,q,\ell)\tilde{\phi}_{kq\ell}^{(+)}  \\
\nonumber
    E \tilde{\phi}_{kq\ell}^{(\pm)} &=  2\epsilon_{\k} \tilde{\phi}_{kq\ell}^{(\pm)}-\frac{kq}{2m}\left( \tilde{\phi}_{kq|\ell-1|}^{(\pm)} + \tilde{\phi}_{kq\ell+1}^{(\pm)} \right)  \\
    &+ \delta_{\pm,+}\tilde{V}(k,q,\ell)\phi_0^{} 
    %
    + \int \frac{dk'k'}{2\pi} V(k,k',\ell) \phi_{k'q\ell}^{(\pm)} \nonumber \\ 
    &- \int \frac{dq'q'}{2\pi} V(q,q',\ell) \phi_{kq'\ell}^{(\pm)}  \; .
\end{align}
\label{eq:eigen_m_pm}%
\end{subequations}
We thus find that the equations for symmetric and antisymmetric dimer-hole wavefunctions decouple. Most importantly, the impurity wavefunction $\phi_0^{}$ only couples to the symmetric dimer-hole wavefunction $\tilde{\phi}_{kq\ell}^{(+)}$. Thus, the antisymmetric solutions will lead to additional eigenvalues for the polaron problem but these will all have zero spectral weight. As such, we can halve the degrees of freedom in the problem by neglecting the antisymmetric sector.

By discretizing the eigenvalue equations on a Gauss-Legendre grid in $k\in (k_F,\infty)$ and $q\in (0,k_F)$ and including a finite number of values $\ell$, we numerically diagonalize Eq.~\eqref{eq:eigen_m_pm} to find eigenvalues and eigenvectors and thus evaluate the polaron spectral function according to Eq.~\eqref{eq:pol_spectral}. We have checked that all our results are numerically converged with respect to the number of points in the momentum grids as well as in the number of $\ell$ values. It is interesting to note that when $\omega<0$ we need only very few, around 4, values of $\ell$ to obtain converged results in the considered  parameter regime, while to describe the scattering states we find that 
$9$ values of $\ell$ are sufficient. 
\begin{figure}
    \centering
    \includegraphics[width=1.0\columnwidth]{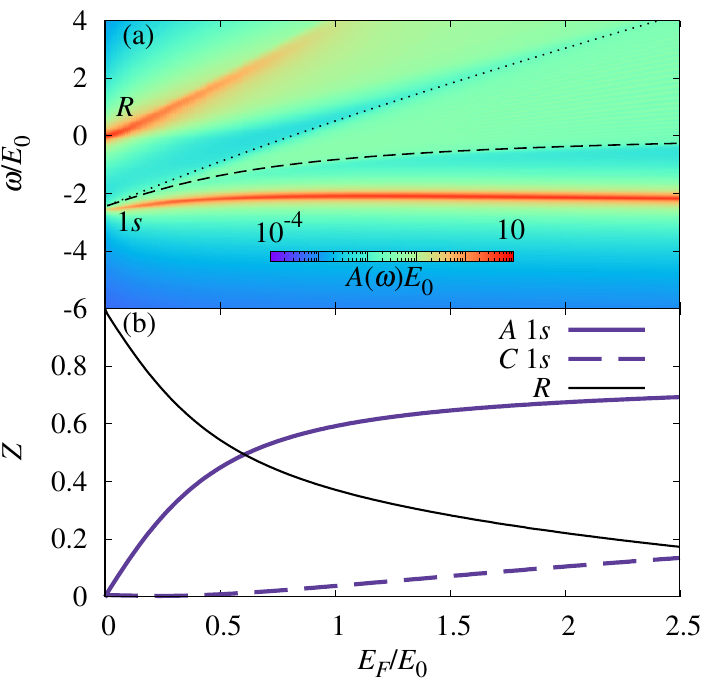}
    \caption{Polaron spectral properties for $U_0=5$. (a) 
    Spectral function $A(\omega)$ as a function of $E_F$ and energy.
    The (black) dashed and dotted lines are the boundaries of the dimer-hole continuum (see text).
    The broadening has been fixed to $\eta = 0.05 E_0$. 
    (b) Evolution of the polaron spectral weights $Z$ as a function of $E_F$ for the attractive ($A$) $1s$ branch, its continuum ($C$), and the repulsive ($R$) branch.}
    \label{fig:polaronU05}
\end{figure}
%
\subsection{Spectral properties}
\begin{figure}    \centering
    \includegraphics[width=1.0\columnwidth]{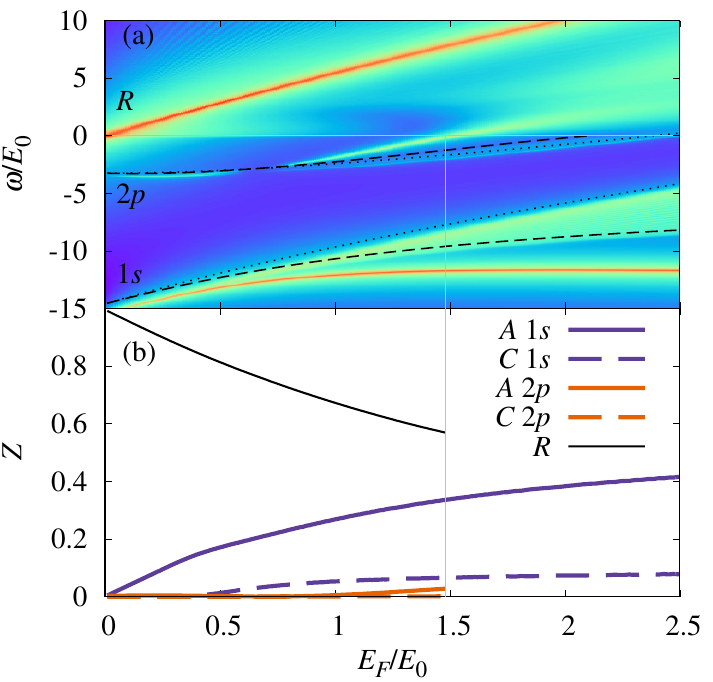}
    \caption{Polaron spectral properties for $U_0=15$. (a) Dipolar Fermi polaron spectrum as a function of $E_F$ and energy.
    The (black) dashed and dotted lines are the boundaries of the $1s$ and $2p$ dimer-hole continua (see text). 
    The broadening has been fixed to $\eta = 0.05 E_0$. 
    (b) Evolution of the polaron spectral weights $Z$ as a function of $E_F$ for the attractive ($A$) $1s$ and $2p$ branches, their continua ($C$), and the polaron repulsive ($R$) branch. The vertical gray line indicates when the $A\ 2p$ branch enters into the continuum at $\omega=0$.}
    \label{fig:polaronU015_Map}
\end{figure}

\begin{figure}
    \centering
    \includegraphics[width=1.0\columnwidth]{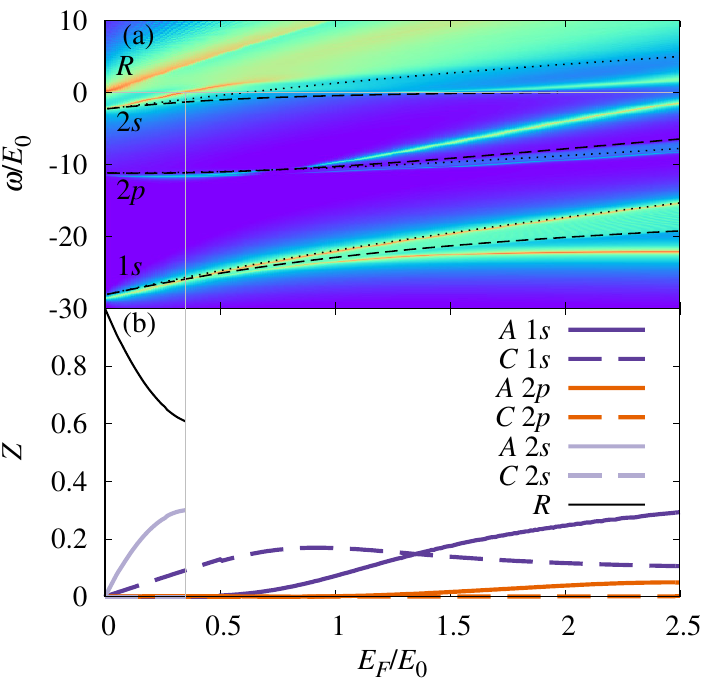}
    \caption{Polaron spectral properties for $U_0=24$. (a) Dipolar Fermi polaron spectrum as a function of $E_F$ and energy.
    The (black) dashed and dotted lines are the boundaries of the $1s$, $2p$, and $2s$ dimer-hole continua (see text). 
    The broadening has been fixed to $\eta = 0.05 E_0$. 
    (b) Evolution of the polaron spectral weights $Z$ as a function of $E_F$ for the attractive ($A$) $1s$, $2p$, and $2s$ branches, their continua ($C$), and the polaron repulsive ($R$) branch. The vertical gray line indicates when the $A\ 2s$ branch enters into the continuum at $\omega=0$.}
    \label{fig:polaronU024_Map}
\end{figure}

We plot the polaron spectral function as a function of energy and Fermi gas density in Figs.~\ref{fig:polaronU05},~\ref{fig:polaronU015_Map}, and~\ref{fig:polaronU024_Map}, for three increasing values of $U_0$. We see that there are strong qualitative differences between the spectra, which originate from the different numbers and types of dimer bound states. We now go through these in detail.

Let us first discuss the case $U_0=5$, corresponding to Fig.~\ref{fig:polaronU05}, for which only the $1s$ two-body dimer state is bound (see Fig.~\ref{fig:molecular_en}). 
The spectrum in Fig.~\ref{fig:polaronU05}(a) is characterized by two polaron branches and a continuum of states. For $\omega<0$, the attractive polaron branch recovers the energy of the $1s$ dimer state in the limit $E_F\to 0$, and we thus label it as the $1s$ attractive branch. This resonance is well separated from the 
continuum, and its energy coincides with the lowest energy-eigenvalue of Eq.~\eqref{eq:eigen_m_pm}. 
The polaron spectrum 
also exhibits another resonance at $\omega\ge 0$, the repulsive polaron, which corresponds to a continuum of states 
rather than being characterized by a single eigenvalue. 
When $E_F\to 0$, the repulsive polaron recovers the energy of the bare impurity at rest, $\omega=0$, while it blueshifts when $E_F$ increases. 
Simultaneously, it broadens and loses spectral weight.
In between the attractive and repulsive branches is a continuum of states where the hole in the $\hat{c}^\dag_{\q-\k,2} \hat{c}^\dag_{ \k,1} \hat{c}_{\q,1}^{}$ complex of the polaron state~\eqref{eq:P3} 
is unbound, while the dimer is bound. The energy of such a state is that of a dimer with center-of-mass momentum $\q$ and a hole at momentum $\q$, and hence the boundaries of this dimer-hole continuum are $E^{(q=0)}$ and $E^{(q=k_F)}-E_F$, respectively, where the energy is computed using ~Eq.~\eqref{eq:symm-a}~\footnote{Note that we only plot the dimer energy at finite center of mass $E^{(q=k_F)}$ corresponding to the symmetric states under the exchange $\ell \mapsto -\ell$ in  Eq.~\eqref{eq:symm-a} since these are the only ones with finite spectral weight. Instead, for zero center of mass momentum, symmetric and antisymmetric solution are degenerate in energy.}. Both energies recover $E_{1s}$ when $E_F\to 0$, as expected.

\begin{figure}
    \centering
    \includegraphics[width=1.0\columnwidth]{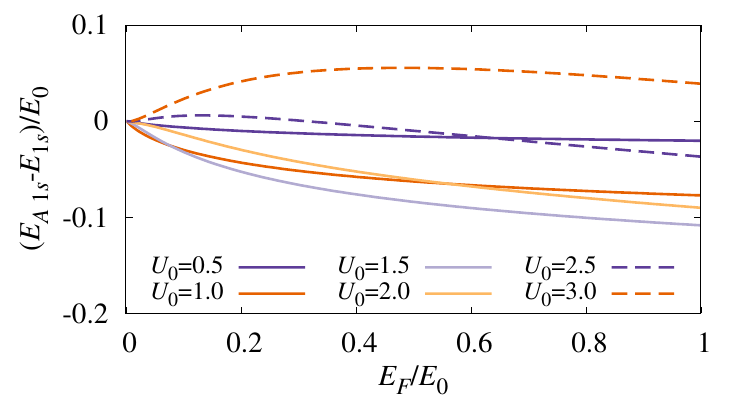}
    \caption{Energy of the $1s$ attractive branch measured from the ground-state energy of the vacuum dimer state, $E_{1s}$, as a function of $E_F$ for different values of $U_0$.}
\label{fig:EA1s_diffU0}
\end{figure}

A distinctive feature of the spectrum in Fig.~\ref{fig:polaronU05}(a) is that the $1s$ attractive branch blueshifts (increases its energy) as we increase $E_F$. This is in contrast to the case of contact interactions~\cite{Schmidt_2012,Ngampruetikorn2012} where the attractive branch always redshifts (lowers its energy) with increasing $E_F$.
The origin of this qualitatively new feature is that the dipole-dipole scattering can be significant at low momenta [see Fig.~\ref{fig:potentials}(b)] and therefore the hole scattering can be strongly enhanced relative to particle scattering despite the reduced phase space. As also discussed in App.~\ref{app:Tmatrix}, we can specifically trace the difference to the term $- \sum_{\q'} V_{|\q-\q'|} \phi_{ \k\q'}^{}$ in Eq.~\eqref{eq:eigen}, which is not present in the previous $T$-matrix treatment of the dipolar polaron problem~\cite{Klawunn-Recati_PRA2013}. This term is negligible for $U_0 \lesssim 2$, in which case the $1s$ attractive branch redshifts when $E_F$ increases, but it becomes important for larger values of $U_0$. The change from redshift to blueshift with increasing $U_0$ is shown in Fig.~\ref{fig:EA1s_diffU0}, where we plot the evolution with $E_F$ of the energy of the $1s$ attractive branch measured from the vacuum dimer energy, $E_{1s}$.

Figure~\ref{fig:polaronU05}(b) shows the spectral weight $Z$ for each of the branches, which is defined as the area of the spectral function under the corresponding peak. Because the eigenvectors of the polaron problem form a complete basis, the spectral function satisfies  the sum rule $\int_{-\infty}^{\infty}d\omega A(\omega)=1 $, and thus the total spectral weight is always 1 for all interaction strengths and densities. When $E_F\sim 0$, the spectral weight belongs entirely to the repulsive polaron branch, which coincides with the non-interacting impurity. When $E_F$ increases, we see that $Z$ is transferred mostly to the attractive branch, first linearly and then sublinearly. Only a small part of the spectral weight is transferred to the dimer-hole continuum.

\begin{figure}
    \centering
    \includegraphics[width=1.0\columnwidth]{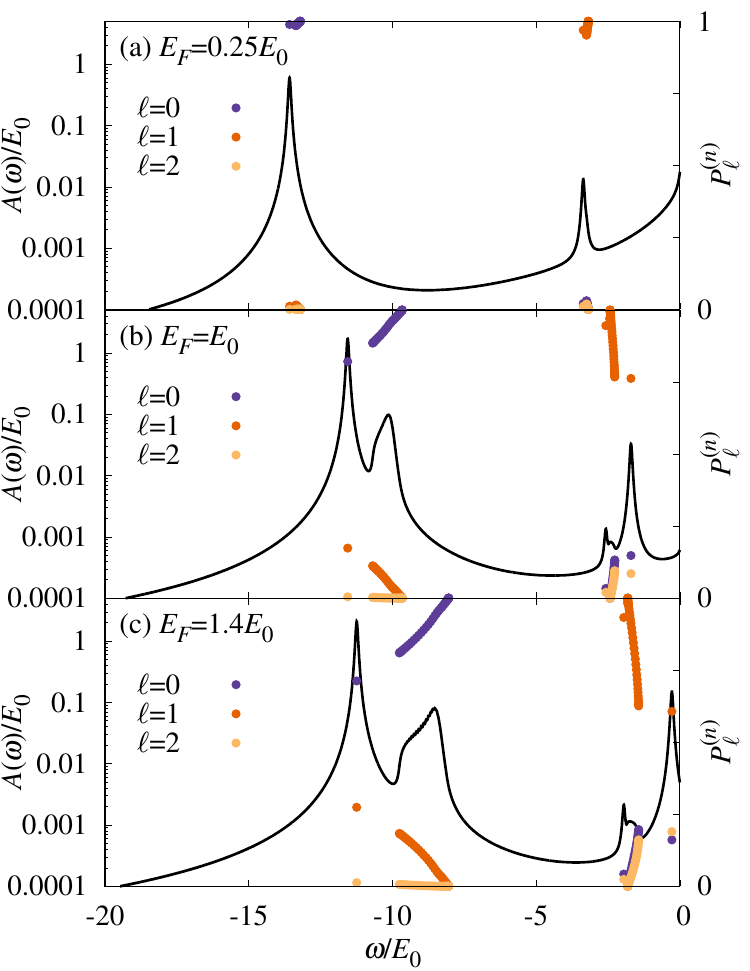}
    \caption{Polaron spectral function as a function of energy for three different values of the Fermi energy $E_F$ and for $U_0=15$.
    Symbols (values on the right $y$-axis) are the probabilities $P_{\ell}^{(n)}$ that the dimer polaron component has a value of the orbital angular momentum equal to $\ell=0, 1, 2$ for a given eigenvalue $n$~\eqref{eq:eigen_m_pm}.}
    \label{fig:polaronU015_Sec}
\end{figure}

In Fig.~\ref{fig:polaronU015_Map} we show the case of stronger interactions $U_0=15$ at which also the dimer state 2$p$  is bound (see Fig.~\ref{fig:molecular_en}). 
Now, the impurity spectrum changes qualitatively from the previous case because there are two attractive polaron branches: when $E_F\to 0$, one recovers, as before, the energy of the  $1s$ dimer state, $E_{1s}$, while the second branch recovers the energy of the $2p$ dimer state, $E_{2p}$. 
Here, we label the attractive polaron resonances by $1s$ and $2p$ even though their orbital angular momentum components evolve with $E_F$ and change in character (see below). 
Both attractive resonances have an associated dimer-hole continuum: as before, the boundaries of these are the energies of the dimer states $E^{(q=0)}$ and $E^{(q=k_F)}-E_F$~\footnote{Note that only for the $1s$ state do we always have $E^{(q=k_F)}-E_F<E^{(q=0)}$.}. %
While the $1s$ dimer-hole continuum is higher in energy than the $1s$ attractive branch, we find that this is not necessarily the case for the $2p$ branch, which for $E_F\gtrsim E_0$ 
clearly appears 
above its dimer-hole continuum (for $E_F<E_0$, the spectral weight is very small and it is difficult to distinguish it from its dimer-hole continuum).
Both the $2p$ attractive branch and the continuum blueshift in energy for increasing $E_F$ and, eventually, move into the continuum of the $q=0$ dimer state  at $\omega=0$. 

Figure~\ref{fig:polaronU015_Map}(a) also shows how the spectral weight of the repulsive branch is predominantly transferred to the $s$-wave-like attractive branch and how, compared with the case of $U_0=5$ in Fig.~\ref{fig:polaronU05}, the 
spectral weight is transferred more slowly 
because the $1s$ state is deeper bound. As a result, the repulsive branch remains brighter and broadens more slowly for increasing  
$E_F$ than in the previous case. This is further analyzed in Fig.~\ref{fig:polaronU015_Map}(b), where we see that the $1s$ attractive branch spectral weight grows linearly with $E_F$ for small $E_F$, while the $2p$ attractive branch spectral weight grows sublinearly and remains very small. Note that, with increasing density, the $2p$ attractive branch enters the continuum (vertical gray line in Fig.~\ref{fig:polaronU015_Map}), in which case the criterion that we employ to evaluate the spectral weight of both the attractive and repulsive branches becomes inapplicable due to the rapid broadening of both modes. 
In Fig.~\ref{fig:polaronU015_Map}(b), we thus stop plotting their spectral weights 
at this point.

To gain further insight into the angular momentum structure of the polarons, we can evaluate the probability that  the hole in a given dimer-hole eigenstate $n$ has an orbital angular momentum component $\ell$~\cite{Tiene_PRB2022}:
\begin{equation}
    P_{\ell}^{(n)} = \frac{1}{1 - |\phi_0^{}|^2} \int \Frac{k\,dk}{2\pi} \Frac{q\,dq}{2\pi}|\tilde{\phi}_{kq\ell}^{(+,n)}|^2  \; .
\label{eq:m-prob}
\end{equation}
Because only the symmetric $+$ states~\eqref{eq:pol-m-symm} have a finite spectral weight, the probability satisfies $P_{-\ell}^{(n)} = P_{\ell}^{(n)}$. Further, the probability is normalized such that $\sum_{\ell\ge 0} P_{\ell}^{(n)} = 1$. 
In Fig.~\ref{fig:polaronU015_Sec}, we plot the polaron spectral function for $U_0=15$ for different values of $E_F$ together with  the probability $P_{\ell}^{(n)}$ for $\ell=0,1,2$ as a function of $\omega$ (with the dots corresponding to discrete eigenvalues). One can clearly see that, for $E_F<E_0$, the eigenvalues $n$ corresponding to the $1s$ attractive branch and its continuum are predominantly $s$-wave, i.e., $P_{0}^{(n)}\sim 1$, while $P_{\ell>0}^{(n)}\sim 0$, and the eigenvalues $n$ corresponding to the $2p$ attractive branch and its continuum are predominantly $p$-wave, i.e. $P_{1}^{(n)}\sim 1$, while $P_{\ell\ne 1}^{(n)}\sim 0$. However, for larger values of $E_F$, there is an evolution of these probabilities, where the $1s$ branch acquires a $p$-wave component, while the $2p$ branch acquires both $s$- and $d$-wave components. 
As far as the dimer-hole continuum is concerned, as already discussed in Sec.~\ref{sec:molecule}, the dimer state at zero center of mass corresponds to a single value of the orbital angular momentum $\ell$. This can be clearly seen in Fig.~\ref{fig:polaronU015_Sec}, where we recognize the energies of the dimer at zero center of mass $E^{(q=0)}$ limiting the dimer-hole continua, as those having $P_\ell^{n}=1$, with either $\ell=0$ or $\ell=1$. However, for the other boundary of the dimer-hole continua, $E^{(q=k_F)}-E_F$, the dimer state at finite center-of-mass momentum is generally not an eigenstate of the orbital angular momentum and involves a mixture of $\ell$ values that evolve with $E_F$.

\begin{figure}
    \centering
    \includegraphics[width=1.0\columnwidth]{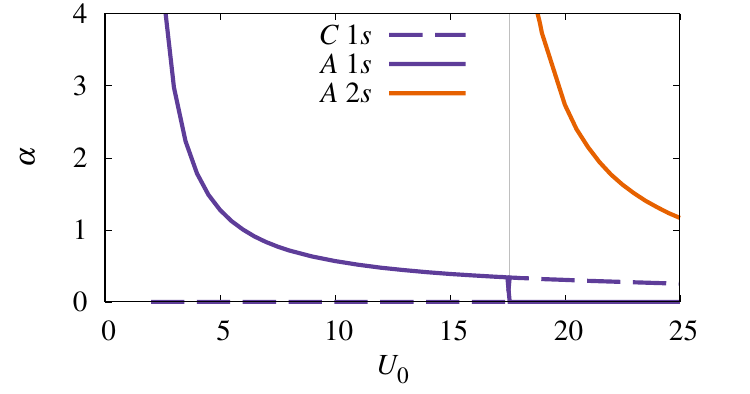}
    \caption{Growth rate of the spectral weight at small densities $\alpha =dZ/d(E_F/E_0)_{E_F=0}$ of the $1s$ and $2s$ attractive (A) branches and of the $1s$ continuum (C) as a function of the dipolar interaction strength $U_0$. The (gray) vertical lines indicates the binding threshold for the $2s$ state $U_{0 2s} \simeq 17.8$ (see Fig.~\ref{fig:molecular_en}).}
    \label{fig:growth}
\end{figure}

In Fig.~\ref{fig:polaronU024_Map} we plot the spectrum for strong dipolar interactions, $U_0=24$, for which also the $2s$ dimer state becomes bound.
Here, the repulsive branch very quickly transfers its spectral weight to the nearby $2s$ attractive branch --- see  Fig.~\ref{fig:polaronU024_Map}(b). 
The behavior of the $2p$ attractive branch is very similar to the case $U_0=15$ analyzed previously. On the other hand, the $1s$ attractive branch now becomes deeply bound and, differently from before, it is the associated dimer-hole continuum that gains spectral weight first, with the $1s$ attractive branch 
only overcoming the spectral weight of the dimer-hole continuum for $E_F \gtrsim E_0$.

The strong distinction between the growth rate of the $1s$ state in the presence and absence of the $2s$ state leads us to define the spectral weight growth rate at small $E_F$,
$\alpha = dZ/d(E_F/E_0)|_{E_F=0}$.  
We show the results for the $1s$ and $2s$ attractive branches in Fig.~\ref{fig:growth} as a function of $U_0$. While for $U_0<17.8$, the $1s$ attractive branch spectral weight grows linearly with $E_F$, when the $2s$ attractive branch appears for $U_0>17.8$, the $1s$ attractive branch spectral weight grows sublinearly and it is instead the spectral weight of its continuum that grows linearly with $E_F$.  
\begin{figure}
    \centering
    \includegraphics[width=1.0\columnwidth]{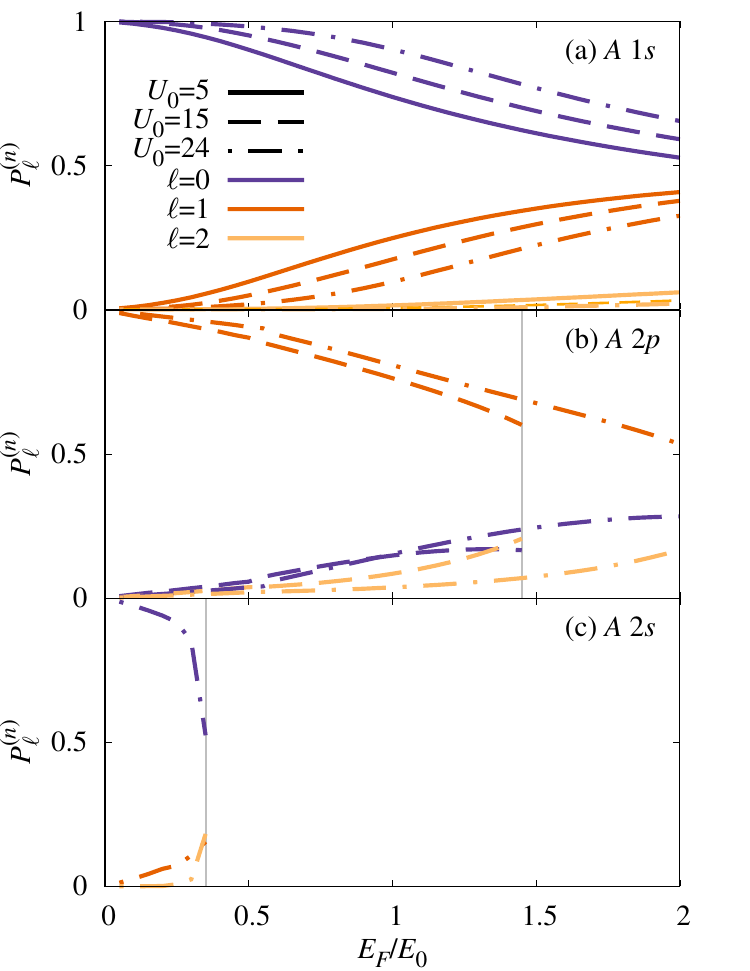}
    \caption{Probability $P_{\ell}^{(n)}$ that the dimer polaron component has a value of the orbital angular momentum equal to $\ell=0, 1, 2$ for those eigenvalue indices $n$ corresponding to  the attractive $1s$ (a), $2p$ (b), and $2s$ (c) branches. The vertical (gray) line in panel (b) indicates when the $A\ 2p$ branch enters the continuum for $U_0=15$ (see Fig.~\ref{fig:polaronU015_Map}), and the one in panel (c) when the $A\ 2s$ branch enters the continuum for $U_0=24$ (see Fig.~\ref{fig:polaronU024_Map}).}
\label{fig:AngComp_vs_EF_U05}
\end{figure}

In Fig.~\ref{fig:AngComp_vs_EF_U05}, we plot the evolution of the probability $P_\ell^{(n)}$ defined in Eq.~\eqref{eq:m-prob} as a function of $E_F$ for those eigenvalues $n$ that correspond to the 
$1s$, $2p$, and $2s$ attractive branches, for the three different values of $U_0$ that correspond to Figs.~\ref{fig:polaronU05},~\ref{fig:polaronU015_Map}, and~\ref{fig:polaronU024_Map}, respectively.  We see that the $1s$ attractive branch orbital character is predominantly $s$-wave at low density, as expected, before becoming more $p$-wave with increasing $E_F$. However, for the attractive $2p$ branch, even though the $p$-wave character dominates over the entire $E_F$ interval studied, there is an exchange with both $s$- and $d$-wave components when $E_F$ increases. Finally, the $2s$ attractive branch quickly loses its $s$-wave character towards both the $p$- and $d$-wave components, before merging with the continuum at $\omega>0$. 
We can in general see that, 
as $U_0$ increases, the exchange of angular momentum components is slower due to the attractive branches moving further apart in energy.

Finally, we note that an alternative way of presenting the polaron spectral properties is one where energy scales are rescaled by $E_F$ and length scales by $1/k_F$. In this case, we fix the dipolar interaction parameter $U_F$~\eqref{eq:new-units} and study the polaron spectrum by varying an additional dimensionless parameter, such as $E_0/E_F$ or $1/k_Fd$:
\begin{equation*}
    \Frac{1}{k_Fd} = \sqrt{\Frac{E_0}{2E_F}} = \frac{U_0}{U_F}\; .
\end{equation*}

\begin{figure}
    \centering
    \includegraphics[width=1.0\columnwidth]{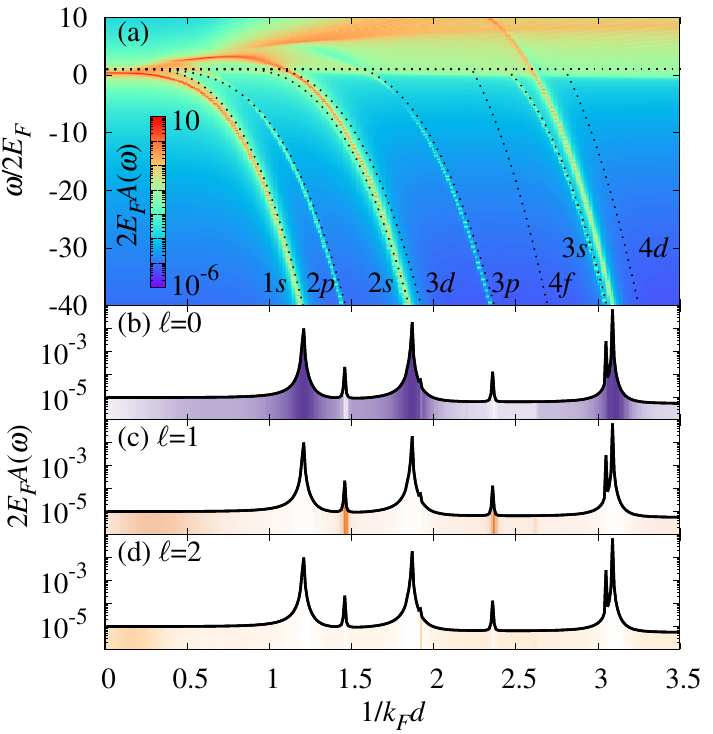}
    \caption{Polaron spectral properties for $U_F=15\sqrt{2}$. (a) Dipolar Fermi polaron spectrum as a function of $1/k_Fd$ and energy.
    The (black) dotted lines are the dimer energies at zero momentum, $E^{(q=0)}$, from the $1s$ state up to the $4d$ one --- unbinding of the zero-momentum dimer occurs at $\omega=2E_F$ (horizontal dotted line). 
    The broadening has been fixed to $\eta = 0.05\times2E_F$. 
    Panels (b-d) are the spectral functions at $\omega = -40 \times 2E_F$ as a function of $1/k_Fd$ and the color maps are the fractions of the spectral function with angular momentum $\ell$, $A_\ell(\omega)/A(\omega)$~\eqref{eq:mom-sf} for $\ell=0,1,2$.}
    \label{fig:Map_UF21}
\end{figure}
We  plot in Fig.~\ref{fig:Map_UF21}  the impurity spectral function obtained by fixing $U_F$ and increasing $1/k_Fd$. In the units previously employed, this corresponds to simultaneously increasing $U_0$ and decreasing $E_F/E_0$, which we see 
leads to the binding of an increasing number of dimer states. As shown in Fig.~\ref{fig:Map_UF21}(a), the branches with stronger spectral weight are those associated with the $s$-wave dimer states. 
In panels (b), (c), and (d) of Fig.~\ref{fig:Map_UF21} we plot the spectral function at a fixed value of energy $\omega<0$ as a function of $1/k_Fd$. As a color map, we plot the fraction of the spectral function with angular momentum $\ell$, defined as the ratio between the angular momentum weighted spectral function
\begin{equation}
    A_\ell(\omega) = -\frac{1}{\pi} \Im \left[\sum_n   \Frac{P_{\ell}^{(n)} |\phi^{(n)}_0|^2}{\omega - E^{(n)}+i\eta} \right]\; ,
\label{eq:mom-sf}
\end{equation}
and $A(\omega)$ --- note that $\sum_{\ell\ge 0} A_\ell(\omega) = A(\omega)$. This allows us to identify the dominant orbital angular momentum component $\ell$ of each attractive branch, which we see is strongly correlated with the angular momentum of the corresponding bound dimer state.

\begin{figure}
    \centering
    \includegraphics[width=1.0\columnwidth]{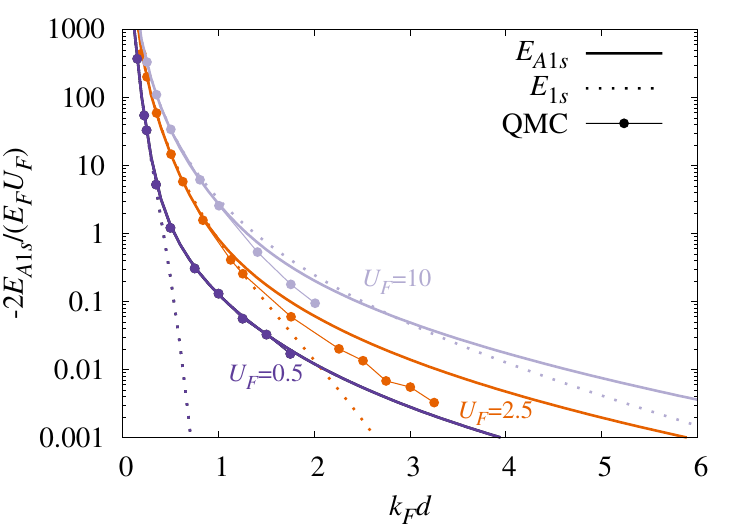}
    \caption{Energy of the  $1s$ attractive branch evaluated as a function of $k_F d$ for different values of $U_F$ (solid lines). We compare our results with those obtained in Ref.~\cite{Matveeva-Giorgini_PRL2013} by QMC methods (symbols) and with the energy of the vacuum dimer state $E_{1s}$ (dotted lines).}
    \label{fig:comp_Matveeva-Giorgini}
\end{figure}

Making use of these units, we compare in Fig.~\ref{fig:comp_Matveeva-Giorgini} the results for the energy of the $1s$ attractive branch obtained within the polaron Ansatz~\eqref{eq:P3} with those obtained in Ref.~\cite{Matveeva-Giorgini_PRL2013} by quantum Monte Carlo (QMC) methods for different values of $U_F$. For $U_F=0.5$, there is excellent agreement between the two theories for any value of $k_Fd$, suggesting that, within this regime, disregarding intralayer interactions is a reliable approximation.
For larger values of $U_F$, perfect agreement is observed only at small $k_Fd$ values, where the polaron energy closely matches that of the vacuum dimer state. For increasing $k_Fd$ values, where layer $1$ approaches the transition to a broken symmetry phase~\cite{Parish-Marchetti_PRL2012,Matveeva2012}, we find that deviations increase, although the qualitative behavior remains similar. 

%

\section{Tunneling rate and impurity spectral function}
\label{sec:observable}
\begin{figure}
    \centering
    \includegraphics[width=0.7\columnwidth]{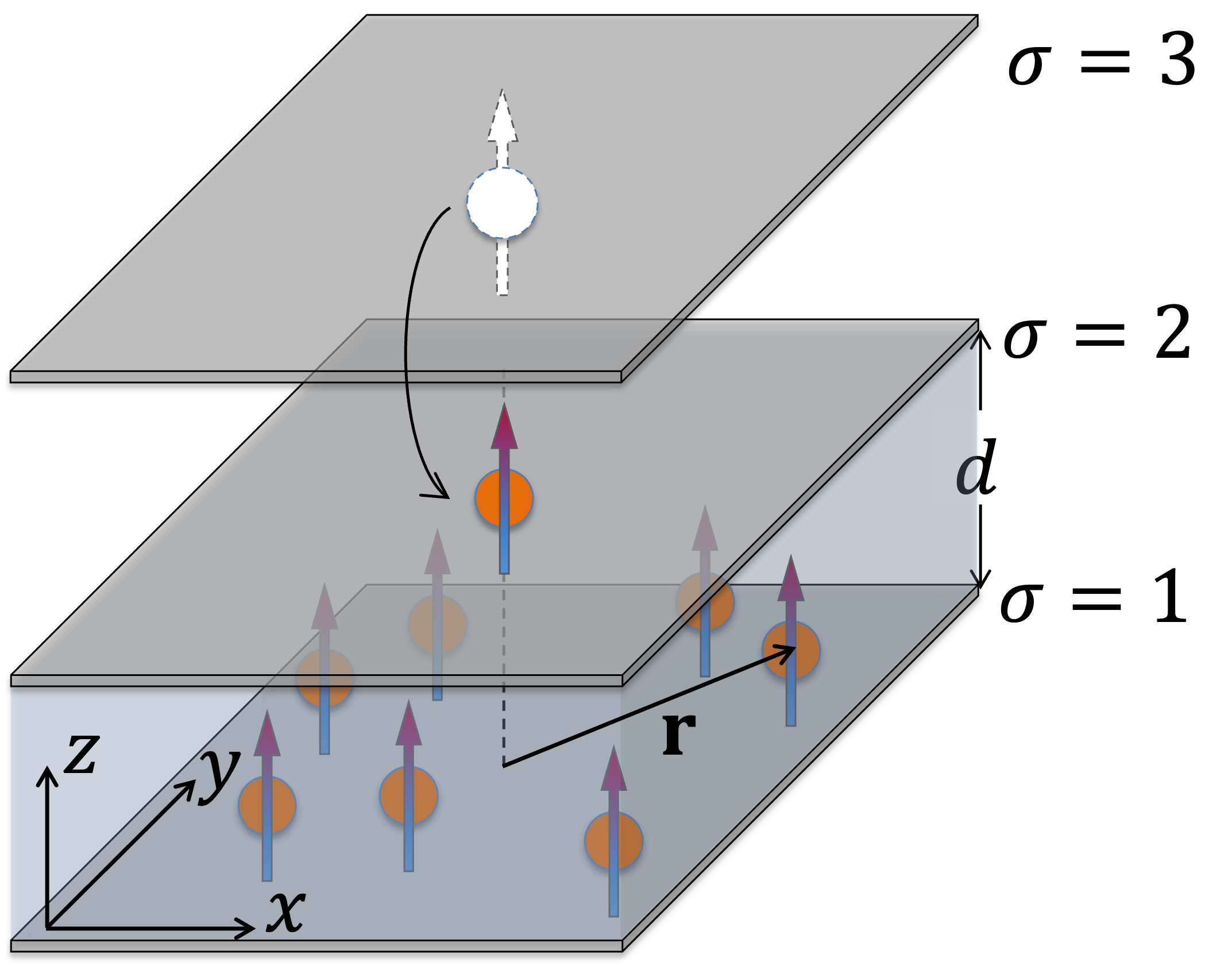}
    \caption{Illustration of how the impurity spectral function can be probed via tunneling from an auxiliary distant layer $\sigma=3$ to the impurity layer $\sigma=2$. Tunneling is suppressed between  layers $1$ and $2$ by a barrier (blue filled region).}
\label{fig:tunneling}
\end{figure}

We now discuss how one can experimentally probe the spectral function. Our proposal is inspired by radiofrequency spectroscopy~\cite{Punk2007}, where one can inject (eject) the impurity by driving transitions from (to) an auxiliary hyperfine state that, in the ideal case, does not interact with the medium. Similarly, we suggest to use an auxiliary layer ($\sigma=3$) that the impurity can tunnel into/from  --- see Fig.~\ref{fig:tunneling}. We furthermore assume that interactions between the dipolar particles can occur only between layers $1$ and $2$, while tunneling can only occur between layers $2$ and $3$. This situation can, e.g., be achieved if a potential barrier is present between layers $1$ and $2$, while layer $3$ is further away from both layers. We also note that, in practice, a small 1-3 interaction can be taken into account by considering the initial state to be a weakly dressed attractive polaron, similarly to what is routinely done in the case of RF spectroscopy for contact interactions, see, e.g., Ref.~\cite{Baym2007}.

We thus have to extend the Hamiltonian~\eqref{eq:hamil} to include  two additional terms, $\hat{H}_3$ and $\hat{H}_t$, that describe, respectively, the kinetic energy of the particles in layer $3$, and the tunneling between layers 2 and $3$:
\begin{subequations}
\label{eq:auxham}
    \begin{align}
        \hat{H}_3 &= \sum_\k (\epsilon_\k + \Delta) \hat{c}_{\k,3}^\dag \hat{c}_{\k,3}^{}\\
        \hat{H}_t &= t \sum_\k \hat{c}_{\k,3}^\dag \hat{c}_{\k,2}^{} + \text{h.c.}\; .
    \end{align}
\end{subequations}
Note that in the kinetic term $\hat{H}_3$ we have included a ``detuning" energy $\Delta$, which represents the difference between the energy minima of the confining potentials in the $z$-direction.

The additional terms in the Hamiltonian exactly match those used in radiofrequency spectroscopy on impurities~\cite{Weizhe_PRA2020}, where $\Delta$ plays the role of the radiofrequency detuning, and $t$ the role of the Rabi coupling. Therefore, this auxiliary layer provides access to the spectral properties. To be specific, within linear response the tunneling rate can be evaluated using Fermi's golden rule between an initial state consisting of one particle in layer $3$ plus a Fermi sea of dipoles in layer $1$, $\hat{c}_{\0,3}^\dag |FS\rangle$, while the final state is the polaron state $|P_3\rangle$~\eqref{eq:P3}. This gives 
\begin{multline}
    \Gamma_{3 \mapsto 2}(\Delta) 
    = 2\pi t^2 \sum_n |\langle P_3^{(n)}| \hat{c}_{\0,2}^{\dag} |FS \rangle|^2 \delta(\Delta - E^{(n)})\\
    = - 2\pi t^2 \Frac{1}{\pi} \Im G(\Delta) = 2\pi t^2 A(\Delta) \; .
\end{multline}
Thus, measuring the tunneling rate as a function of the energy difference $\Delta$ between the two lowest eigenenergies of the $z$-confining lattice potentials is equivalent to measuring the polaron spectral function $A(\omega)$~\eqref{eq:imp_sf} evaluated in this work.

\section{Conclusions and perspectives}
\label{sec:concl} 
In this work we have studied a gas of dipolar fermions in a bilayer geometry in the limit of extreme imbalance, i.e., a single dipole in one layer interacting with a Fermi sea of dipoles in a different layer. We analyze two different yet connected solutions of this problem. We first investigate the properties of the interlayer  dimer bound state, generalized to include the Pauli blocking effect from the inert Fermi sea. We find a series of bound states, characterized by the orbital angular momentum and the principal quantum number,  binding for increasing dipolar interaction strength or decreasing bilayer distance.  
For the dimer ground state, we determine the spontaneous emergence of a finite center of mass momentum when increasing $E_F$ above a threshold value. The finite momentum dimer corresponds to the large imbalance limit of the FFLO state  studied in Ref.~\cite{Lee-Shlyapnikov_PRA2017}. We find that this state has a mixed partial wave character, including $s$-, $p$-, and $d$-wave contributions. 

The other solution we analyze is the many-body polaron  state, where the presence of the impurity in one layer leads to particle-hole excitations of the Fermi sea in the other layer. We derive the polaron spectral properties by employing a single particle-hole variational ansatz, and we propose that the tunneling rate of the impurity from an additional auxiliary layer can be employed to experimentally access the spectrum. We find that the polaron spectrum is characterized by a series of attractive polaron branches which we trace back to the dimer bound states. At small $E_F$, the polaron energies and their orbital character recover those of the dimer states. However, both energies and orbital angular momentum components evolve and interchange with $E_F$. We find that the energy of the ground-state $1s$ polaron branch evolves with $E_F$ in a qualitative different way depending on the value of $U_0$. We explain this distinctive property of finite-range dipole-dipole interactions in terms of whether hole scattering is negligible or not in the polaron formation. 
We characterize the transfer of oscillator strength from the repulsive branch to the series of attractive branches in terms of their partial wave character and their distance in energy from the repulsive branch.

In our model, we neglect the repulsive interactions between particles in the Fermi sea. 
As such, we neglect the possibility of strong intralayer correlations that could lead, at very low temperatures, to the spontaneous appearance of density modulated phases such as stripes~\cite{Parish-Marchetti_PRL2012} and Wigner crystals~\cite{Matveeva2012}, which are predicted to occur 
for either strong enough dipolar interactions or large enough Fermi densities.
While beyond the scope of our study, an exciting perspective of our work is the generalization of the polaron formalism to include the possibility of the impurity interacting with such 
strongly correlated phases~\cite{Matveeva-Giorgini_PRL2013},  
which could potentially leave signatures in the polaron spectral 
response. 
Indeed, this would mirror very recent experiments on exciton polarons in doped 2D semiconductor monolayers which have probed strongly correlated states of 2D electron gases, such as Wigner crystals~\cite{Smolenski_Nature2021, Shimazaki_PRX2021}, fractional quantum Hall states in proximal graphene layers~\cite{Popert_NanoLett2022} and correlated-Mott states of electrons in a moiré superlattice~\cite{Schwartz_science2021}.

Another interesting perspective of our work would involve considering a configuration where the alignment of the dipoles is tilted at a slight angle relative to the normal direction. In this case, the anisotropy induced by the dipole-dipole interaction results in a distorted Fermi surface~\cite{Bruun-Taylor_PRL2008,Yamaguchi-Miyakawa_PRA2010}, consequently influencing the properties of the Fermi polaron and giving rise to spatial anisotropies, as illustrated in a three-dimensional setup by Ref.~\cite{Nishimura-Yabu_PRA2021}. Indeed, deformations 
of the Fermi surface have already been experimentally observed in a three-dimensional degenerate dipolar Fermi gas composed of Er atoms~\cite{Aikawa-Ferlaino_science2014}. 
Furthermore, a tilted configuration is expected to stabilize the FFLO state (see, e.g., Ref.~\cite{Kawamura-Ohashi_PRA2022}) since it breaks the continuous rotational symmetry and thus suppresses the pairing fluctuations that destroy FFLO long-range order~\cite{Shimahara_JPSJ1988}. 

The research data underpinning this publication can be accessed at Ref.~\cite{Tiene_PRA2024_dataset}. 

\acknowledgments
We would like to thank Stefano Giorgini for fruitful discussions and for letting us use the data from Ref.~\cite{Matveeva-Giorgini_PRL2013}. AT and FMM acknowledge financial support from the Ministerio de Ciencia e Innovaci\'on (MICINN), project No.~AEI/10.13039/501100011033 PID2020-113415RB-C22 (2DEnLight).
FMM acknowledges financial support from the Proyecto Sinérgico CAM 2020 Y2020/TCS-6545 (NanoQuCo-CM).
JL and MMP are supported through Australian Research Council Future Fellowships FT160100244 and FT200100619, respectively. JL and MMP also acknowledge support from the Australian Research Council Centre of Excellence in Future Low-Energy Electronics Technologies (CE170100039).

\appendix

\begin{figure}
    \centering
    \includegraphics[width=1.0\columnwidth]{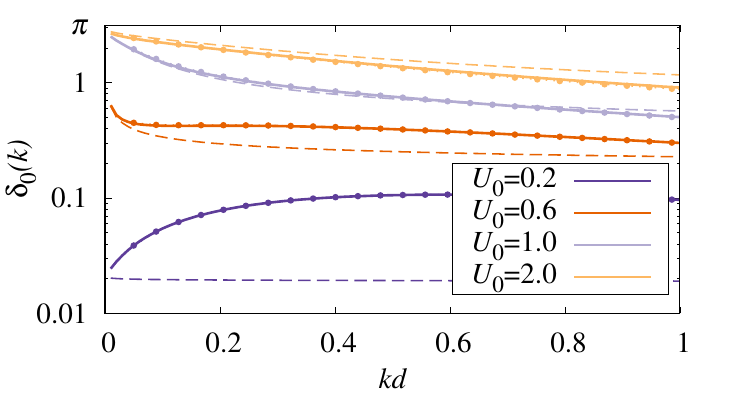}
    \caption{Phase shift $\delta_0(k)$ as a function of momentum for different values of $U_0$. Numerical solutions obtained with the variable-phase method (solid) are compared with the analytical expression~\eqref{eq:ph-Klawunn} derived in Ref.~\cite{Klawunn-Santos_PRA2010} (symbols) and the universal low-energy expression~\eqref{eq:contact} (dashed).
    \label{fig:phase-shift}}
\end{figure}
%
\section{Scattering phase shift}
\label{app:phase-sh}
In this appendix we evaluate the $s$-wave energy-dependent scattering phase shift $\delta_0(k)$, where $E=k^2/2\mu = k^2/m$, with $\mu=m/2$ being the reduced mass. 
We use the  variable-phase method~\cite{Morse1933} which allows us to evaluate the phase shift produced by a potential that vanishes for all $r>R$, i.e., we define $V_R(r)=V(r)\Theta(R-r)$. In this sense, the phase shift $\delta_\ell(R)$ can be viewed as the accumulated phase shift at position $R$ in the true potential $V(r)$. 
In 2D, $\delta_\ell(R)$ satisfies the following first-order, non-linear differential
equation~\cite{Portnoi_SSC97}:
\begin{multline}
    \Frac{d\delta_\ell(R)}{dR} = -\frac{\pi}{2} 2\mu V(R) R\\ \times \left[J_\ell(kR) \cos \delta_\ell(R) - Y_\ell(kR) \sin \delta_\ell(R)\right]^2\; ,
\label{eq:var-phase-m}    
\end{multline}
with boundary condition $\delta_\ell(0)=0$.
In this expression, $J_\ell(x)$ and $Y_\ell(x)$ are Bessel functions of the first and second kinds, respectively (the latter are also called Neumann functions), and $\ell$ is again the orbital angular momentum. Thus, for $s$-wave, $\ell=0$. Finally, the scattering phase shift in the true potential $V(r)$ is given by the limit $\delta_\ell(k) = \lim_{R\to \infty} \delta_\ell(R)$, and it is a function of the scattering energy through the momentum $k$ in Eq.~\eqref{eq:var-phase-m}.

The scattering phase shift for the interlayer dipolar potential~\eqref{eq:interd-real} has been previously evaluated in Ref.~\cite{Klawunn-Santos_PRA2010}, where the following approximate analytical expression for the $s$-wave scattering phase shift was obtained for small values of $U_0$:
\begin{equation}
    \tan \delta_0(k) \simeq \Frac{- \frac{\pi}{2} I_{JJ} (k) - \frac{\pi^2}{4} [I_{JJJY} (k) - I_{JYJJ} (k)]}{1-\frac{\pi}{2} I_{JY} (k)- \frac{\pi^2}{4} [I_{JJYY} (k) - \frac{1}{2}I_{JY}^2 (k)]}\; ,
\label{eq:ph-Klawunn}
\end{equation}
where $J \mapsto J_0(x)$, $Y \mapsto Y_0(x)$, and 
\begin{align*}
    I_{FG}(k) &= \int_{0}^{\infty} dr \,r V(r) F(kr) G(kr)\\
    I_{FGPQ}(k) &= \int_0^\infty dr \,r V(r) F(kr) G(kr)\\
    &\times \int_r^\infty ds \, s V(s) P(ks) Q(ks)\; .
\end{align*}
Using Eq.~\eqref{eq:approx2}, a small-$k$ expansion of this expression 
allows us to recover the universal low-energy expression of the phase shift for a short-range potential with a dimer state with binding energy $|E_{1s}|$:
\begin{equation}
    \cot \delta_0 (k) = \Frac{1}{\pi} \ln \left( \Frac{k^2/2 \mu}{|E_{1s}|}\right) 
    \; .
\label{eq:contact}    
\end{equation}
We note that this is the true phase shift for a contact potential; however other potentials will generally have corrections of $O(k^2)$~\cite{Levinsen2Dreview}.

In  Fig.~\ref{fig:phase-shift} we compare the numerical results for the scattering phase shift obtained by solving Eq.~\eqref{eq:var-phase-m} with the approximation~\eqref{eq:ph-Klawunn} derived in Ref.~\cite{Klawunn-Santos_PRA2010} and the universal low-energy expression in Eq.~\eqref{eq:contact}. While we see that Eq.~\eqref{eq:ph-Klawunn} is a good approximation for values of $U_0\lesssim 2$,  the phase shift  for a contact potential is a good approximation  only when 
 $kd\lesssim1$ and $U_0 \sim 1$. 
For $U_0 < 0.6$, as also argued by Ref.~\cite{Klawunn-Santos_PRA2010},  the binding energy of the $1s$ state becomes anomalously small and 
thus the expression~\eqref{eq:contact} cannot, in practice, be considered the leading term
for the low-energy scattering. However, when $U_0 \gtrsim 2$, both approximations become increasingly inaccurate since then there are other dimer states that are close to becoming bound.

\section{Relation between $T$-matrix approach and the variational ansatz}
\label{app:Tmatrix}
We now discuss the relationship between the $T$-matrix approach used to obtain the lowest energy attractive polaron branch in Ref.~\cite{Klawunn-Recati_PRA2013} and the variational ansatz~\eqref{eq:P3} that we employ in this work. 

\begin{figure}
    \centering
    \includegraphics[width=1\columnwidth]{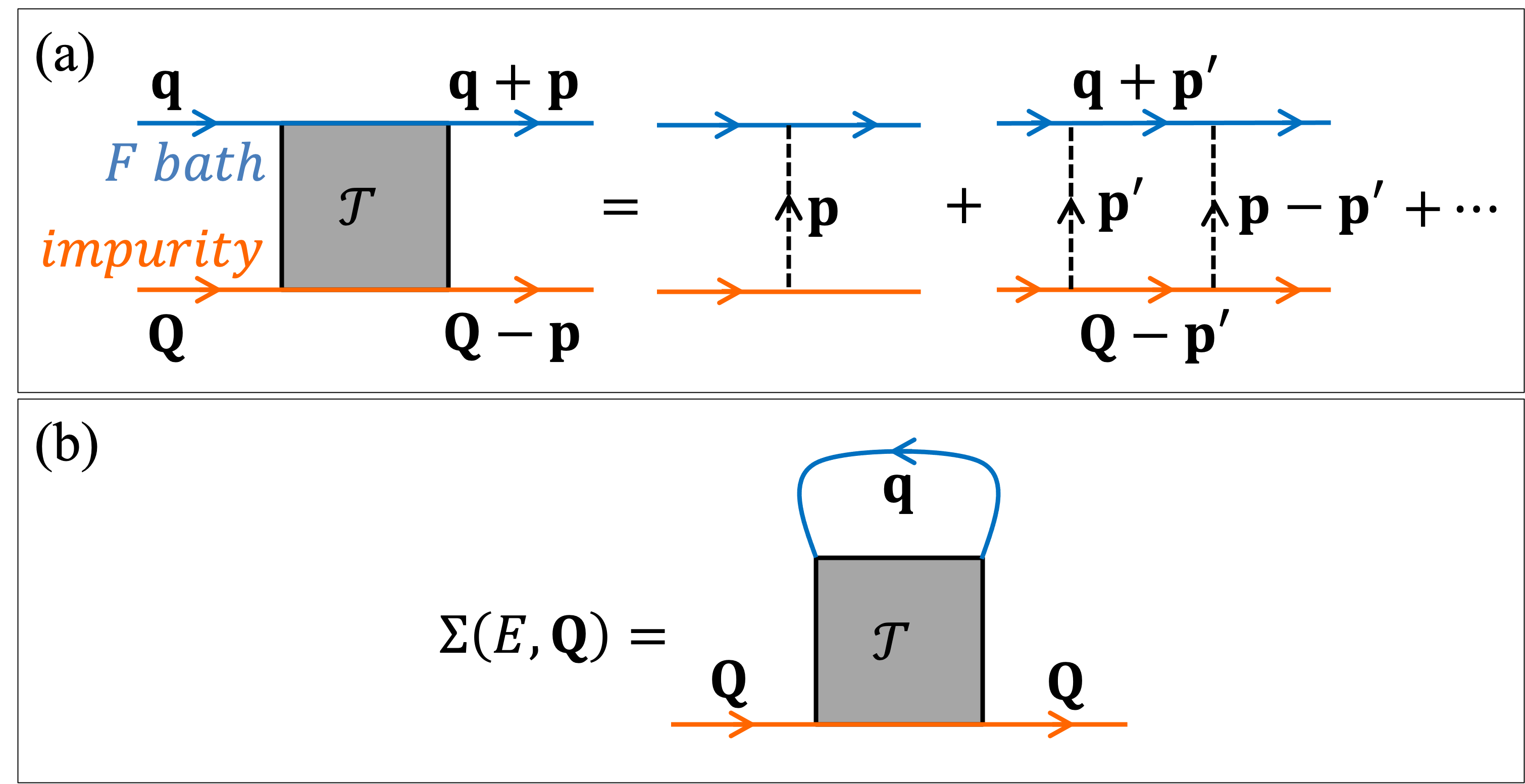}
    \caption{(a) Ladder diagrams for the $T$ matrix describing the scattering between an impurity with momentum $\Q$ and a particle of the bath with momentum $\q$, which are exchanging a momentum $\p$. (b) Impurity self-energy in terms of the $T$-matrix.}
    \label{fig:T-matrix}
\end{figure}

For a finite-range scattering potential $V_{\p}$, the $T$ matrix  
describes the scattering between an incoming impurity with momentum $\Q$ and a bath fermion with momentum $\q$ which are exchanging a momentum $\p$. By using a diagrammatic expansion within the ladder approximation (see Fig.~\ref{fig:T-matrix}(a)), all terms can be re-summed to give an implicit equation for the $T$ matrix. Considering, for simplicity, the case of an impurity with zero momentum $Q=0$, one obtains the following implicit equation for the $T$ matrix:
\begin{equation}
    \mathcal{T}_{\q\p}(E)  = V_p
    + \sum_{\p'} \Frac{(1-f_{\q+\p'}) V_{|\p-\p'|} \mathcal{T}_{\q\p'}(E)}{E + \epsilon_\q  - \epsilon_{\p'} - \epsilon_{\q+\p'}} \; .
\label{eq:T-matrix-def}
\end{equation}
Here, $f_\k$ is the Fermi-Dirac distribution, i.e., at zero temperature $f_\k = \Theta(k_F - k)$, and $E + \epsilon_\q$ is the initial energy of the scattering process.
\begin{figure}
    \centering
    \includegraphics[width=1.0\columnwidth]{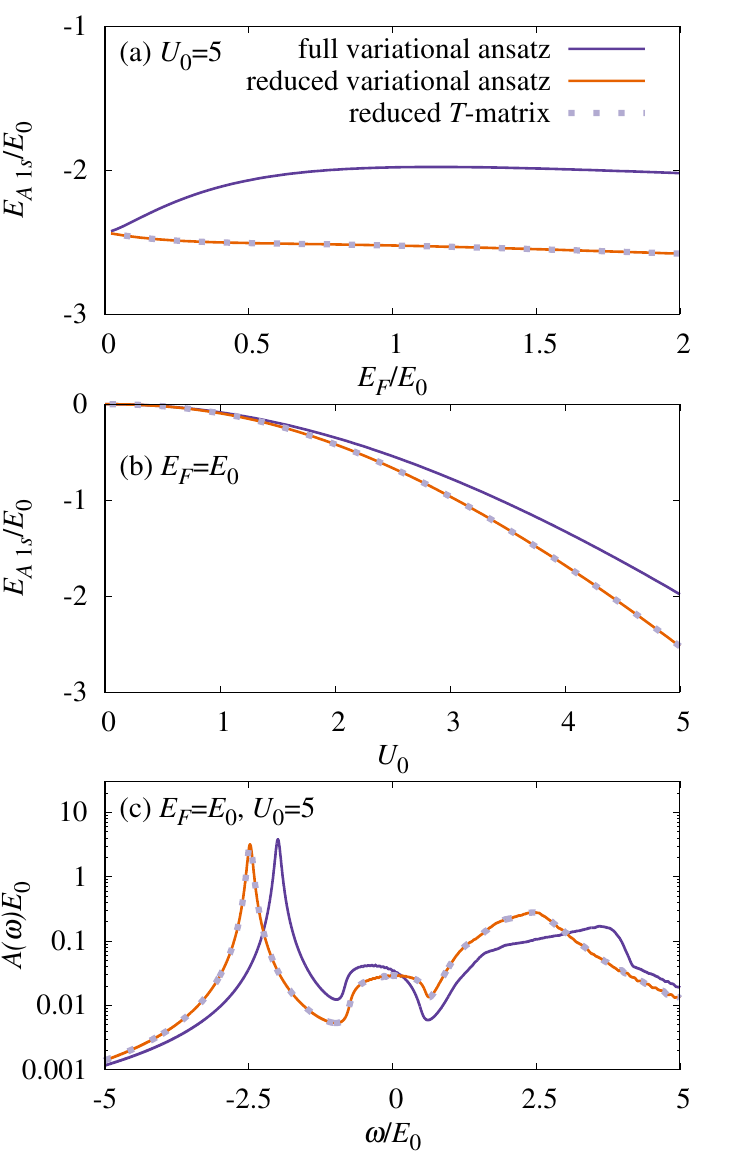}
    \caption{Polaron properties evaluated in three different ways: (1) by solving the polaron eigenvalue problem~\eqref{eq:eigen} (full variational ansatz), (2) by neglecting the $-\sum_{\q'} V_{|\q-\q'|} \phi_{ \k\q'}^{}$ term describing the scattering with holes (reduced variational ansatz), and (3) by solving the $T$-matrix equation~\eqref{eq:T-matrix-def} by inversion. 
    (a,b) Energy of the polaron $1s$ attractive branch for (a) $U_0=5$ as a function of $E_F$ and (b) $E_F=E_0$ as a function of $U_0$. (c) Entire polaron spectrum as a function of $\omega$ for $E_F=E_0$ and $U_0=5$.
    }
    \label{fig:comp-eig-T2}
\end{figure}

Starting from the $T$ matrix, one can evaluate the entire polaron  spectrum by evaluating the impurity Green's function in terms of the self-energy --- see Fig.~\ref{fig:T-matrix}(b):
\begin{subequations}
    \begin{align}
        \label{eq:imp-green} 
        G(\omega) &= \Frac{1}{\omega - \Sigma(\omega)}\;,\\
        \Sigma(\omega) &= \sum_{\q} f_{\q} \mathcal{T}_{\q,\p=\0}(\omega+i\eta)\; .
    \label{eq:Self_En}
    \end{align}
\end{subequations}
The spectral function is  then obtained from the Green's function as in Eq.~\eqref{eq:imp_sf}.

The $T$ matrix can also be obtained directly from the polaron eigenvalue equations~\eqref{eq:eigen}. Defining 
\begin{equation}
\label{eq:EigProb_Tfunc}
    \mathcal{S}_{\k\q} (E) \equiv (E -E_{\k\q}) \frac{\phi_{\k\q}^{}}{\phi_{0}^{}} \;,
\end{equation}
Eq.~\eqref{eq:EigProb_phikq} becomes
\begin{multline} 
   \mathcal{S}_{\k\q}(E) =  V_{|\k-\q|}  + \sum_{\k'} (1- f_{\k'})V_{|\k-\k'|} \frac{\mathcal{S}_{\k'\q}(E)}{E -E_{\k'\q}}\\
   - \sum_{\q'} f_{\q'} V_{|\q-\q'|} \frac{\mathcal{S}_{\k\q'}(E)}{E -E_{\k\q'}}\; .
\label{eq:S-matrix}
\end{multline}
If one neglects the last term,  this equation coincides with Eq.~\eqref{eq:T-matrix-def} by changing variable $\k'= \q+\p'$ and redefining 
\begin{equation}
    \mathcal{S}_{\q+\p',\q}(E) = \mathcal{T}_{\q\p'} (E)\; .
\end{equation}
The last term in Eq.~\eqref{eq:S-matrix} describes the scattering of the impurity with a hole of the fermionic bath and 
corresponds to  
the $-\sum_{\q'} V_{|\q-\q'|} \phi_{ \k\q'}^{}$ term in the 
polaron eigenvalue equations~\eqref{eq:eigen}.
For a contact interaction potential, the terms $-\sum_{\q'} V_{|\q-\q'|} \phi_{ \k\q'}^{}$ can be safely neglected~\cite{Chevy_PRA2006} because the phase space for hole scattering is small. However, this is not always true for a longer-range potential such as the interlayer dipolar potential. 

In order to show this, we plot in Fig.~\ref{fig:comp-eig-T2} the lowest polaron energy as well as the entire polaron spectrum, comparing the results of three methods: (1) by solving the polaron eigenvalue problem~\eqref{eq:eigen}, (2) by neglecting the term $-\sum_{\q'} V_{|\q-\q'|} \phi_{ \k\q'}^{}$ in Eq.~\eqref{eq:eigen}, and (3) by numerically solving the $T$-matrix equation~\eqref{eq:T-matrix-def} (see below). Methods (2) and (3) give exactly the same results, as they should. However, there is a non-negligible shift compared with the results obtained with method (1) if either $E_F$  is large, $E_F\gg E_0$, or if $U_0>1$. 
This coincides with the regime where the interlayer dipolar potential gives a very different scattering phase shift to that of the contact potential 
--- see App.~\ref{app:phase-sh}.
In particular, the solution of the full problem without neglecting the $-\sum_{\q'} V_{|\q-\q'|} \phi_{ \k\q'}^{}$ term is always blueshifted in energy compared to the case where one neglects the scattering of the impurity with the holes in the Fermi sea. This shift 
increases both as a function of $U_0$ and $E_F$. If $U_0$ is large, $U_0>1$, as in Fig.~\ref{fig:comp-eig-T2}, then the overall redshift of the $1s$ attractive polaron branch as a function of $E_F$ can be changed into a blueshift. 

To numerically solve for the $T$ matrix, we can assume that it depends only on a single angle, the one between $\q$ and $\p$ ($s$-wave ansatz), so that 
the implicit equation~\eqref{eq:T-matrix-def} can be easily solved by direct inversion. If we define vector indices by $i=(p, \varphi)$ and $i'=(p',\varphi')$, the $T$ matrix becomes
\begin{equation*}
    \mathcal{T}_{i} (E,q) = \sum_{i'}\left[\mathbb{I} - \mathbb{K} (E,q)\right]^{-1}_{ii'} V_{i'}\; ,
\end{equation*}
where $V_i = V_p$, and the matrix kernel is
\begin{equation*}
    \mathbb{K}_{ii'} (E,q) = \Frac{dp' p'}{2\pi} \Frac{d\varphi'}{2\pi} \Frac{(1-f_{\q+\p'}) V_{|\p-\p'|}}{E +\epsilon_\q - \epsilon_{\p'} - \epsilon_{\q+\p'}} \; .
\end{equation*}

Similarly, Eq.~\eqref{eq:S-matrix} can also  be solved by inversion, with the difference that now the vector space has a larger dimension. If we define the vector index as $i=(k,q,\varphi)$, where $\varphi$ is the angle between $\k$ and $\q$, and if we use the notation $k, k'>k_F$ and $q, q'<k_F$,  we find that the $T$ matrix $\mathcal{S}_{\k\q} (E)$ can be evaluated as 
\begin{equation}
    \mathcal{S}_i (E) = \sum_{i'} [\mathbb{I} -\tilde{\mathbb{K}} (E)+\tilde{\mathbb{W}} (E)]^{-1}_{ii'} V_{i'}\; ,
\end{equation}
where $V_{i} = V_{|\k-\q|}$, and where the two kernels are
\begin{align*}
    \tilde{\mathbb{K}}_{ii'} (E) &= \Frac{dk'k'}{2\pi} \Frac{d\varphi'}{2\pi} dq' \delta(q-q')  \frac{V_{|\k-\k'|}}{E -E_{\k'\q}}\; ,\\
    \tilde{\mathbb{W}}_{ii'} (E) &= \Frac{dq'q'}{2\pi} \Frac{d\varphi'}{2\pi} dk' \delta(k-k')  \frac{V_{|\q-\q'|}}{E -E_{\k\q'}}\; .
\end{align*}

The variational ansatz thus allows repeated impurity-hole scattering, unlike the $T$-matrix formulation. Note that this is qualitatively different from screening effects such as the Gork’ov--Melik-Barkhudarov particle-hole screening of the particle-particle correlations responsible for superfluidity in the Bose-Einstein-condensate-Bardeen-Cooper-Schrieffer (BEC-BCS) crossover~\cite{Pisani2018}. In particular, an additional excitation would be required to screen the interactions between the impurity and a fermion from the medium. This would be an interesting future direction of research, but is beyond the scope of the current work.


\begin{thebibliography}{91}%
\makeatletter
\providecommand \@ifxundefined [1]{%
 \@ifx{#1\undefined}
}%
\providecommand \@ifnum [1]{%
 \ifnum #1\expandafter \@firstoftwo
 \else \expandafter \@secondoftwo
 \fi
}%
\providecommand \@ifx [1]{%
 \ifx #1\expandafter \@firstoftwo
 \else \expandafter \@secondoftwo
 \fi
}%
\providecommand \natexlab [1]{#1}%
\providecommand \enquote  [1]{``#1''}%
\providecommand \bibnamefont  [1]{#1}%
\providecommand \bibfnamefont [1]{#1}%
\providecommand \citenamefont [1]{#1}%
\providecommand \href@noop [0]{\@secondoftwo}%
\providecommand \href [0]{\begingroup \@sanitize@url \@href}%
\providecommand \@href[1]{\@@startlink{#1}\@@href}%
\providecommand \@@href[1]{\endgroup#1\@@endlink}%
\providecommand \@sanitize@url [0]{\catcode `\\12\catcode `\$12\catcode
  `\&12\catcode `\#12\catcode `\^12\catcode `\_12\catcode `\%12\relax}%
\providecommand \@@startlink[1]{}%
\providecommand \@@endlink[0]{}%
\providecommand \url  [0]{\begingroup\@sanitize@url \@url }%
\providecommand \@url [1]{\endgroup\@href {#1}{\urlprefix }}%
\providecommand \urlprefix  [0]{URL }%
\providecommand \Eprint [0]{\href }%
\providecommand \doibase [0]{http://dx.doi.org/}%
\providecommand \selectlanguage [0]{\@gobble}%
\providecommand \bibinfo  [0]{\@secondoftwo}%
\providecommand \bibfield  [0]{\@secondoftwo}%
\providecommand \translation [1]{[#1]}%
\providecommand \BibitemOpen [0]{}%
\providecommand \bibitemStop [0]{}%
\providecommand \bibitemNoStop [0]{.\EOS\space}%
\providecommand \EOS [0]{\spacefactor3000\relax}%
\providecommand \BibitemShut  [1]{\csname bibitem#1\endcsname}%
\let\auto@bib@innerbib\@empty
\bibitem [{\citenamefont {Lahaye}\ \emph {et~al.}(2009)\citenamefont {Lahaye},
  \citenamefont {Menotti}, \citenamefont {Santos}, \citenamefont {Lewenstein},\
  and\ \citenamefont {Pfau}}]{Lahaye_RPP2009}%
  \BibitemOpen
  \bibfield  {author} {\bibinfo {author} {\bibfnamefont {T.}~\bibnamefont
  {Lahaye}}, \bibinfo {author} {\bibfnamefont {C.}~\bibnamefont {Menotti}},
  \bibinfo {author} {\bibfnamefont {L.}~\bibnamefont {Santos}}, \bibinfo
  {author} {\bibfnamefont {M.}~\bibnamefont {Lewenstein}}, \ and\ \bibinfo
  {author} {\bibfnamefont {T.}~\bibnamefont {Pfau}},\ }\bibfield  {title}
  {\bibinfo {title} {\emph {The physics of dipolar bosonic quantum gases}},\
  }\href {\doibase 10.1088/0034-4885/72/12/126401} {\bibfield  {journal}
  {\bibinfo  {journal} {Rep. Prog. Phys.}\ }\textbf {\bibinfo {volume} {72}},\
  \bibinfo {pages} {126401} (\bibinfo {year} {2009})}\BibitemShut {NoStop}%
\bibitem [{\citenamefont {Baranov}\ \emph {et~al.}(2012)\citenamefont
  {Baranov}, \citenamefont {Dalmonte}, \citenamefont {Pupillo},\ and\
  \citenamefont {Zoller}}]{Baranov-Zoller_ChemReviews2012}%
  \BibitemOpen
  \bibfield  {author} {\bibinfo {author} {\bibfnamefont {M.~A.}\ \bibnamefont
  {Baranov}}, \bibinfo {author} {\bibfnamefont {M.}~\bibnamefont {Dalmonte}},
  \bibinfo {author} {\bibfnamefont {G.}~\bibnamefont {Pupillo}}, \ and\
  \bibinfo {author} {\bibfnamefont {P.}~\bibnamefont {Zoller}},\ }\bibfield
  {title} {\bibinfo {title} {\emph {Condensed Matter Theory of Dipolar Quantum
  Gases}},\ }\href {\doibase 10.1021/cr2003568} {\bibfield  {journal} {\bibinfo
   {journal} {Chem. Rev.}\ }\textbf {\bibinfo {volume} {112}},\ \bibinfo
  {pages} {5012} (\bibinfo {year} {2012})}\BibitemShut {NoStop}%
\bibitem [{\citenamefont {Chomaz}\ \emph {et~al.}(2022)\citenamefont {Chomaz},
  \citenamefont {Ferrier-Barbut}, \citenamefont {Ferlaino}, \citenamefont
  {Laburthe-Tolra}, \citenamefont {Lev},\ and\ \citenamefont
  {Pfau}}]{Chomaz_RPP2023}%
  \BibitemOpen
  \bibfield  {author} {\bibinfo {author} {\bibfnamefont {L.}~\bibnamefont
  {Chomaz}}, \bibinfo {author} {\bibfnamefont {I.}~\bibnamefont
  {Ferrier-Barbut}}, \bibinfo {author} {\bibfnamefont {F.}~\bibnamefont
  {Ferlaino}}, \bibinfo {author} {\bibfnamefont {B.}~\bibnamefont
  {Laburthe-Tolra}}, \bibinfo {author} {\bibfnamefont {B.~L.}\ \bibnamefont
  {Lev}}, \ and\ \bibinfo {author} {\bibfnamefont {T.}~\bibnamefont {Pfau}},\
  }\bibfield  {title} {\bibinfo {title} {\emph {Dipolar physics: a review of
  experiments with magnetic quantum gases}},\ }\href {\doibase
  10.1088/1361-6633/aca814} {\bibfield  {journal} {\bibinfo  {journal} {Rep.
  Prog. Phys.}\ }\textbf {\bibinfo {volume} {86}},\ \bibinfo {pages} {026401}
  (\bibinfo {year} {2022})}\BibitemShut {NoStop}%
\bibitem [{\citenamefont {Griesmaier}\ \emph {et~al.}(2005)\citenamefont
  {Griesmaier}, \citenamefont {Werner}, \citenamefont {Hensler}, \citenamefont
  {Stuhler},\ and\ \citenamefont {Pfau}}]{Griesmaier-Pfau_PRL2005}%
  \BibitemOpen
  \bibfield  {author} {\bibinfo {author} {\bibfnamefont {A.}~\bibnamefont
  {Griesmaier}}, \bibinfo {author} {\bibfnamefont {J.}~\bibnamefont {Werner}},
  \bibinfo {author} {\bibfnamefont {S.}~\bibnamefont {Hensler}}, \bibinfo
  {author} {\bibfnamefont {J.}~\bibnamefont {Stuhler}}, \ and\ \bibinfo
  {author} {\bibfnamefont {T.}~\bibnamefont {Pfau}},\ }\bibfield  {title}
  {\bibinfo {title} {\emph {Bose-Einstein Condensation of Chromium}},\ }\href
  {\doibase 10.1103/PhysRevLett.94.160401} {\bibfield  {journal} {\bibinfo
  {journal} {Phys. Rev. Lett.}\ }\textbf {\bibinfo {volume} {94}},\ \bibinfo
  {pages} {160401} (\bibinfo {year} {2005})}\BibitemShut {NoStop}%
\bibitem [{\citenamefont {Stuhler}\ \emph {et~al.}(2005)\citenamefont
  {Stuhler}, \citenamefont {Griesmaier}, \citenamefont {Koch}, \citenamefont
  {Fattori}, \citenamefont {Pfau}, \citenamefont {Giovanazzi}, \citenamefont
  {Pedri},\ and\ \citenamefont {Santos}}]{Stuhler_PRL2005}%
  \BibitemOpen
  \bibfield  {author} {\bibinfo {author} {\bibfnamefont {J.}~\bibnamefont
  {Stuhler}}, \bibinfo {author} {\bibfnamefont {A.}~\bibnamefont {Griesmaier}},
  \bibinfo {author} {\bibfnamefont {T.}~\bibnamefont {Koch}}, \bibinfo {author}
  {\bibfnamefont {M.}~\bibnamefont {Fattori}}, \bibinfo {author} {\bibfnamefont
  {T.}~\bibnamefont {Pfau}}, \bibinfo {author} {\bibfnamefont {S.}~\bibnamefont
  {Giovanazzi}}, \bibinfo {author} {\bibfnamefont {P.}~\bibnamefont {Pedri}}, \
  and\ \bibinfo {author} {\bibfnamefont {L.}~\bibnamefont {Santos}},\
  }\bibfield  {title} {\bibinfo {title} {\emph {Observation of Dipole-Dipole
  Interaction in a Degenerate Quantum Gas}},\ }\href {\doibase
  10.1103/PhysRevLett.95.150406} {\bibfield  {journal} {\bibinfo  {journal}
  {Phys. Rev. Lett.}\ }\textbf {\bibinfo {volume} {95}},\ \bibinfo {pages}
  {150406} (\bibinfo {year} {2005})}\BibitemShut {NoStop}%
\bibitem [{\citenamefont {Lu}\ \emph {et~al.}(2011)\citenamefont {Lu},
  \citenamefont {Burdick}, \citenamefont {Youn},\ and\ \citenamefont
  {Lev}}]{Lu_PRL2011}%
  \BibitemOpen
  \bibfield  {author} {\bibinfo {author} {\bibfnamefont {M.}~\bibnamefont
  {Lu}}, \bibinfo {author} {\bibfnamefont {N.~Q.}\ \bibnamefont {Burdick}},
  \bibinfo {author} {\bibfnamefont {S.~H.}\ \bibnamefont {Youn}}, \ and\
  \bibinfo {author} {\bibfnamefont {B.~L.}\ \bibnamefont {Lev}},\ }\bibfield
  {title} {\bibinfo {title} {\emph {Strongly Dipolar Bose-Einstein Condensate
  of Dysprosium}},\ }\href {\doibase 10.1103/PhysRevLett.107.190401} {\bibfield
   {journal} {\bibinfo  {journal} {Phys. Rev. Lett.}\ }\textbf {\bibinfo
  {volume} {107}},\ \bibinfo {pages} {190401} (\bibinfo {year}
  {2011})}\BibitemShut {NoStop}%
\bibitem [{\citenamefont {Lu}\ \emph {et~al.}(2012)\citenamefont {Lu},
  \citenamefont {Burdick},\ and\ \citenamefont {Lev}}]{Lu-Lev_PRL2012}%
  \BibitemOpen
  \bibfield  {author} {\bibinfo {author} {\bibfnamefont {M.}~\bibnamefont
  {Lu}}, \bibinfo {author} {\bibfnamefont {N.~Q.}\ \bibnamefont {Burdick}}, \
  and\ \bibinfo {author} {\bibfnamefont {B.~L.}\ \bibnamefont {Lev}},\
  }\bibfield  {title} {\bibinfo {title} {\emph {Quantum Degenerate Dipolar
  Fermi Gas}},\ }\href {\doibase 10.1103/PhysRevLett.108.215301} {\bibfield
  {journal} {\bibinfo  {journal} {Phys. Rev. Lett.}\ }\textbf {\bibinfo
  {volume} {108}},\ \bibinfo {pages} {215301} (\bibinfo {year}
  {2012})}\BibitemShut {NoStop}%
\bibitem [{\citenamefont {Aikawa}\ \emph {et~al.}(2012)\citenamefont {Aikawa},
  \citenamefont {Frisch}, \citenamefont {Mark}, \citenamefont {Baier},
  \citenamefont {Rietzler}, \citenamefont {Grimm},\ and\ \citenamefont
  {Ferlaino}}]{Aikawa_PRL2012}%
  \BibitemOpen
  \bibfield  {author} {\bibinfo {author} {\bibfnamefont {K.}~\bibnamefont
  {Aikawa}}, \bibinfo {author} {\bibfnamefont {A.}~\bibnamefont {Frisch}},
  \bibinfo {author} {\bibfnamefont {M.}~\bibnamefont {Mark}}, \bibinfo {author}
  {\bibfnamefont {S.}~\bibnamefont {Baier}}, \bibinfo {author} {\bibfnamefont
  {A.}~\bibnamefont {Rietzler}}, \bibinfo {author} {\bibfnamefont
  {R.}~\bibnamefont {Grimm}}, \ and\ \bibinfo {author} {\bibfnamefont
  {F.}~\bibnamefont {Ferlaino}},\ }\bibfield  {title} {\bibinfo {title} {\emph
  {Bose-Einstein Condensation of Erbium}},\ }\href {\doibase
  10.1103/PhysRevLett.108.210401} {\bibfield  {journal} {\bibinfo  {journal}
  {Phys. Rev. Lett.}\ }\textbf {\bibinfo {volume} {108}},\ \bibinfo {pages}
  {210401} (\bibinfo {year} {2012})}\BibitemShut {NoStop}%
\bibitem [{\citenamefont {Aikawa}\ \emph
  {et~al.}(2014{\natexlab{a}})\citenamefont {Aikawa}, \citenamefont {Frisch},
  \citenamefont {Mark}, \citenamefont {Baier}, \citenamefont {Grimm},
  \citenamefont {Bohn}, \citenamefont {Jin}, \citenamefont {Bruun},\ and\
  \citenamefont {Ferlaino}}]{Aikawa-Ferlaino_PRL2014}%
  \BibitemOpen
  \bibfield  {author} {\bibinfo {author} {\bibfnamefont {K.}~\bibnamefont
  {Aikawa}}, \bibinfo {author} {\bibfnamefont {A.}~\bibnamefont {Frisch}},
  \bibinfo {author} {\bibfnamefont {M.}~\bibnamefont {Mark}}, \bibinfo {author}
  {\bibfnamefont {S.}~\bibnamefont {Baier}}, \bibinfo {author} {\bibfnamefont
  {R.}~\bibnamefont {Grimm}}, \bibinfo {author} {\bibfnamefont {J.~L.}\
  \bibnamefont {Bohn}}, \bibinfo {author} {\bibfnamefont {D.~S.}\ \bibnamefont
  {Jin}}, \bibinfo {author} {\bibfnamefont {G.~M.}\ \bibnamefont {Bruun}}, \
  and\ \bibinfo {author} {\bibfnamefont {F.}~\bibnamefont {Ferlaino}},\
  }\bibfield  {title} {\bibinfo {title} {\emph {Anisotropic Relaxation Dynamics
  in a Dipolar Fermi Gas Driven Out of Equilibrium}},\ }\href {\doibase
  10.1103/PhysRevLett.113.263201} {\bibfield  {journal} {\bibinfo  {journal}
  {Phys. Rev. Lett.}\ }\textbf {\bibinfo {volume} {113}},\ \bibinfo {pages}
  {263201} (\bibinfo {year} {2014}{\natexlab{a}})}\BibitemShut {NoStop}%
\bibitem [{\citenamefont {Böttcher}\ \emph {et~al.}(2020)\citenamefont
  {Böttcher}, \citenamefont {Schmidt}, \citenamefont {Hertkorn}, \citenamefont
  {Ng}, \citenamefont {Graham}, \citenamefont {Guo}, \citenamefont {Langen},\
  and\ \citenamefont {Pfau}}]{Bottcher_2021}%
  \BibitemOpen
  \bibfield  {author} {\bibinfo {author} {\bibfnamefont {F.}~\bibnamefont
  {Böttcher}}, \bibinfo {author} {\bibfnamefont {J.-N.}\ \bibnamefont
  {Schmidt}}, \bibinfo {author} {\bibfnamefont {J.}~\bibnamefont {Hertkorn}},
  \bibinfo {author} {\bibfnamefont {K.~S.~H.}\ \bibnamefont {Ng}}, \bibinfo
  {author} {\bibfnamefont {S.~D.}\ \bibnamefont {Graham}}, \bibinfo {author}
  {\bibfnamefont {M.}~\bibnamefont {Guo}}, \bibinfo {author} {\bibfnamefont
  {T.}~\bibnamefont {Langen}}, \ and\ \bibinfo {author} {\bibfnamefont
  {T.}~\bibnamefont {Pfau}},\ }\bibfield  {title} {\bibinfo {title} {\emph {New
  states of matter with fine-tuned interactions: quantum droplets and dipolar
  supersolids}},\ }\href {\doibase 10.1088/1361-6633/abc9ab} {\bibfield
  {journal} {\bibinfo  {journal} {Rep. Prog. Phys.}\ }\textbf {\bibinfo
  {volume} {84}},\ \bibinfo {pages} {012403} (\bibinfo {year}
  {2020})}\BibitemShut {NoStop}%
\bibitem [{\citenamefont {Löw}\ \emph {et~al.}(2012)\citenamefont {Löw},
  \citenamefont {Weimer}, \citenamefont {Nipper}, \citenamefont {Balewski},
  \citenamefont {Butscher}, \citenamefont {Büchler},\ and\ \citenamefont
  {Pfau}}]{Loew-Pfau_JPB2012}%
  \BibitemOpen
  \bibfield  {author} {\bibinfo {author} {\bibfnamefont {R.}~\bibnamefont
  {Löw}}, \bibinfo {author} {\bibfnamefont {H.}~\bibnamefont {Weimer}},
  \bibinfo {author} {\bibfnamefont {J.}~\bibnamefont {Nipper}}, \bibinfo
  {author} {\bibfnamefont {J.~B.}\ \bibnamefont {Balewski}}, \bibinfo {author}
  {\bibfnamefont {B.}~\bibnamefont {Butscher}}, \bibinfo {author}
  {\bibfnamefont {H.~P.}\ \bibnamefont {Büchler}}, \ and\ \bibinfo {author}
  {\bibfnamefont {T.}~\bibnamefont {Pfau}},\ }\bibfield  {title} {\bibinfo
  {title} {\emph {An experimental and theoretical guide to strongly interacting
  Rydberg gases}},\ }\href {\doibase 10.1088/0953-4075/45/11/113001} {\bibfield
   {journal} {\bibinfo  {journal} {Journal of Physics B: Atomic, Molecular and
  Optical Physics}\ }\textbf {\bibinfo {volume} {45}},\ \bibinfo {pages}
  {113001} (\bibinfo {year} {2012})}\BibitemShut {NoStop}%
\bibitem [{\citenamefont {Moses}\ \emph {et~al.}(2017)\citenamefont {Moses},
  \citenamefont {Covey}, \citenamefont {Miecnikowski}, \citenamefont {Jin},\
  and\ \citenamefont {Ye}}]{Moses_NatPhys2017}%
  \BibitemOpen
  \bibfield  {author} {\bibinfo {author} {\bibfnamefont {S.}~\bibnamefont
  {Moses}}, \bibinfo {author} {\bibfnamefont {J.}~\bibnamefont {Covey}},
  \bibinfo {author} {\bibfnamefont {M.}~\bibnamefont {Miecnikowski}}, \bibinfo
  {author} {\bibfnamefont {D.}~\bibnamefont {Jin}}, \ and\ \bibinfo {author}
  {\bibfnamefont {J.}~\bibnamefont {Ye}},\ }\bibfield  {title} {\bibinfo
  {title} {\emph {New frontiers for quantum gases of polar molecules}},\ }\href
  {\doibase 10.1038/nphys3985} {\bibfield  {journal} {\bibinfo  {journal}
  {Nature Physics}\ }\textbf {\bibinfo {volume} {13}},\ \bibinfo {pages} {13}
  (\bibinfo {year} {2017})}\BibitemShut {NoStop}%
\bibitem [{\citenamefont {Shaffer}\ \emph {et~al.}(2018)\citenamefont
  {Shaffer}, \citenamefont {Rittenhouse},\ and\ \citenamefont
  {Sadeghpour}}]{Shaffer_NatComm2018}%
  \BibitemOpen
  \bibfield  {author} {\bibinfo {author} {\bibfnamefont {J.~P.}\ \bibnamefont
  {Shaffer}}, \bibinfo {author} {\bibfnamefont {S.~T.}\ \bibnamefont
  {Rittenhouse}}, \ and\ \bibinfo {author} {\bibfnamefont {H.~R.}\ \bibnamefont
  {Sadeghpour}},\ }\bibfield  {title} {\bibinfo {title} {\emph {Ultracold
  Rydberg molecules}},\ }\href {\doibase 10.1038/s41467-018-04135-6} {\bibfield
   {journal} {\bibinfo  {journal} {Nature Communications}\ }\textbf {\bibinfo
  {volume} {9}},\ \bibinfo {pages} {1965} (\bibinfo {year} {2018})}\BibitemShut
  {NoStop}%
\bibitem [{\citenamefont {Valtolina}\ \emph {et~al.}(2020)\citenamefont
  {Valtolina}, \citenamefont {Matsuda}, \citenamefont {Tobias}, \citenamefont
  {Li}, \citenamefont {De~Marco},\ and\ \citenamefont
  {Ye}}]{Valtolina_Nature2020}%
  \BibitemOpen
  \bibfield  {author} {\bibinfo {author} {\bibfnamefont {G.}~\bibnamefont
  {Valtolina}}, \bibinfo {author} {\bibfnamefont {K.}~\bibnamefont {Matsuda}},
  \bibinfo {author} {\bibfnamefont {W.~G.}\ \bibnamefont {Tobias}}, \bibinfo
  {author} {\bibfnamefont {J.-R.}\ \bibnamefont {Li}}, \bibinfo {author}
  {\bibfnamefont {L.}~\bibnamefont {De~Marco}}, \ and\ \bibinfo {author}
  {\bibfnamefont {J.}~\bibnamefont {Ye}},\ }\bibfield  {title} {\bibinfo
  {title} {\emph {Dipolar evaporation of reactive molecules to below the Fermi
  temperature}},\ }\href {\doibase 10.1038/s41586-020-2980-7} {\bibfield
  {journal} {\bibinfo  {journal} {Nature}\ }\textbf {\bibinfo {volume} {588}},\
  \bibinfo {pages} {239} (\bibinfo {year} {2020})}\BibitemShut {NoStop}%
\bibitem [{\citenamefont {Marco}\ \emph {et~al.}(2019)\citenamefont {Marco},
  \citenamefont {Valtolina}, \citenamefont {Matsuda}, \citenamefont {Tobias},
  \citenamefont {Covey},\ and\ \citenamefont {Ye}}]{DeMarco-Jun_Science2019}%
  \BibitemOpen
  \bibfield  {author} {\bibinfo {author} {\bibfnamefont {L.~D.}\ \bibnamefont
  {Marco}}, \bibinfo {author} {\bibfnamefont {G.}~\bibnamefont {Valtolina}},
  \bibinfo {author} {\bibfnamefont {K.}~\bibnamefont {Matsuda}}, \bibinfo
  {author} {\bibfnamefont {W.~G.}\ \bibnamefont {Tobias}}, \bibinfo {author}
  {\bibfnamefont {J.~P.}\ \bibnamefont {Covey}}, \ and\ \bibinfo {author}
  {\bibfnamefont {J.}~\bibnamefont {Ye}},\ }\bibfield  {title} {\bibinfo
  {title} {\emph {A degenerate Fermi gas of polar molecules}},\ }\href
  {\doibase 10.1126/science.aau7230} {\bibfield  {journal} {\bibinfo  {journal}
  {Science}\ }\textbf {\bibinfo {volume} {363}},\ \bibinfo {pages} {853}
  (\bibinfo {year} {2019})}\BibitemShut {NoStop}%
\bibitem [{\citenamefont {Tobias}\ \emph {et~al.}(2020)\citenamefont {Tobias},
  \citenamefont {Matsuda}, \citenamefont {Valtolina}, \citenamefont {De~Marco},
  \citenamefont {Li},\ and\ \citenamefont {Ye}}]{Tobias-Ye_PRL2020}%
  \BibitemOpen
  \bibfield  {author} {\bibinfo {author} {\bibfnamefont {W.~G.}\ \bibnamefont
  {Tobias}}, \bibinfo {author} {\bibfnamefont {K.}~\bibnamefont {Matsuda}},
  \bibinfo {author} {\bibfnamefont {G.}~\bibnamefont {Valtolina}}, \bibinfo
  {author} {\bibfnamefont {L.}~\bibnamefont {De~Marco}}, \bibinfo {author}
  {\bibfnamefont {J.-R.}\ \bibnamefont {Li}}, \ and\ \bibinfo {author}
  {\bibfnamefont {J.}~\bibnamefont {Ye}},\ }\bibfield  {title} {\bibinfo
  {title} {\emph {Thermalization and Sub-Poissonian Density Fluctuations in a
  Degenerate Molecular Fermi Gas}},\ }\href {\doibase
  10.1103/PhysRevLett.124.033401} {\bibfield  {journal} {\bibinfo  {journal}
  {Phys. Rev. Lett.}\ }\textbf {\bibinfo {volume} {124}},\ \bibinfo {pages}
  {033401} (\bibinfo {year} {2020})}\BibitemShut {NoStop}%
\bibitem [{\citenamefont {Schindewolf}\ \emph {et~al.}(2022)\citenamefont
  {Schindewolf}, \citenamefont {Bause}, \citenamefont {Chen}, \citenamefont
  {Duda}, \citenamefont {Karman}, \citenamefont {Bloch},\ and\ \citenamefont
  {Luo}}]{Schindewolf-Bloch_luo_Nature2022}%
  \BibitemOpen
  \bibfield  {author} {\bibinfo {author} {\bibfnamefont {A.}~\bibnamefont
  {Schindewolf}}, \bibinfo {author} {\bibfnamefont {R.}~\bibnamefont {Bause}},
  \bibinfo {author} {\bibfnamefont {X.-Y.}\ \bibnamefont {Chen}}, \bibinfo
  {author} {\bibfnamefont {M.}~\bibnamefont {Duda}}, \bibinfo {author}
  {\bibfnamefont {T.}~\bibnamefont {Karman}}, \bibinfo {author} {\bibfnamefont
  {I.}~\bibnamefont {Bloch}}, \ and\ \bibinfo {author} {\bibfnamefont {X.-Y.}\
  \bibnamefont {Luo}},\ }\bibfield  {title} {\bibinfo {title} {\emph
  {Evaporation of microwave-shielded polar molecules to quantum degeneracy}},\
  }\href {\doibase 10.1038/s41586-022-04900-0} {\bibfield  {journal} {\bibinfo
  {journal} {Nature}\ }\textbf {\bibinfo {volume} {607}},\ \bibinfo {pages}
  {677} (\bibinfo {year} {2022})}\BibitemShut {NoStop}%
\bibitem [{\citenamefont {Duda}\ \emph {et~al.}(2023)\citenamefont {Duda},
  \citenamefont {Chen}, \citenamefont {Schindewolf}, \citenamefont {Bause},
  \citenamefont {von Milczewski}, \citenamefont {Schmidt}, \citenamefont
  {Bloch},\ and\ \citenamefont {Luo}}]{Duda-Bloch-Lou_NP2023}%
  \BibitemOpen
  \bibfield  {author} {\bibinfo {author} {\bibfnamefont {M.}~\bibnamefont
  {Duda}}, \bibinfo {author} {\bibfnamefont {X.-Y.}\ \bibnamefont {Chen}},
  \bibinfo {author} {\bibfnamefont {A.}~\bibnamefont {Schindewolf}}, \bibinfo
  {author} {\bibfnamefont {R.}~\bibnamefont {Bause}}, \bibinfo {author}
  {\bibfnamefont {J.}~\bibnamefont {von Milczewski}}, \bibinfo {author}
  {\bibfnamefont {R.}~\bibnamefont {Schmidt}}, \bibinfo {author} {\bibfnamefont
  {I.}~\bibnamefont {Bloch}}, \ and\ \bibinfo {author} {\bibfnamefont {X.-Y.}\
  \bibnamefont {Luo}},\ }\bibfield  {title} {\bibinfo {title} {\emph
  {Transition from a polaronic condensate to a degenerate Fermi gas of
  heteronuclear molecules}},\ }\href {\doibase 10.1038/s41567-023-01948-1}
  {\bibfield  {journal} {\bibinfo  {journal} {Nature Physics}\ }\textbf
  {\bibinfo {volume} {19}},\ \bibinfo {pages} {720} (\bibinfo {year}
  {2023})}\BibitemShut {NoStop}%
\bibitem [{\citenamefont {Deiglmayr}\ \emph {et~al.}(2008)\citenamefont
  {Deiglmayr}, \citenamefont {Grochola}, \citenamefont {Repp}, \citenamefont
  {M\"ortlbauer}, \citenamefont {Gl\"uck}, \citenamefont {Lange}, \citenamefont
  {Dulieu}, \citenamefont {Wester},\ and\ \citenamefont
  {Weidem\"uller}}]{Deiglmayr_PRL2008}%
  \BibitemOpen
  \bibfield  {author} {\bibinfo {author} {\bibfnamefont {J.}~\bibnamefont
  {Deiglmayr}}, \bibinfo {author} {\bibfnamefont {A.}~\bibnamefont {Grochola}},
  \bibinfo {author} {\bibfnamefont {M.}~\bibnamefont {Repp}}, \bibinfo {author}
  {\bibfnamefont {K.}~\bibnamefont {M\"ortlbauer}}, \bibinfo {author}
  {\bibfnamefont {C.}~\bibnamefont {Gl\"uck}}, \bibinfo {author} {\bibfnamefont
  {J.}~\bibnamefont {Lange}}, \bibinfo {author} {\bibfnamefont
  {O.}~\bibnamefont {Dulieu}}, \bibinfo {author} {\bibfnamefont
  {R.}~\bibnamefont {Wester}}, \ and\ \bibinfo {author} {\bibfnamefont
  {M.}~\bibnamefont {Weidem\"uller}},\ }\bibfield  {title} {\bibinfo {title}
  {\emph {Formation of Ultracold Polar Molecules in the Rovibrational Ground
  State}},\ }\href {\doibase 10.1103/PhysRevLett.101.133004} {\bibfield
  {journal} {\bibinfo  {journal} {Phys. Rev. Lett.}\ }\textbf {\bibinfo
  {volume} {101}},\ \bibinfo {pages} {133004} (\bibinfo {year}
  {2008})}\BibitemShut {NoStop}%
\bibitem [{\citenamefont {Repp}\ \emph {et~al.}(2013)\citenamefont {Repp},
  \citenamefont {Pires}, \citenamefont {Ulmanis}, \citenamefont {Heck},
  \citenamefont {Kuhnle}, \citenamefont {Weidem\"uller},\ and\ \citenamefont
  {Tiemann}}]{Repp_PRA2013}%
  \BibitemOpen
  \bibfield  {author} {\bibinfo {author} {\bibfnamefont {M.}~\bibnamefont
  {Repp}}, \bibinfo {author} {\bibfnamefont {R.}~\bibnamefont {Pires}},
  \bibinfo {author} {\bibfnamefont {J.}~\bibnamefont {Ulmanis}}, \bibinfo
  {author} {\bibfnamefont {R.}~\bibnamefont {Heck}}, \bibinfo {author}
  {\bibfnamefont {E.~D.}\ \bibnamefont {Kuhnle}}, \bibinfo {author}
  {\bibfnamefont {M.}~\bibnamefont {Weidem\"uller}}, \ and\ \bibinfo {author}
  {\bibfnamefont {E.}~\bibnamefont {Tiemann}},\ }\bibfield  {title} {\bibinfo
  {title} {\emph {Observation of interspecies ${}^{6}$Li-${}^{133}$Cs Feshbach
  resonances}},\ }\href {\doibase 10.1103/PhysRevA.87.010701} {\bibfield
  {journal} {\bibinfo  {journal} {Phys. Rev. A}\ }\textbf {\bibinfo {volume}
  {87}},\ \bibinfo {pages} {010701} (\bibinfo {year} {2013})}\BibitemShut
  {NoStop}%
\bibitem [{\citenamefont {Park}\ \emph {et~al.}(2023)\citenamefont {Park},
  \citenamefont {Lu}, \citenamefont {Jamison}, \citenamefont {Tscherbul},\ and\
  \citenamefont {Ketterle}}]{Park-Ketterle_Nature2023}%
  \BibitemOpen
  \bibfield  {author} {\bibinfo {author} {\bibfnamefont {J.~J.}\ \bibnamefont
  {Park}}, \bibinfo {author} {\bibfnamefont {Y.-K.}\ \bibnamefont {Lu}},
  \bibinfo {author} {\bibfnamefont {A.~O.}\ \bibnamefont {Jamison}}, \bibinfo
  {author} {\bibfnamefont {T.~V.}\ \bibnamefont {Tscherbul}}, \ and\ \bibinfo
  {author} {\bibfnamefont {W.}~\bibnamefont {Ketterle}},\ }\bibfield  {title}
  {\bibinfo {title} {\emph {A Feshbach resonance in collisions between triplet
  ground-state molecules}},\ }\href {\doibase 10.1038/s41586-022-05635-8}
  {\bibfield  {journal} {\bibinfo  {journal} {Nature}\ }\textbf {\bibinfo
  {volume} {614}},\ \bibinfo {pages} {54} (\bibinfo {year} {2023})}\BibitemShut
  {NoStop}%
\bibitem [{\citenamefont {Tobias}\ \emph {et~al.}(2022)\citenamefont {Tobias},
  \citenamefont {Matsuda}, \citenamefont {Li}, \citenamefont {Miller},
  \citenamefont {Carroll}, \citenamefont {Bilitewski}, \citenamefont {Rey},\
  and\ \citenamefont {Ye}}]{Tobias-Ye_Science2022}%
  \BibitemOpen
  \bibfield  {author} {\bibinfo {author} {\bibfnamefont {W.~G.}\ \bibnamefont
  {Tobias}}, \bibinfo {author} {\bibfnamefont {K.}~\bibnamefont {Matsuda}},
  \bibinfo {author} {\bibfnamefont {J.-R.}\ \bibnamefont {Li}}, \bibinfo
  {author} {\bibfnamefont {C.}~\bibnamefont {Miller}}, \bibinfo {author}
  {\bibfnamefont {A.~N.}\ \bibnamefont {Carroll}}, \bibinfo {author}
  {\bibfnamefont {T.}~\bibnamefont {Bilitewski}}, \bibinfo {author}
  {\bibfnamefont {A.~M.}\ \bibnamefont {Rey}}, \ and\ \bibinfo {author}
  {\bibfnamefont {J.}~\bibnamefont {Ye}},\ }\bibfield  {title} {\bibinfo
  {title} {\emph {Reactions between layer-resolved molecules mediated by
  dipolar spin exchange}},\ }\href {\doibase 10.1126/science.abn8525}
  {\bibfield  {journal} {\bibinfo  {journal} {Science}\ }\textbf {\bibinfo
  {volume} {375}},\ \bibinfo {pages} {1299} (\bibinfo {year}
  {2022})}\BibitemShut {NoStop}%
\bibitem [{\citenamefont {Cooper}\ and\ \citenamefont
  {Shlyapnikov}(2009)}]{Cooper-Shlyapnikov_PRL2009}%
  \BibitemOpen
  \bibfield  {author} {\bibinfo {author} {\bibfnamefont {N.~R.}\ \bibnamefont
  {Cooper}}\ and\ \bibinfo {author} {\bibfnamefont {G.~V.}\ \bibnamefont
  {Shlyapnikov}},\ }\bibfield  {title} {\bibinfo {title} {\emph {Stable
  Topological Superfluid Phase of Ultracold Polar Fermionic Molecules}},\
  }\href {\doibase 10.1103/PhysRevLett.103.155302} {\bibfield  {journal}
  {\bibinfo  {journal} {Phys. Rev. Lett.}\ }\textbf {\bibinfo {volume} {103}},\
  \bibinfo {pages} {155302} (\bibinfo {year} {2009})}\BibitemShut {NoStop}%
\bibitem [{\citenamefont {Levinsen}\ \emph {et~al.}(2011)\citenamefont
  {Levinsen}, \citenamefont {Cooper},\ and\ \citenamefont
  {Shlyapnikov}}]{Levinsen-Cooper_PRA2011}%
  \BibitemOpen
  \bibfield  {author} {\bibinfo {author} {\bibfnamefont {J.}~\bibnamefont
  {Levinsen}}, \bibinfo {author} {\bibfnamefont {N.~R.}\ \bibnamefont
  {Cooper}}, \ and\ \bibinfo {author} {\bibfnamefont {G.~V.}\ \bibnamefont
  {Shlyapnikov}},\ }\bibfield  {title} {\bibinfo {title} {\emph {Topological
  ${p}_{x}+{\mathit{ip}}_{y}$ superfluid phase of fermionic polar molecules}},\
  }\href {\doibase 10.1103/PhysRevA.84.013603} {\bibfield  {journal} {\bibinfo
  {journal} {Phys. Rev. A}\ }\textbf {\bibinfo {volume} {84}},\ \bibinfo
  {pages} {013603} (\bibinfo {year} {2011})}\BibitemShut {NoStop}%
\bibitem [{\citenamefont {Bruun}\ and\ \citenamefont
  {Taylor}(2008)}]{Bruun-Taylor_PRL2008}%
  \BibitemOpen
  \bibfield  {author} {\bibinfo {author} {\bibfnamefont {G.~M.}\ \bibnamefont
  {Bruun}}\ and\ \bibinfo {author} {\bibfnamefont {E.}~\bibnamefont {Taylor}},\
  }\bibfield  {title} {\bibinfo {title} {\emph {Quantum Phases of a
  Two-Dimensional Dipolar Fermi Gas}},\ }\href {\doibase
  10.1103/PhysRevLett.101.245301} {\bibfield  {journal} {\bibinfo  {journal}
  {Phys. Rev. Lett.}\ }\textbf {\bibinfo {volume} {101}},\ \bibinfo {pages}
  {245301} (\bibinfo {year} {2008})}\BibitemShut {NoStop}%
\bibitem [{\citenamefont {Yamaguchi}\ \emph {et~al.}(2010)\citenamefont
  {Yamaguchi}, \citenamefont {Sogo}, \citenamefont {Ito},\ and\ \citenamefont
  {Miyakawa}}]{Yamaguchi-Miyakawa_PRA2010}%
  \BibitemOpen
  \bibfield  {author} {\bibinfo {author} {\bibfnamefont {Y.}~\bibnamefont
  {Yamaguchi}}, \bibinfo {author} {\bibfnamefont {T.}~\bibnamefont {Sogo}},
  \bibinfo {author} {\bibfnamefont {T.}~\bibnamefont {Ito}}, \ and\ \bibinfo
  {author} {\bibfnamefont {T.}~\bibnamefont {Miyakawa}},\ }\bibfield  {title}
  {\bibinfo {title} {\emph {Density-wave instability in a two-dimensional
  dipolar Fermi gas}},\ }\href {\doibase 10.1103/PhysRevA.82.013643} {\bibfield
   {journal} {\bibinfo  {journal} {Phys. Rev. A}\ }\textbf {\bibinfo {volume}
  {82}},\ \bibinfo {pages} {013643} (\bibinfo {year} {2010})}\BibitemShut
  {NoStop}%
\bibitem [{\citenamefont {Sun}\ \emph {et~al.}(2010)\citenamefont {Sun},
  \citenamefont {Wu},\ and\ \citenamefont {Das~Sarma}}]{Sun-DasSarma_PRB2010}%
  \BibitemOpen
  \bibfield  {author} {\bibinfo {author} {\bibfnamefont {K.}~\bibnamefont
  {Sun}}, \bibinfo {author} {\bibfnamefont {C.}~\bibnamefont {Wu}}, \ and\
  \bibinfo {author} {\bibfnamefont {S.}~\bibnamefont {Das~Sarma}},\ }\bibfield
  {title} {\bibinfo {title} {\emph {Spontaneous inhomogeneous phases in
  ultracold dipolar Fermi gases}},\ }\href {\doibase
  10.1103/PhysRevB.82.075105} {\bibfield  {journal} {\bibinfo  {journal} {Phys.
  Rev. B}\ }\textbf {\bibinfo {volume} {82}},\ \bibinfo {pages} {075105}
  (\bibinfo {year} {2010})}\BibitemShut {NoStop}%
\bibitem [{\citenamefont {Parish}\ and\ \citenamefont
  {Marchetti}(2012)}]{Parish-Marchetti_PRL2012}%
  \BibitemOpen
  \bibfield  {author} {\bibinfo {author} {\bibfnamefont {M.~M.}\ \bibnamefont
  {Parish}}\ and\ \bibinfo {author} {\bibfnamefont {F.~M.}\ \bibnamefont
  {Marchetti}},\ }\bibfield  {title} {\bibinfo {title} {\emph {Density
  Instabilities in a Two-Dimensional Dipolar Fermi Gas}},\ }\href {\doibase
  10.1103/PhysRevLett.108.145304} {\bibfield  {journal} {\bibinfo  {journal}
  {Phys. Rev. Lett.}\ }\textbf {\bibinfo {volume} {108}},\ \bibinfo {pages}
  {145304} (\bibinfo {year} {2012})}\BibitemShut {NoStop}%
\bibitem [{\citenamefont {Matveeva}\ and\ \citenamefont
  {Giorgini}(2012)}]{Matveeva2012}%
  \BibitemOpen
  \bibfield  {author} {\bibinfo {author} {\bibfnamefont {N.}~\bibnamefont
  {Matveeva}}\ and\ \bibinfo {author} {\bibfnamefont {S.}~\bibnamefont
  {Giorgini}},\ }\bibfield  {title} {\bibinfo {title} {\emph {Liquid and
  Crystal Phases of Dipolar Fermions in Two Dimensions}},\ }\href {\doibase
  10.1103/PhysRevLett.109.200401} {\bibfield  {journal} {\bibinfo  {journal}
  {Phys. Rev. Lett.}\ }\textbf {\bibinfo {volume} {109}},\ \bibinfo {pages}
  {200401} (\bibinfo {year} {2012})}\BibitemShut {NoStop}%
\bibitem [{\citenamefont {Marchetti}\ and\ \citenamefont
  {Parish}(2013)}]{Marchetti-Parish_PRB2013}%
  \BibitemOpen
  \bibfield  {author} {\bibinfo {author} {\bibfnamefont {F.~M.}\ \bibnamefont
  {Marchetti}}\ and\ \bibinfo {author} {\bibfnamefont {M.~M.}\ \bibnamefont
  {Parish}},\ }\bibfield  {title} {\bibinfo {title} {\emph {Density-wave phases
  of dipolar fermions in a bilayer}},\ }\href {\doibase
  10.1103/PhysRevB.87.045110} {\bibfield  {journal} {\bibinfo  {journal} {Phys.
  Rev. B}\ }\textbf {\bibinfo {volume} {87}},\ \bibinfo {pages} {045110}
  (\bibinfo {year} {2013})}\BibitemShut {NoStop}%
\bibitem [{\citenamefont {Klawunn}\ \emph
  {et~al.}(2010{\natexlab{a}})\citenamefont {Klawunn}, \citenamefont {Duhme},\
  and\ \citenamefont {Santos}}]{Klawunn-Duhme-Santos_PRA2010}%
  \BibitemOpen
  \bibfield  {author} {\bibinfo {author} {\bibfnamefont {M.}~\bibnamefont
  {Klawunn}}, \bibinfo {author} {\bibfnamefont {J.}~\bibnamefont {Duhme}}, \
  and\ \bibinfo {author} {\bibfnamefont {L.}~\bibnamefont {Santos}},\
  }\bibfield  {title} {\bibinfo {title} {\emph {Bose-Fermi mixtures of
  self-assembled filaments of fermionic polar molecules}},\ }\href {\doibase
  10.1103/PhysRevA.81.013604} {\bibfield  {journal} {\bibinfo  {journal} {Phys.
  Rev. A}\ }\textbf {\bibinfo {volume} {81}},\ \bibinfo {pages} {013604}
  (\bibinfo {year} {2010}{\natexlab{a}})}\BibitemShut {NoStop}%
\bibitem [{\citenamefont {Potter}\ \emph {et~al.}(2010)\citenamefont {Potter},
  \citenamefont {Berg}, \citenamefont {Wang}, \citenamefont {Halperin},\ and\
  \citenamefont {Demler}}]{Potter-Demler_PRL2010}%
  \BibitemOpen
  \bibfield  {author} {\bibinfo {author} {\bibfnamefont {A.~C.}\ \bibnamefont
  {Potter}}, \bibinfo {author} {\bibfnamefont {E.}~\bibnamefont {Berg}},
  \bibinfo {author} {\bibfnamefont {D.-W.}\ \bibnamefont {Wang}}, \bibinfo
  {author} {\bibfnamefont {B.~I.}\ \bibnamefont {Halperin}}, \ and\ \bibinfo
  {author} {\bibfnamefont {E.}~\bibnamefont {Demler}},\ }\bibfield  {title}
  {\bibinfo {title} {\emph {Superfluidity and Dimerization in a Multilayered
  System of Fermionic Polar Molecules}},\ }\href {\doibase
  10.1103/PhysRevLett.105.220406} {\bibfield  {journal} {\bibinfo  {journal}
  {Phys. Rev. Lett.}\ }\textbf {\bibinfo {volume} {105}},\ \bibinfo {pages}
  {220406} (\bibinfo {year} {2010})}\BibitemShut {NoStop}%
\bibitem [{\citenamefont {Pikovski}\ \emph {et~al.}(2010)\citenamefont
  {Pikovski}, \citenamefont {Klawunn}, \citenamefont {Shlyapnikov},\ and\
  \citenamefont {Santos}}]{Pikovski-Santos_PRL2010}%
  \BibitemOpen
  \bibfield  {author} {\bibinfo {author} {\bibfnamefont {A.}~\bibnamefont
  {Pikovski}}, \bibinfo {author} {\bibfnamefont {M.}~\bibnamefont {Klawunn}},
  \bibinfo {author} {\bibfnamefont {G.~V.}\ \bibnamefont {Shlyapnikov}}, \ and\
  \bibinfo {author} {\bibfnamefont {L.}~\bibnamefont {Santos}},\ }\bibfield
  {title} {\bibinfo {title} {\emph {Interlayer Superfluidity in Bilayer Systems
  of Fermionic Polar Molecules}},\ }\href {\doibase
  10.1103/PhysRevLett.105.215302} {\bibfield  {journal} {\bibinfo  {journal}
  {Phys. Rev. Lett.}\ }\textbf {\bibinfo {volume} {105}},\ \bibinfo {pages}
  {215302} (\bibinfo {year} {2010})}\BibitemShut {NoStop}%
\bibitem [{\citenamefont {Klawunn}\ \emph
  {et~al.}(2010{\natexlab{b}})\citenamefont {Klawunn}, \citenamefont
  {Pikovski},\ and\ \citenamefont {Santos}}]{Klawunn-Santos_PRA2010}%
  \BibitemOpen
  \bibfield  {author} {\bibinfo {author} {\bibfnamefont {M.}~\bibnamefont
  {Klawunn}}, \bibinfo {author} {\bibfnamefont {A.}~\bibnamefont {Pikovski}}, \
  and\ \bibinfo {author} {\bibfnamefont {L.}~\bibnamefont {Santos}},\
  }\bibfield  {title} {\bibinfo {title} {\emph {Two-dimensional scattering and
  bound states of polar molecules in bilayers}},\ }\href {\doibase
  10.1103/PhysRevA.82.044701} {\bibfield  {journal} {\bibinfo  {journal} {Phys.
  Rev. A}\ }\textbf {\bibinfo {volume} {82}},\ \bibinfo {pages} {044701}
  (\bibinfo {year} {2010}{\natexlab{b}})}\BibitemShut {NoStop}%
\bibitem [{\citenamefont {Mazloom}\ and\ \citenamefont
  {Abedinpour}(2018)}]{Mazloom-Abedinpour_PRB2018}%
  \BibitemOpen
  \bibfield  {author} {\bibinfo {author} {\bibfnamefont {A.}~\bibnamefont
  {Mazloom}}\ and\ \bibinfo {author} {\bibfnamefont {S.~H.}\ \bibnamefont
  {Abedinpour}},\ }\bibfield  {title} {\bibinfo {title} {\emph {Interplay of
  interlayer pairing and many-body screening in a bilayer of dipolar
  fermions}},\ }\href {\doibase 10.1103/PhysRevB.98.014513} {\bibfield
  {journal} {\bibinfo  {journal} {Phys. Rev. B}\ }\textbf {\bibinfo {volume}
  {98}},\ \bibinfo {pages} {014513} (\bibinfo {year} {2018})}\BibitemShut
  {NoStop}%
\bibitem [{\citenamefont {Mazloom}\ and\ \citenamefont
  {Abedinpour}(2017)}]{Mazloom-Abedinpour_PRB2017}%
  \BibitemOpen
  \bibfield  {author} {\bibinfo {author} {\bibfnamefont {A.}~\bibnamefont
  {Mazloom}}\ and\ \bibinfo {author} {\bibfnamefont {S.~H.}\ \bibnamefont
  {Abedinpour}},\ }\bibfield  {title} {\bibinfo {title} {\emph {Superfluidity
  in density imbalanced bilayers of dipolar fermions}},\ }\href {\doibase
  10.1103/PhysRevB.96.064513} {\bibfield  {journal} {\bibinfo  {journal} {Phys.
  Rev. B}\ }\textbf {\bibinfo {volume} {96}},\ \bibinfo {pages} {064513}
  (\bibinfo {year} {2017})}\BibitemShut {NoStop}%
\bibitem [{\citenamefont {Lee}\ \emph {et~al.}(2017)\citenamefont {Lee},
  \citenamefont {Matveenko}, \citenamefont {Wang},\ and\ \citenamefont
  {Shlyapnikov}}]{Lee-Shlyapnikov_PRA2017}%
  \BibitemOpen
  \bibfield  {author} {\bibinfo {author} {\bibfnamefont {H.}~\bibnamefont
  {Lee}}, \bibinfo {author} {\bibfnamefont {S.~I.}\ \bibnamefont {Matveenko}},
  \bibinfo {author} {\bibfnamefont {D.-W.}\ \bibnamefont {Wang}}, \ and\
  \bibinfo {author} {\bibfnamefont {G.~V.}\ \bibnamefont {Shlyapnikov}},\
  }\bibfield  {title} {\bibinfo {title} {\emph
  {Fulde-Ferrell-Larkin-Ovchinnikov state in bilayer dipolar systems}},\ }\href
  {\doibase 10.1103/PhysRevA.96.061602} {\bibfield  {journal} {\bibinfo
  {journal} {Phys. Rev. A}\ }\textbf {\bibinfo {volume} {96}},\ \bibinfo
  {pages} {061602} (\bibinfo {year} {2017})}\BibitemShut {NoStop}%
\bibitem [{\citenamefont {Klawunn}\ and\ \citenamefont
  {Recati}(2013)}]{Klawunn-Recati_PRA2013}%
  \BibitemOpen
  \bibfield  {author} {\bibinfo {author} {\bibfnamefont {M.}~\bibnamefont
  {Klawunn}}\ and\ \bibinfo {author} {\bibfnamefont {A.}~\bibnamefont
  {Recati}},\ }\bibfield  {title} {\bibinfo {title} {\emph {Polar molecules in
  bilayers with high population imbalance}},\ }\href {\doibase
  10.1103/PhysRevA.88.013633} {\bibfield  {journal} {\bibinfo  {journal} {Phys.
  Rev. A}\ }\textbf {\bibinfo {volume} {88}},\ \bibinfo {pages} {013633}
  (\bibinfo {year} {2013})}\BibitemShut {NoStop}%
\bibitem [{\citenamefont {Matveeva}\ and\ \citenamefont
  {Giorgini}(2013)}]{Matveeva-Giorgini_PRL2013}%
  \BibitemOpen
  \bibfield  {author} {\bibinfo {author} {\bibfnamefont {N.}~\bibnamefont
  {Matveeva}}\ and\ \bibinfo {author} {\bibfnamefont {S.}~\bibnamefont
  {Giorgini}},\ }\bibfield  {title} {\bibinfo {title} {\emph {Impurity Problem
  in a Bilayer System of Dipoles}},\ }\href {\doibase
  10.1103/PhysRevLett.111.220405} {\bibfield  {journal} {\bibinfo  {journal}
  {Phys. Rev. Lett.}\ }\textbf {\bibinfo {volume} {111}},\ \bibinfo {pages}
  {220405} (\bibinfo {year} {2013})}\BibitemShut {NoStop}%
\bibitem [{Note1()}]{Note1}%
  \BibitemOpen
  \bibinfo {note} {The ``repulsive Fermi polaron'' problem with dipolar
  fermions in a single layer geometry has been considered in Ref.~\cite
  {Bombin-Boronat_PRA2019}.}\BibitemShut {Stop}%
\bibitem [{\citenamefont {Z\"ollner}\ \emph {et~al.}(2011)\citenamefont
  {Z\"ollner}, \citenamefont {Bruun},\ and\ \citenamefont
  {Pethick}}]{Zollner2011}%
  \BibitemOpen
  \bibfield  {author} {\bibinfo {author} {\bibfnamefont {S.}~\bibnamefont
  {Z\"ollner}}, \bibinfo {author} {\bibfnamefont {G.~M.}\ \bibnamefont
  {Bruun}}, \ and\ \bibinfo {author} {\bibfnamefont {C.~J.}\ \bibnamefont
  {Pethick}},\ }\bibfield  {title} {\bibinfo {title} {\emph {Polarons and
  molecules in a two-dimensional Fermi gas}},\ }\href {\doibase
  10.1103/PhysRevA.83.021603} {\bibfield  {journal} {\bibinfo  {journal} {Phys.
  Rev. A}\ }\textbf {\bibinfo {volume} {83}},\ \bibinfo {pages} {021603}
  (\bibinfo {year} {2011})}\BibitemShut {NoStop}%
\bibitem [{\citenamefont {Parish}(2011)}]{Parish_PRA11}%
  \BibitemOpen
  \bibfield  {author} {\bibinfo {author} {\bibfnamefont {M.~M.}\ \bibnamefont
  {Parish}},\ }\bibfield  {title} {\bibinfo {title} {\emph {Polaron-molecule
  transitions in a two-dimensional Fermi gas}},\ }\href {\doibase
  10.1103/PhysRevA.83.051603} {\bibfield  {journal} {\bibinfo  {journal} {Phys.
  Rev. A}\ }\textbf {\bibinfo {volume} {83}},\ \bibinfo {pages} {051603}
  (\bibinfo {year} {2011})}\BibitemShut {NoStop}%
\bibitem [{\citenamefont {Schmidt}\ \emph
  {et~al.}(2012{\natexlab{a}})\citenamefont {Schmidt}, \citenamefont {Enss},
  \citenamefont {Pietil\"a},\ and\ \citenamefont {Demler}}]{Schmidt_PRA2012}%
  \BibitemOpen
  \bibfield  {author} {\bibinfo {author} {\bibfnamefont {R.}~\bibnamefont
  {Schmidt}}, \bibinfo {author} {\bibfnamefont {T.}~\bibnamefont {Enss}},
  \bibinfo {author} {\bibfnamefont {V.}~\bibnamefont {Pietil\"a}}, \ and\
  \bibinfo {author} {\bibfnamefont {E.}~\bibnamefont {Demler}},\ }\bibfield
  {title} {\bibinfo {title} {\emph {Fermi polarons in two dimensions}},\ }\href
  {\doibase 10.1103/PhysRevA.85.021602} {\bibfield  {journal} {\bibinfo
  {journal} {Phys. Rev. A}\ }\textbf {\bibinfo {volume} {85}},\ \bibinfo
  {pages} {021602} (\bibinfo {year} {2012}{\natexlab{a}})}\BibitemShut
  {NoStop}%
\bibitem [{\citenamefont {Ngampruetikorn}\ \emph {et~al.}(2012)\citenamefont
  {Ngampruetikorn}, \citenamefont {Levinsen},\ and\ \citenamefont
  {Parish}}]{Ngampruetikorn2012}%
  \BibitemOpen
  \bibfield  {author} {\bibinfo {author} {\bibfnamefont {V.}~\bibnamefont
  {Ngampruetikorn}}, \bibinfo {author} {\bibfnamefont {J.}~\bibnamefont
  {Levinsen}}, \ and\ \bibinfo {author} {\bibfnamefont {M.~M.}\ \bibnamefont
  {Parish}},\ }\bibfield  {title} {\bibinfo {title} {\emph {Repulsive polarons
  in two-dimensional Fermi gases}},\ }\href {\doibase
  10.1209/0295-5075/98/30005} {\bibfield  {journal} {\bibinfo  {journal}
  {Europhysics Letters}\ }\textbf {\bibinfo {volume} {98}},\ \bibinfo {pages}
  {30005} (\bibinfo {year} {2012})}\BibitemShut {NoStop}%
\bibitem [{\citenamefont {Parish}\ and\ \citenamefont
  {Levinsen}(2013)}]{Parish-Levinsen_PRA2013}%
  \BibitemOpen
  \bibfield  {author} {\bibinfo {author} {\bibfnamefont {M.~M.}\ \bibnamefont
  {Parish}}\ and\ \bibinfo {author} {\bibfnamefont {J.}~\bibnamefont
  {Levinsen}},\ }\bibfield  {title} {\bibinfo {title} {\emph {Highly polarized
  Fermi gases in two dimensions}},\ }\href {\doibase
  10.1103/PhysRevA.87.033616} {\bibfield  {journal} {\bibinfo  {journal} {Phys.
  Rev. A}\ }\textbf {\bibinfo {volume} {87}},\ \bibinfo {pages} {033616}
  (\bibinfo {year} {2013})}\BibitemShut {NoStop}%
\bibitem [{\citenamefont {Tajima}\ \emph {et~al.}(2021)\citenamefont {Tajima},
  \citenamefont {Takahashi}, \citenamefont {Mistakidis}, \citenamefont
  {Nakano},\ and\ \citenamefont {Iida}}]{Tajima_review2021}%
  \BibitemOpen
  \bibfield  {author} {\bibinfo {author} {\bibfnamefont {H.}~\bibnamefont
  {Tajima}}, \bibinfo {author} {\bibfnamefont {J.}~\bibnamefont {Takahashi}},
  \bibinfo {author} {\bibfnamefont {S.~I.}\ \bibnamefont {Mistakidis}},
  \bibinfo {author} {\bibfnamefont {E.}~\bibnamefont {Nakano}}, \ and\ \bibinfo
  {author} {\bibfnamefont {K.}~\bibnamefont {Iida}},\ }\bibfield  {title}
  {\bibinfo {title} {\emph {Polaron Problems in Ultracold Atoms: Role of a
  Fermi Sea across Different Spatial Dimensions and Quantum Fluctuations of a
  Bose Medium}},\ }\href {https://www.mdpi.com/2218-2004/9/1/18} {\bibfield
  {journal} {\bibinfo  {journal} {Atoms}\ }\textbf {\bibinfo {volume} {9}},\
  \bibinfo {pages} {18} (\bibinfo {year} {2021})}\BibitemShut {NoStop}%
\bibitem [{\citenamefont {Sidler}\ \emph {et~al.}(2016)\citenamefont {Sidler},
  \citenamefont {Back}, \citenamefont {Cotlet}, \citenamefont {Srivastava},
  \citenamefont {Fink}, \citenamefont {Kroner}, \citenamefont {Demler},\ and\
  \citenamefont {Imamoglu}}]{Sidler_2016}%
  \BibitemOpen
  \bibfield  {author} {\bibinfo {author} {\bibfnamefont {M.}~\bibnamefont
  {Sidler}}, \bibinfo {author} {\bibfnamefont {P.}~\bibnamefont {Back}},
  \bibinfo {author} {\bibfnamefont {O.}~\bibnamefont {Cotlet}}, \bibinfo
  {author} {\bibfnamefont {A.}~\bibnamefont {Srivastava}}, \bibinfo {author}
  {\bibfnamefont {T.}~\bibnamefont {Fink}}, \bibinfo {author} {\bibfnamefont
  {M.}~\bibnamefont {Kroner}}, \bibinfo {author} {\bibfnamefont
  {E.}~\bibnamefont {Demler}}, \ and\ \bibinfo {author} {\bibfnamefont
  {A.}~\bibnamefont {Imamoglu}},\ }\bibfield  {title} {\bibinfo {title} {\emph
  {Fermi polaron-polaritons in charge-tunable atomically thin
  semiconductors}},\ }\href {\doibase 10.1038/nphys3949} {\bibfield  {journal}
  {\bibinfo  {journal} {Nature Physics}\ }\textbf {\bibinfo {volume} {13}},\
  \bibinfo {pages} {255} (\bibinfo {year} {2016})}\BibitemShut {NoStop}%
\bibitem [{\citenamefont {Efimkin}\ \emph {et~al.}(2021)\citenamefont
  {Efimkin}, \citenamefont {Laird}, \citenamefont {Levinsen}, \citenamefont
  {Parish},\ and\ \citenamefont {MacDonald}}]{Efimkin_PRB_2021}%
  \BibitemOpen
  \bibfield  {author} {\bibinfo {author} {\bibfnamefont {D.~K.}\ \bibnamefont
  {Efimkin}}, \bibinfo {author} {\bibfnamefont {E.~K.}\ \bibnamefont {Laird}},
  \bibinfo {author} {\bibfnamefont {J.}~\bibnamefont {Levinsen}}, \bibinfo
  {author} {\bibfnamefont {M.~M.}\ \bibnamefont {Parish}}, \ and\ \bibinfo
  {author} {\bibfnamefont {A.~H.}\ \bibnamefont {MacDonald}},\ }\bibfield
  {title} {\bibinfo {title} {\emph {Electron-exciton interactions in the
  exciton-polaron problem}},\ }\href {\doibase 10.1103/PhysRevB.103.075417}
  {\bibfield  {journal} {\bibinfo  {journal} {Phys. Rev. B}\ }\textbf {\bibinfo
  {volume} {103}},\ \bibinfo {pages} {075417} (\bibinfo {year}
  {2021})}\BibitemShut {NoStop}%
\bibitem [{\citenamefont {Tiene}\ \emph {et~al.}(2022)\citenamefont {Tiene},
  \citenamefont {Levinsen}, \citenamefont {Keeling}, \citenamefont {Parish},\
  and\ \citenamefont {Marchetti}}]{Tiene_PRB2022}%
  \BibitemOpen
  \bibfield  {author} {\bibinfo {author} {\bibfnamefont {A.}~\bibnamefont
  {Tiene}}, \bibinfo {author} {\bibfnamefont {J.}~\bibnamefont {Levinsen}},
  \bibinfo {author} {\bibfnamefont {J.}~\bibnamefont {Keeling}}, \bibinfo
  {author} {\bibfnamefont {M.~M.}\ \bibnamefont {Parish}}, \ and\ \bibinfo
  {author} {\bibfnamefont {F.~M.}\ \bibnamefont {Marchetti}},\ }\bibfield
  {title} {\bibinfo {title} {\emph {Effect of fermion indistinguishability on
  optical absorption of doped two-dimensional semiconductors}},\ }\href
  {\doibase 10.1103/PhysRevB.105.125404} {\bibfield  {journal} {\bibinfo
  {journal} {Phys. Rev. B}\ }\textbf {\bibinfo {volume} {105}},\ \bibinfo
  {pages} {125404} (\bibinfo {year} {2022})}\BibitemShut {NoStop}%
\bibitem [{\citenamefont {Huang}\ \emph {et~al.}(2023)\citenamefont {Huang},
  \citenamefont {Sampson}, \citenamefont {Ni}, \citenamefont {Liu},
  \citenamefont {Liang}, \citenamefont {Watanabe}, \citenamefont {Taniguchi},
  \citenamefont {Li}, \citenamefont {Martin}, \citenamefont {Levinsen},
  \citenamefont {Parish}, \citenamefont {Tutuc}, \citenamefont {Efimkin},\ and\
  \citenamefont {Li}}]{Huang_PRX23}%
  \BibitemOpen
  \bibfield  {author} {\bibinfo {author} {\bibfnamefont {D.}~\bibnamefont
  {Huang}}, \bibinfo {author} {\bibfnamefont {K.}~\bibnamefont {Sampson}},
  \bibinfo {author} {\bibfnamefont {Y.}~\bibnamefont {Ni}}, \bibinfo {author}
  {\bibfnamefont {Z.}~\bibnamefont {Liu}}, \bibinfo {author} {\bibfnamefont
  {D.}~\bibnamefont {Liang}}, \bibinfo {author} {\bibfnamefont
  {K.}~\bibnamefont {Watanabe}}, \bibinfo {author} {\bibfnamefont
  {T.}~\bibnamefont {Taniguchi}}, \bibinfo {author} {\bibfnamefont
  {H.}~\bibnamefont {Li}}, \bibinfo {author} {\bibfnamefont {E.}~\bibnamefont
  {Martin}}, \bibinfo {author} {\bibfnamefont {J.}~\bibnamefont {Levinsen}},
  \bibinfo {author} {\bibfnamefont {M.~M.}\ \bibnamefont {Parish}}, \bibinfo
  {author} {\bibfnamefont {E.}~\bibnamefont {Tutuc}}, \bibinfo {author}
  {\bibfnamefont {D.~K.}\ \bibnamefont {Efimkin}}, \ and\ \bibinfo {author}
  {\bibfnamefont {X.}~\bibnamefont {Li}},\ }\bibfield  {title} {\bibinfo
  {title} {\emph {Quantum Dynamics of Attractive and Repulsive Polarons in a
  Doped ${\mathrm{MoSe}}_{2}$ Monolayer}},\ }\href {\doibase
  10.1103/PhysRevX.13.011029} {\bibfield  {journal} {\bibinfo  {journal} {Phys.
  Rev. X}\ }\textbf {\bibinfo {volume} {13}},\ \bibinfo {pages} {011029}
  (\bibinfo {year} {2023})}\BibitemShut {NoStop}%
\bibitem [{\citenamefont {Yang}\ \emph {et~al.}(1991)\citenamefont {Yang},
  \citenamefont {Guo}, \citenamefont {Chan}, \citenamefont {Wong},\ and\
  \citenamefont {Ching}}]{Yang_PRA1991}%
  \BibitemOpen
  \bibfield  {author} {\bibinfo {author} {\bibfnamefont {X.~L.}\ \bibnamefont
  {Yang}}, \bibinfo {author} {\bibfnamefont {S.~H.}\ \bibnamefont {Guo}},
  \bibinfo {author} {\bibfnamefont {F.~T.}\ \bibnamefont {Chan}}, \bibinfo
  {author} {\bibfnamefont {K.~W.}\ \bibnamefont {Wong}}, \ and\ \bibinfo
  {author} {\bibfnamefont {W.~Y.}\ \bibnamefont {Ching}},\ }\bibfield  {title}
  {\bibinfo {title} {\emph {Analytic solution of a two-dimensional hydrogen
  atom. I. Nonrelativistic theory}},\ }\href {\doibase
  10.1103/PhysRevA.43.1186} {\bibfield  {journal} {\bibinfo  {journal} {Phys.
  Rev. A}\ }\textbf {\bibinfo {volume} {43}},\ \bibinfo {pages} {1186}
  (\bibinfo {year} {1991})}\BibitemShut {NoStop}%
\bibitem [{\citenamefont {Massignan}\ \emph {et~al.}(2014)\citenamefont
  {Massignan}, \citenamefont {Zaccanti},\ and\ \citenamefont
  {Bruun}}]{Massignan_2014}%
  \BibitemOpen
  \bibfield  {author} {\bibinfo {author} {\bibfnamefont {P.}~\bibnamefont
  {Massignan}}, \bibinfo {author} {\bibfnamefont {M.}~\bibnamefont {Zaccanti}},
  \ and\ \bibinfo {author} {\bibfnamefont {G.~M.}\ \bibnamefont {Bruun}},\
  }\bibfield  {title} {\bibinfo {title} {\emph {Polarons, dressed molecules and
  itinerant ferromagnetism in ultracold Fermi gases}},\ }\href {\doibase
  10.1088/0034-4885/77/3/034401} {\bibfield  {journal} {\bibinfo  {journal}
  {Rep. Prog. Phys.}\ }\textbf {\bibinfo {volume} {77}},\ \bibinfo {pages}
  {034401} (\bibinfo {year} {2014})}\BibitemShut {NoStop}%
\bibitem [{\citenamefont {Scazza}\ \emph {et~al.}(2022)\citenamefont {Scazza},
  \citenamefont {Zaccanti}, \citenamefont {Massignan}, \citenamefont {Parish},\
  and\ \citenamefont {Levinsen}}]{Scazza_2022}%
  \BibitemOpen
  \bibfield  {author} {\bibinfo {author} {\bibfnamefont {F.}~\bibnamefont
  {Scazza}}, \bibinfo {author} {\bibfnamefont {M.}~\bibnamefont {Zaccanti}},
  \bibinfo {author} {\bibfnamefont {P.}~\bibnamefont {Massignan}}, \bibinfo
  {author} {\bibfnamefont {M.~M.}\ \bibnamefont {Parish}}, \ and\ \bibinfo
  {author} {\bibfnamefont {J.}~\bibnamefont {Levinsen}},\ }\bibfield  {title}
  {\bibinfo {title} {\emph {Repulsive Fermi and Bose Polarons in Quantum
  Gases}},\ }\href {\doibase 10.3390/atoms10020055} {\bibfield  {journal}
  {\bibinfo  {journal} {Atoms}\ }\textbf {\bibinfo {volume} {10}},\ \bibinfo
  {pages} {55} (\bibinfo {year} {2022})}\BibitemShut {NoStop}%
\bibitem [{\citenamefont {Li}\ \emph {et~al.}(2010)\citenamefont {Li},
  \citenamefont {Hwang},\ and\ \citenamefont
  {Das~Sarma}}]{Li-Das-Sarma_PRB2010}%
  \BibitemOpen
  \bibfield  {author} {\bibinfo {author} {\bibfnamefont {Q.}~\bibnamefont
  {Li}}, \bibinfo {author} {\bibfnamefont {E.~H.}\ \bibnamefont {Hwang}}, \
  and\ \bibinfo {author} {\bibfnamefont {S.}~\bibnamefont {Das~Sarma}},\
  }\bibfield  {title} {\bibinfo {title} {\emph {Collective modes of monolayer,
  bilayer, and multilayer fermionic dipolar liquid}},\ }\href {\doibase
  10.1103/PhysRevB.82.235126} {\bibfield  {journal} {\bibinfo  {journal} {Phys.
  Rev. B}\ }\textbf {\bibinfo {volume} {82}},\ \bibinfo {pages} {235126}
  (\bibinfo {year} {2010})}\BibitemShut {NoStop}%
\bibitem [{Note2()}]{Note2}%
  \BibitemOpen
  \bibinfo {note} {Note that Refs.~\cite {Lahaye_RPP2009,Chomaz_RPP2023} use a
  different definition of $C_{dd}$ but the same definition of $a_{dd}$:
  $C_{dd}^{\protect \text {Lehaye}} = 4\pi C_{dd}^{\protect \text {Chomaz}} =
  4\pi D^2 = 12\pi \protect \frac {a_{dd}}{m}$.}\BibitemShut {Stop}%
\bibitem [{\citenamefont {Du}\ \emph {et~al.}(2023)\citenamefont {Du},
  \citenamefont {Barral}, \citenamefont {Cantara}, \citenamefont {de~Hond},
  \citenamefont {Lu},\ and\ \citenamefont {Ketterle}}]{Du-Ketterle_arxiv2023}%
  \BibitemOpen
  \bibfield  {author} {\bibinfo {author} {\bibfnamefont {L.}~\bibnamefont
  {Du}}, \bibinfo {author} {\bibfnamefont {P.}~\bibnamefont {Barral}}, \bibinfo
  {author} {\bibfnamefont {M.}~\bibnamefont {Cantara}}, \bibinfo {author}
  {\bibfnamefont {J.}~\bibnamefont {de~Hond}}, \bibinfo {author} {\bibfnamefont
  {Y.-K.}\ \bibnamefont {Lu}}, \ and\ \bibinfo {author} {\bibfnamefont
  {W.}~\bibnamefont {Ketterle}},\ }\href@noop {} {\bibinfo {title} {\emph
  {Atomic physics on a 50 nm scale: Realization of a bilayer system of dipolar
  atoms}}} (\bibinfo {year} {2023}),\ \Eprint {http://arxiv.org/abs/2302.07209}
  {arXiv:2302.07209 [cond-mat.quant-gas]} \BibitemShut {NoStop}%
\bibitem [{\citenamefont {Carr}\ \emph {et~al.}(2009)\citenamefont {Carr},
  \citenamefont {DeMille}, \citenamefont {Krems},\ and\ \citenamefont
  {Ye}}]{Carr_NJP2009}%
  \BibitemOpen
  \bibfield  {author} {\bibinfo {author} {\bibfnamefont {L.~D.}\ \bibnamefont
  {Carr}}, \bibinfo {author} {\bibfnamefont {D.}~\bibnamefont {DeMille}},
  \bibinfo {author} {\bibfnamefont {R.~V.}\ \bibnamefont {Krems}}, \ and\
  \bibinfo {author} {\bibfnamefont {J.}~\bibnamefont {Ye}},\ }\bibfield
  {title} {\bibinfo {title} {\emph {Cold and ultracold molecules: science,
  technology and applications}},\ }\href {\doibase
  10.1088/1367-2630/11/5/055049} {\bibfield  {journal} {\bibinfo  {journal}
  {New Journal of Physics}\ }\textbf {\bibinfo {volume} {11}},\ \bibinfo
  {pages} {055049} (\bibinfo {year} {2009})}\BibitemShut {NoStop}%
\bibitem [{Note3()}]{Note3}%
  \BibitemOpen
  \bibinfo {note} {Note that, as mentioned previously, the Schr{\"o}dinger
  equation becomes diagonal in $\ell $ when either $Q=0$ or $E_F=0$, such that
  symmetric and antisymmetric solutions become degenerate.}\BibitemShut {Stop}%
\bibitem [{\citenamefont {Yudson}\ \emph {et~al.}(1997)\citenamefont {Yudson},
  \citenamefont {Rozman},\ and\ \citenamefont {Reineker}}]{Yudson_PRB1997}%
  \BibitemOpen
  \bibfield  {author} {\bibinfo {author} {\bibfnamefont {V.~I.}\ \bibnamefont
  {Yudson}}, \bibinfo {author} {\bibfnamefont {M.~G.}\ \bibnamefont {Rozman}},
  \ and\ \bibinfo {author} {\bibfnamefont {P.}~\bibnamefont {Reineker}},\
  }\bibfield  {title} {\bibinfo {title} {\emph {Bound states of two particles
  confined to parallel two-dimensional layers and interacting via dipole-dipole
  or dipole-charge laws}},\ }\href {\doibase 10.1103/PhysRevB.55.5214}
  {\bibfield  {journal} {\bibinfo  {journal} {Phys. Rev. B}\ }\textbf {\bibinfo
  {volume} {55}},\ \bibinfo {pages} {5214} (\bibinfo {year}
  {1997})}\BibitemShut {NoStop}%
\bibitem [{\citenamefont {Parish}\ \emph {et~al.}(2011)\citenamefont {Parish},
  \citenamefont {Marchetti},\ and\ \citenamefont {Littlewood}}]{Parish_EPL11}%
  \BibitemOpen
  \bibfield  {author} {\bibinfo {author} {\bibfnamefont {M.~M.}\ \bibnamefont
  {Parish}}, \bibinfo {author} {\bibfnamefont {F.~M.}\ \bibnamefont
  {Marchetti}}, \ and\ \bibinfo {author} {\bibfnamefont {P.~B.}\ \bibnamefont
  {Littlewood}},\ }\bibfield  {title} {\bibinfo {title} {\emph {Supersolidity
  in electron-hole bilayers with a large density imbalance}},\ }\href {\doibase
  10.1209/0295-5075/95/27007} {\bibfield  {journal} {\bibinfo  {journal}
  {Europhysics Letters}\ }\textbf {\bibinfo {volume} {95}},\ \bibinfo {pages}
  {27007} (\bibinfo {year} {2011})}\BibitemShut {NoStop}%
\bibitem [{\citenamefont {Cotlet}\ \emph {et~al.}(2020)\citenamefont {Cotlet},
  \citenamefont {Wild}, \citenamefont {Lukin},\ and\ \citenamefont
  {Imamoglu}}]{Cotlet_PRB20}%
  \BibitemOpen
  \bibfield  {author} {\bibinfo {author} {\bibfnamefont {O.}~\bibnamefont
  {Cotlet}}, \bibinfo {author} {\bibfnamefont {D.~S.}\ \bibnamefont {Wild}},
  \bibinfo {author} {\bibfnamefont {M.~D.}\ \bibnamefont {Lukin}}, \ and\
  \bibinfo {author} {\bibfnamefont {A.}~\bibnamefont {Imamoglu}},\ }\bibfield
  {title} {\bibinfo {title} {\emph {Rotons in optical excitation spectra of
  monolayer semiconductors}},\ }\href {\doibase 10.1103/PhysRevB.101.205409}
  {\bibfield  {journal} {\bibinfo  {journal} {Phys. Rev. B}\ }\textbf {\bibinfo
  {volume} {101}},\ \bibinfo {pages} {205409} (\bibinfo {year}
  {2020})}\BibitemShut {NoStop}%
\bibitem [{\citenamefont {Tiene}\ \emph {et~al.}(2020)\citenamefont {Tiene},
  \citenamefont {Levinsen}, \citenamefont {Parish}, \citenamefont {MacDonald},
  \citenamefont {Keeling},\ and\ \citenamefont {Marchetti}}]{Tiene_PRR20}%
  \BibitemOpen
  \bibfield  {author} {\bibinfo {author} {\bibfnamefont {A.}~\bibnamefont
  {Tiene}}, \bibinfo {author} {\bibfnamefont {J.}~\bibnamefont {Levinsen}},
  \bibinfo {author} {\bibfnamefont {M.~M.}\ \bibnamefont {Parish}}, \bibinfo
  {author} {\bibfnamefont {A.~H.}\ \bibnamefont {MacDonald}}, \bibinfo {author}
  {\bibfnamefont {J.}~\bibnamefont {Keeling}}, \ and\ \bibinfo {author}
  {\bibfnamefont {F.~M.}\ \bibnamefont {Marchetti}},\ }\bibfield  {title}
  {\bibinfo {title} {\emph {Extremely imbalanced two-dimensional
  electron-hole-photon systems}},\ }\href {\doibase
  10.1103/PhysRevResearch.2.023089} {\bibfield  {journal} {\bibinfo  {journal}
  {Phys. Rev. Res.}\ }\textbf {\bibinfo {volume} {2}},\ \bibinfo {pages}
  {023089} (\bibinfo {year} {2020})}\BibitemShut {NoStop}%
\bibitem [{\citenamefont {Fulde}\ and\ \citenamefont
  {Ferrell}(1964)}]{FuldeFerrell_PR64}%
  \BibitemOpen
  \bibfield  {author} {\bibinfo {author} {\bibfnamefont {P.}~\bibnamefont
  {Fulde}}\ and\ \bibinfo {author} {\bibfnamefont {R.~A.}\ \bibnamefont
  {Ferrell}},\ }\bibfield  {title} {\bibinfo {title} {\emph {Superconductivity
  in a Strong Spin-Exchange Field}},\ }\href {\doibase
  10.1103/PhysRev.135.A550} {\bibfield  {journal} {\bibinfo  {journal} {Phys.
  Rev.}\ }\textbf {\bibinfo {volume} {135}},\ \bibinfo {pages} {A550} (\bibinfo
  {year} {1964})}\BibitemShut {NoStop}%
\bibitem [{\citenamefont {Larkin}\ and\ \citenamefont
  {Ovchinnikov}(1964)}]{Larkin_64}%
  \BibitemOpen
  \bibfield  {author} {\bibinfo {author} {\bibfnamefont {A.~I.}\ \bibnamefont
  {Larkin}}\ and\ \bibinfo {author} {\bibfnamefont {Y.~N.}\ \bibnamefont
  {Ovchinnikov}},\ }\bibfield  {title} {\bibinfo {title} {\emph {{Nonuniform
  state of superconductors}}},\ }\href@noop {} {\bibfield  {journal} {\bibinfo
  {journal} {Zh. Eksp. Teor. Fiz.}\ }\textbf {\bibinfo {volume} {47}},\
  \bibinfo {pages} {1136} (\bibinfo {year} {1964})}\BibitemShut {NoStop}%
\bibitem [{\citenamefont {Casalbuoni}\ and\ \citenamefont
  {Nardulli}(2004)}]{Casalbuoni_RMP04}%
  \BibitemOpen
  \bibfield  {author} {\bibinfo {author} {\bibfnamefont {R.}~\bibnamefont
  {Casalbuoni}}\ and\ \bibinfo {author} {\bibfnamefont {G.}~\bibnamefont
  {Nardulli}},\ }\bibfield  {title} {\bibinfo {title} {\emph {Inhomogeneous
  superconductivity in condensed matter and QCD}},\ }\href {\doibase
  10.1103/RevModPhys.76.263} {\bibfield  {journal} {\bibinfo  {journal} {Rev.
  Mod. Phys.}\ }\textbf {\bibinfo {volume} {76}},\ \bibinfo {pages} {263}
  (\bibinfo {year} {2004})}\BibitemShut {NoStop}%
\bibitem [{\citenamefont {Radzihovsky}\ and\ \citenamefont
  {Sheehy}(2010)}]{Radzihovsky_RPP10}%
  \BibitemOpen
  \bibfield  {author} {\bibinfo {author} {\bibfnamefont {L.}~\bibnamefont
  {Radzihovsky}}\ and\ \bibinfo {author} {\bibfnamefont {D.~E.}\ \bibnamefont
  {Sheehy}},\ }\bibfield  {title} {\bibinfo {title} {\emph {Imbalanced
  Feshbach-resonant Fermi gases}},\ }\href {\doibase
  10.1088/0034-4885/73/7/076501} {\bibfield  {journal} {\bibinfo  {journal}
  {Rep. Prog. Phys.}\ }\textbf {\bibinfo {volume} {73}},\ \bibinfo {pages}
  {076501} (\bibinfo {year} {2010})}\BibitemShut {NoStop}%
\bibitem [{\citenamefont {Chevy}(2006)}]{Chevy_PRA2006}%
  \BibitemOpen
  \bibfield  {author} {\bibinfo {author} {\bibfnamefont {F.}~\bibnamefont
  {Chevy}},\ }\bibfield  {title} {\bibinfo {title} {\emph {Universal phase
  diagram of a strongly interacting Fermi gas with unbalanced spin
  populations}},\ }\href {\doibase 10.1103/PhysRevA.74.063628} {\bibfield
  {journal} {\bibinfo  {journal} {Phys. Rev. A}\ }\textbf {\bibinfo {volume}
  {74}},\ \bibinfo {pages} {063628} (\bibinfo {year} {2006})}\BibitemShut
  {NoStop}%
\bibitem [{\citenamefont {Combescot}\ and\ \citenamefont
  {Giraud}(2008)}]{Combescot2008}%
  \BibitemOpen
  \bibfield  {author} {\bibinfo {author} {\bibfnamefont {R.}~\bibnamefont
  {Combescot}}\ and\ \bibinfo {author} {\bibfnamefont {S.}~\bibnamefont
  {Giraud}},\ }\bibfield  {title} {\bibinfo {title} {\emph {Normal State of
  Highly Polarized Fermi Gases: Full Many-Body Treatment}},\ }\href {\doibase
  10.1103/PhysRevLett.101.050404} {\bibfield  {journal} {\bibinfo  {journal}
  {Phys. Rev. Lett.}\ }\textbf {\bibinfo {volume} {101}},\ \bibinfo {pages}
  {050404} (\bibinfo {year} {2008})}\BibitemShut {NoStop}%
\bibitem [{\citenamefont {Combescot}\ \emph {et~al.}(2007)\citenamefont
  {Combescot}, \citenamefont {Recati}, \citenamefont {Lobo},\ and\
  \citenamefont {Chevy}}]{Combescot2007}%
  \BibitemOpen
  \bibfield  {author} {\bibinfo {author} {\bibfnamefont {R.}~\bibnamefont
  {Combescot}}, \bibinfo {author} {\bibfnamefont {A.}~\bibnamefont {Recati}},
  \bibinfo {author} {\bibfnamefont {C.}~\bibnamefont {Lobo}}, \ and\ \bibinfo
  {author} {\bibfnamefont {F.}~\bibnamefont {Chevy}},\ }\bibfield  {title}
  {\bibinfo {title} {\emph {Normal State of Highly Polarized Fermi Gases:
  Simple Many-Body Approaches}},\ }\href {\doibase
  10.1103/PhysRevLett.98.180402} {\bibfield  {journal} {\bibinfo  {journal}
  {Phys. Rev. Lett.}\ }\textbf {\bibinfo {volume} {98}},\ \bibinfo {pages}
  {180402} (\bibinfo {year} {2007})}\BibitemShut {NoStop}%
\bibitem [{\citenamefont {Baranov}\ \emph {et~al.}(2011)\citenamefont
  {Baranov}, \citenamefont {Micheli}, \citenamefont {Ronen},\ and\
  \citenamefont {Zoller}}]{baranov2011}%
  \BibitemOpen
  \bibfield  {author} {\bibinfo {author} {\bibfnamefont {M.~A.}\ \bibnamefont
  {Baranov}}, \bibinfo {author} {\bibfnamefont {A.}~\bibnamefont {Micheli}},
  \bibinfo {author} {\bibfnamefont {S.}~\bibnamefont {Ronen}}, \ and\ \bibinfo
  {author} {\bibfnamefont {P.}~\bibnamefont {Zoller}},\ }\bibfield  {title}
  {\bibinfo {title} {\emph {Bilayer superfluidity of fermionic polar molecules:
  Many-body effects}},\ }\href {\doibase 10.1103/PhysRevA.83.043602} {\bibfield
   {journal} {\bibinfo  {journal} {Phys. Rev. A}\ }\textbf {\bibinfo {volume}
  {83}},\ \bibinfo {pages} {043602} (\bibinfo {year} {2011})}\BibitemShut
  {NoStop}%
\bibitem [{\citenamefont {Mahan}(2000)}]{Mahan2000book}%
  \BibitemOpen
  \bibfield  {author} {\bibinfo {author} {\bibfnamefont {G.}~\bibnamefont
  {Mahan}},\ }\href {https://books.google.es/books?id=v8du6cp0vUAC} {\emph
  {\bibinfo {title} {Many-Particle Physics}}}\ (\bibinfo  {publisher} {Kluwer
  Academic/Plenum Publishers},\ \bibinfo {year} {2000})\BibitemShut {NoStop}%
\bibitem [{Note4()}]{Note4}%
  \BibitemOpen
  \bibinfo {note} {Note that we only plot the dimer energy at finite center of
  mass $E^{(q=k_F)}$ corresponding to the symmetric states under the exchange
  $\ell \DOTSB \mapstochar \rightarrow -\ell $ in Eq.~\protect \textup {\hbox
  {\mathsurround \z@ \protect \normalfont (\ignorespaces \ref
  {eq:symm-a}\unskip \@@italiccorr )}} since these are the only ones with
  finite spectral weight. Instead, for zero center of mass momentum, symmetric
  and antisymmetric solution are degenerate in energy.}\BibitemShut {Stop}%
\bibitem [{\citenamefont {Schmidt}\ \emph
  {et~al.}(2012{\natexlab{b}})\citenamefont {Schmidt}, \citenamefont {Enss},
  \citenamefont {Pietilä},\ and\ \citenamefont {Demler}}]{Schmidt_2012}%
  \BibitemOpen
  \bibfield  {author} {\bibinfo {author} {\bibfnamefont {R.}~\bibnamefont
  {Schmidt}}, \bibinfo {author} {\bibfnamefont {T.}~\bibnamefont {Enss}},
  \bibinfo {author} {\bibfnamefont {V.}~\bibnamefont {Pietilä}}, \ and\
  \bibinfo {author} {\bibfnamefont {E.}~\bibnamefont {Demler}},\ }\bibfield
  {title} {\bibinfo {title} {\emph {Fermi polarons in two dimensions}},\ }\href
  {https://doi.org/10.1103%2Fphysreva.85.021602} {\bibfield  {journal}
  {\bibinfo  {journal} {Phys. Rev. A}\ }\textbf {\bibinfo {volume} {85}}
  (\bibinfo {year} {2012}{\natexlab{b}})}\BibitemShut {NoStop}%
\bibitem [{Note5()}]{Note5}%
  \BibitemOpen
  \bibinfo {note} {Note that only for the $1s$ state do we always have
  $E^{(q=k_F)}-E_F<E^{(q=0)}$.}\BibitemShut {Stop}%
\bibitem [{\citenamefont {Punk}\ and\ \citenamefont
  {Zwerger}(2007)}]{Punk2007}%
  \BibitemOpen
  \bibfield  {author} {\bibinfo {author} {\bibfnamefont {M.}~\bibnamefont
  {Punk}}\ and\ \bibinfo {author} {\bibfnamefont {W.}~\bibnamefont {Zwerger}},\
  }\bibfield  {title} {\bibinfo {title} {\emph {Theory of rf-Spectroscopy of
  Strongly Interacting Fermions}},\ }\href {\doibase
  10.1103/PhysRevLett.99.170404} {\bibfield  {journal} {\bibinfo  {journal}
  {Phys. Rev. Lett.}\ }\textbf {\bibinfo {volume} {99}},\ \bibinfo {pages}
  {170404} (\bibinfo {year} {2007})}\BibitemShut {NoStop}%
\bibitem [{\citenamefont {Baym}\ \emph {et~al.}(2007)\citenamefont {Baym},
  \citenamefont {Pethick}, \citenamefont {Yu},\ and\ \citenamefont
  {Zwierlein}}]{Baym2007}%
  \BibitemOpen
  \bibfield  {author} {\bibinfo {author} {\bibfnamefont {G.}~\bibnamefont
  {Baym}}, \bibinfo {author} {\bibfnamefont {C.~J.}\ \bibnamefont {Pethick}},
  \bibinfo {author} {\bibfnamefont {Z.}~\bibnamefont {Yu}}, \ and\ \bibinfo
  {author} {\bibfnamefont {M.~W.}\ \bibnamefont {Zwierlein}},\ }\bibfield
  {title} {\bibinfo {title} {\emph {Coherence and Clock Shifts in Ultracold
  Fermi Gases with Resonant Interactions}},\ }\href {\doibase
  10.1103/PhysRevLett.99.190407} {\bibfield  {journal} {\bibinfo  {journal}
  {Phys. Rev. Lett.}\ }\textbf {\bibinfo {volume} {99}},\ \bibinfo {pages}
  {190407} (\bibinfo {year} {2007})}\BibitemShut {NoStop}%
\bibitem [{\citenamefont {Liu}\ \emph {et~al.}(2020)\citenamefont {Liu},
  \citenamefont {Shi}, \citenamefont {Parish},\ and\ \citenamefont
  {Levinsen}}]{Weizhe_PRA2020}%
  \BibitemOpen
  \bibfield  {author} {\bibinfo {author} {\bibfnamefont {W.~E.}\ \bibnamefont
  {Liu}}, \bibinfo {author} {\bibfnamefont {Z.-Y.}\ \bibnamefont {Shi}},
  \bibinfo {author} {\bibfnamefont {M.~M.}\ \bibnamefont {Parish}}, \ and\
  \bibinfo {author} {\bibfnamefont {J.}~\bibnamefont {Levinsen}},\ }\bibfield
  {title} {\bibinfo {title} {\emph {Theory of radio-frequency spectroscopy of
  impurities in quantum gases}},\ }\href {\doibase 10.1103/PhysRevA.102.023304}
  {\bibfield  {journal} {\bibinfo  {journal} {Phys. Rev. A}\ }\textbf {\bibinfo
  {volume} {102}},\ \bibinfo {pages} {023304} (\bibinfo {year}
  {2020})}\BibitemShut {NoStop}%
\bibitem [{\citenamefont {Smole{\'{n}}ski}\ \emph {et~al.}(2021)\citenamefont
  {Smole{\'{n}}ski}, \citenamefont {Dolgirev}, \citenamefont {Kuhlenkamp},
  \citenamefont {Popert}, \citenamefont {Shimazaki}, \citenamefont {Back},
  \citenamefont {Lu}, \citenamefont {Kroner}, \citenamefont {Watanabe},
  \citenamefont {Taniguchi}, \citenamefont {Esterlis}, \citenamefont {Demler},\
  and\ \citenamefont {Imamo{\u{g}}lu}}]{Smolenski_Nature2021}%
  \BibitemOpen
  \bibfield  {author} {\bibinfo {author} {\bibfnamefont {T.}~\bibnamefont
  {Smole{\'{n}}ski}}, \bibinfo {author} {\bibfnamefont {P.~E.}\ \bibnamefont
  {Dolgirev}}, \bibinfo {author} {\bibfnamefont {C.}~\bibnamefont
  {Kuhlenkamp}}, \bibinfo {author} {\bibfnamefont {A.}~\bibnamefont {Popert}},
  \bibinfo {author} {\bibfnamefont {Y.}~\bibnamefont {Shimazaki}}, \bibinfo
  {author} {\bibfnamefont {P.}~\bibnamefont {Back}}, \bibinfo {author}
  {\bibfnamefont {X.}~\bibnamefont {Lu}}, \bibinfo {author} {\bibfnamefont
  {M.}~\bibnamefont {Kroner}}, \bibinfo {author} {\bibfnamefont
  {K.}~\bibnamefont {Watanabe}}, \bibinfo {author} {\bibfnamefont
  {T.}~\bibnamefont {Taniguchi}}, \bibinfo {author} {\bibfnamefont
  {I.}~\bibnamefont {Esterlis}}, \bibinfo {author} {\bibfnamefont
  {E.}~\bibnamefont {Demler}}, \ and\ \bibinfo {author} {\bibfnamefont
  {A.}~\bibnamefont {Imamo{\u{g}}lu}},\ }\bibfield  {title} {\bibinfo {title}
  {\emph {Signatures of Wigner crystal of electrons in a monolayer
  semiconductor}},\ }\href {\doibase 10.1038/s41586-021-03590-4} {\bibfield
  {journal} {\bibinfo  {journal} {Nature}\ }\textbf {\bibinfo {volume} {595}},\
  \bibinfo {pages} {53} (\bibinfo {year} {2021})}\BibitemShut {NoStop}%
\bibitem [{\citenamefont {Shimazaki}\ \emph {et~al.}(2021)\citenamefont
  {Shimazaki}, \citenamefont {Kuhlenkamp}, \citenamefont {Schwartz},
  \citenamefont {Smole\ifmmode~\acute{n}\else \'{n}\fi{}ski}, \citenamefont
  {Watanabe}, \citenamefont {Taniguchi}, \citenamefont {Kroner}, \citenamefont
  {Schmidt}, \citenamefont {Knap},\ and\ \citenamefont
  {Imamo\ifmmode~\breve{g}\else \u{g}\fi{}lu}}]{Shimazaki_PRX2021}%
  \BibitemOpen
  \bibfield  {author} {\bibinfo {author} {\bibfnamefont {Y.}~\bibnamefont
  {Shimazaki}}, \bibinfo {author} {\bibfnamefont {C.}~\bibnamefont
  {Kuhlenkamp}}, \bibinfo {author} {\bibfnamefont {I.}~\bibnamefont
  {Schwartz}}, \bibinfo {author} {\bibfnamefont {T.}~\bibnamefont
  {Smole\ifmmode~\acute{n}\else \'{n}\fi{}ski}}, \bibinfo {author}
  {\bibfnamefont {K.}~\bibnamefont {Watanabe}}, \bibinfo {author}
  {\bibfnamefont {T.}~\bibnamefont {Taniguchi}}, \bibinfo {author}
  {\bibfnamefont {M.}~\bibnamefont {Kroner}}, \bibinfo {author} {\bibfnamefont
  {R.}~\bibnamefont {Schmidt}}, \bibinfo {author} {\bibfnamefont
  {M.}~\bibnamefont {Knap}}, \ and\ \bibinfo {author} {\bibfnamefont
  {A.}~\bibnamefont {Imamo\ifmmode~\breve{g}\else \u{g}\fi{}lu}},\ }\bibfield
  {title} {\bibinfo {title} {\emph {Optical Signatures of Periodic Charge
  Distribution in a Mott-like Correlated Insulator State}},\ }\href {\doibase
  10.1103/PhysRevX.11.021027} {\bibfield  {journal} {\bibinfo  {journal} {Phys.
  Rev. X}\ }\textbf {\bibinfo {volume} {11}},\ \bibinfo {pages} {021027}
  (\bibinfo {year} {2021})}\BibitemShut {NoStop}%
\bibitem [{\citenamefont {Popert}\ \emph {et~al.}(2022)\citenamefont {Popert},
  \citenamefont {Shimazaki}, \citenamefont {Kroner}, \citenamefont {Watanabe},
  \citenamefont {Taniguchi}, \citenamefont {Imamoğlu},\ and\ \citenamefont
  {Smoleński}}]{Popert_NanoLett2022}%
  \BibitemOpen
  \bibfield  {author} {\bibinfo {author} {\bibfnamefont {A.}~\bibnamefont
  {Popert}}, \bibinfo {author} {\bibfnamefont {Y.}~\bibnamefont {Shimazaki}},
  \bibinfo {author} {\bibfnamefont {M.}~\bibnamefont {Kroner}}, \bibinfo
  {author} {\bibfnamefont {K.}~\bibnamefont {Watanabe}}, \bibinfo {author}
  {\bibfnamefont {T.}~\bibnamefont {Taniguchi}}, \bibinfo {author}
  {\bibfnamefont {A.}~\bibnamefont {Imamoğlu}}, \ and\ \bibinfo {author}
  {\bibfnamefont {T.}~\bibnamefont {Smoleński}},\ }\bibfield  {title}
  {\bibinfo {title} {\emph {Optical Sensing of Fractional Quantum Hall Effect
  in Graphene}},\ }\href {\doibase 10.1021/acs.nanolett.2c02000} {\bibfield
  {journal} {\bibinfo  {journal} {Nano Letters}\ }\textbf {\bibinfo {volume}
  {22}},\ \bibinfo {pages} {7363} (\bibinfo {year} {2022})}\BibitemShut
  {NoStop}%
\bibitem [{\citenamefont {Schwartz}\ \emph {et~al.}(2021)\citenamefont
  {Schwartz}, \citenamefont {Shimazaki}, \citenamefont {Kuhlenkamp},
  \citenamefont {Watanabe}, \citenamefont {Taniguchi}, \citenamefont {Kroner},\
  and\ \citenamefont {Imamo{\u g}lu}}]{Schwartz_science2021}%
  \BibitemOpen
  \bibfield  {author} {\bibinfo {author} {\bibfnamefont {I.}~\bibnamefont
  {Schwartz}}, \bibinfo {author} {\bibfnamefont {Y.}~\bibnamefont {Shimazaki}},
  \bibinfo {author} {\bibfnamefont {C.}~\bibnamefont {Kuhlenkamp}}, \bibinfo
  {author} {\bibfnamefont {K.}~\bibnamefont {Watanabe}}, \bibinfo {author}
  {\bibfnamefont {T.}~\bibnamefont {Taniguchi}}, \bibinfo {author}
  {\bibfnamefont {M.}~\bibnamefont {Kroner}}, \ and\ \bibinfo {author}
  {\bibfnamefont {A.}~\bibnamefont {Imamo{\u g}lu}},\ }\bibfield  {title}
  {\bibinfo {title} {\emph {Electrically tunable Feshbach resonances in twisted
  bilayer semiconductors}},\ }\href {\doibase 10.1126/science.abj3831}
  {\bibfield  {journal} {\bibinfo  {journal} {Science}\ }\textbf {\bibinfo
  {volume} {374}},\ \bibinfo {pages} {336} (\bibinfo {year}
  {2021})}\BibitemShut {NoStop}%
\bibitem [{\citenamefont {Nishimura}\ \emph {et~al.}(2021)\citenamefont
  {Nishimura}, \citenamefont {Nakano}, \citenamefont {Iida}, \citenamefont
  {Tajima}, \citenamefont {Miyakawa},\ and\ \citenamefont
  {Yabu}}]{Nishimura-Yabu_PRA2021}%
  \BibitemOpen
  \bibfield  {author} {\bibinfo {author} {\bibfnamefont {K.}~\bibnamefont
  {Nishimura}}, \bibinfo {author} {\bibfnamefont {E.}~\bibnamefont {Nakano}},
  \bibinfo {author} {\bibfnamefont {K.}~\bibnamefont {Iida}}, \bibinfo {author}
  {\bibfnamefont {H.}~\bibnamefont {Tajima}}, \bibinfo {author} {\bibfnamefont
  {T.}~\bibnamefont {Miyakawa}}, \ and\ \bibinfo {author} {\bibfnamefont
  {H.}~\bibnamefont {Yabu}},\ }\bibfield  {title} {\bibinfo {title} {\emph
  {Ground state of the polaron in an ultracold dipolar Fermi gas}},\ }\href
  {\doibase 10.1103/PhysRevA.103.033324} {\bibfield  {journal} {\bibinfo
  {journal} {Phys. Rev. A}\ }\textbf {\bibinfo {volume} {103}},\ \bibinfo
  {pages} {033324} (\bibinfo {year} {2021})}\BibitemShut {NoStop}%
\bibitem [{\citenamefont {Aikawa}\ \emph
  {et~al.}(2014{\natexlab{b}})\citenamefont {Aikawa}, \citenamefont {Baier},
  \citenamefont {Frisch}, \citenamefont {Mark}, \citenamefont {Ravensbergen},\
  and\ \citenamefont {Ferlaino}}]{Aikawa-Ferlaino_science2014}%
  \BibitemOpen
  \bibfield  {author} {\bibinfo {author} {\bibfnamefont {K.}~\bibnamefont
  {Aikawa}}, \bibinfo {author} {\bibfnamefont {S.}~\bibnamefont {Baier}},
  \bibinfo {author} {\bibfnamefont {A.}~\bibnamefont {Frisch}}, \bibinfo
  {author} {\bibfnamefont {M.}~\bibnamefont {Mark}}, \bibinfo {author}
  {\bibfnamefont {C.}~\bibnamefont {Ravensbergen}}, \ and\ \bibinfo {author}
  {\bibfnamefont {F.}~\bibnamefont {Ferlaino}},\ }\bibfield  {title} {\bibinfo
  {title} {\emph {Observation of Fermi surface deformation in a dipolar quantum
  gas}},\ }\href {\doibase 10.1126/science.1255259} {\bibfield  {journal}
  {\bibinfo  {journal} {Science}\ }\textbf {\bibinfo {volume} {345}},\ \bibinfo
  {pages} {1484} (\bibinfo {year} {2014}{\natexlab{b}})}\BibitemShut {NoStop}%
\bibitem [{\citenamefont {Kawamura}\ and\ \citenamefont
  {Ohashi}(2022)}]{Kawamura-Ohashi_PRA2022}%
  \BibitemOpen
  \bibfield  {author} {\bibinfo {author} {\bibfnamefont {T.}~\bibnamefont
  {Kawamura}}\ and\ \bibinfo {author} {\bibfnamefont {Y.}~\bibnamefont
  {Ohashi}},\ }\bibfield  {title} {\bibinfo {title} {\emph {Feasibility of a
  Fulde-Ferrell-Larkin-Ovchinnikov superfluid Fermi atomic gas}},\ }\href
  {\doibase 10.1103/PhysRevA.106.033320} {\bibfield  {journal} {\bibinfo
  {journal} {Phys. Rev. A}\ }\textbf {\bibinfo {volume} {106}},\ \bibinfo
  {pages} {033320} (\bibinfo {year} {2022})}\BibitemShut {NoStop}%
\bibitem [{\citenamefont {Shimahara}(1998)}]{Shimahara_JPSJ1988}%
  \BibitemOpen
  \bibfield  {author} {\bibinfo {author} {\bibfnamefont {H.}~\bibnamefont
  {Shimahara}},\ }\bibfield  {title} {\bibinfo {title} {\emph {Phase
  Fluctuations and Kosterlitz-Thouless Transition in Two-Dimensional
  Fulde-Ferrell-Larkin-Ovchinnikov Superconductors}},\ }\href {\doibase
  10.1143/JPSJ.67.1872} {\bibfield  {journal} {\bibinfo  {journal} {Journal of
  the Physical Society of Japan}\ }\textbf {\bibinfo {volume} {67}},\ \bibinfo
  {pages} {1872} (\bibinfo {year} {1998})}\BibitemShut {NoStop}%
\bibitem [{\citenamefont {Tiene}\ \emph {et~al.}(2024)\citenamefont {Tiene},
  \citenamefont {Tamargo~Bracho}, \citenamefont {Parish}, \citenamefont
  {Levinsen},\ and\ \citenamefont {Marchetti}}]{Tiene_PRA2024_dataset}%
  \BibitemOpen
  \bibfield  {author} {\bibinfo {author} {\bibfnamefont {A.}~\bibnamefont
  {Tiene}}, \bibinfo {author} {\bibfnamefont {A.}~\bibnamefont
  {Tamargo~Bracho}}, \bibinfo {author} {\bibfnamefont {M.~M.}\ \bibnamefont
  {Parish}}, \bibinfo {author} {\bibfnamefont {J.}~\bibnamefont {Levinsen}}, \
  and\ \bibinfo {author} {\bibfnamefont {F.~M.}\ \bibnamefont {Marchetti}},\
  }\href {\doibase 10.21950/UTHVWP} {\bibinfo {title} {\emph {{Multiple polaron
  quasiparticles with dipolar fermions in a bilayer geometry (dataset)}}}}
  (\bibinfo {year} {2024})\BibitemShut {NoStop}%
\bibitem [{\citenamefont {Morse}\ and\ \citenamefont
  {Allis}(1933)}]{Morse1933}%
  \BibitemOpen
  \bibfield  {author} {\bibinfo {author} {\bibfnamefont {P.~M.}\ \bibnamefont
  {Morse}}\ and\ \bibinfo {author} {\bibfnamefont {W.~P.}\ \bibnamefont
  {Allis}},\ }\bibfield  {title} {\bibinfo {title} {\emph {The Effect of
  Exchange on the Scattering of Slow Electrons from Atoms}},\ }\href {\doibase
  10.1103/PhysRev.44.269} {\bibfield  {journal} {\bibinfo  {journal} {Phys.
  Rev.}\ }\textbf {\bibinfo {volume} {44}},\ \bibinfo {pages} {269} (\bibinfo
  {year} {1933})}\BibitemShut {NoStop}%
\bibitem [{\citenamefont {Portnoi}\ and\ \citenamefont
  {Galbraith}(1997)}]{Portnoi_SSC97}%
  \BibitemOpen
  \bibfield  {author} {\bibinfo {author} {\bibfnamefont {M.}~\bibnamefont
  {Portnoi}}\ and\ \bibinfo {author} {\bibfnamefont {I.}~\bibnamefont
  {Galbraith}},\ }\bibfield  {title} {\bibinfo {title} {\emph {Variable-phase
  method and Levinson's theorem in two dimensions: Application to a screened
  Coulomb potential}},\ }\href {\doibase
  https://doi.org/10.1016/S0038-1098(97)00203-2} {\bibfield  {journal}
  {\bibinfo  {journal} {Solid State Communications}\ }\textbf {\bibinfo
  {volume} {103}},\ \bibinfo {pages} {325} (\bibinfo {year}
  {1997})}\BibitemShut {NoStop}%
\bibitem [{\citenamefont {Levinsen}\ and\ \citenamefont
  {Parish}(2015)}]{Levinsen2Dreview}%
  \BibitemOpen
  \bibfield  {author} {\bibinfo {author} {\bibfnamefont {J.}~\bibnamefont
  {Levinsen}}\ and\ \bibinfo {author} {\bibfnamefont {M.~M.}\ \bibnamefont
  {Parish}},\ }\bibfield  {title} {\bibinfo {title} {\emph {{Strongly
  interacting two-dimensional Fermi gases}}},\ }\href
  {https://doi.org/10.1142/9789814667746_0001} {\bibfield  {journal} {\bibinfo
  {journal} {Annual Review of Cold Atoms and Molecules}\ }\textbf {\bibinfo
  {volume} {3}},\ \bibinfo {pages} {1} (\bibinfo {year} {2015})}\BibitemShut
  {NoStop}%
\bibitem [{\citenamefont {Pisani}\ \emph {et~al.}(2018)\citenamefont {Pisani},
  \citenamefont {Perali}, \citenamefont {Pieri},\ and\ \citenamefont
  {Strinati}}]{Pisani2018}%
  \BibitemOpen
  \bibfield  {author} {\bibinfo {author} {\bibfnamefont {L.}~\bibnamefont
  {Pisani}}, \bibinfo {author} {\bibfnamefont {A.}~\bibnamefont {Perali}},
  \bibinfo {author} {\bibfnamefont {P.}~\bibnamefont {Pieri}}, \ and\ \bibinfo
  {author} {\bibfnamefont {G.~C.}\ \bibnamefont {Strinati}},\ }\bibfield
  {title} {\bibinfo {title} {\emph {Entanglement between pairing and screening
  in the Gorkov-Melik-Barkhudarov correction to the critical temperature
  throughout the BCS-BEC crossover}},\ }\href {\doibase
  10.1103/PhysRevB.97.014528} {\bibfield  {journal} {\bibinfo  {journal} {Phys.
  Rev. B}\ }\textbf {\bibinfo {volume} {97}},\ \bibinfo {pages} {014528}
  (\bibinfo {year} {2018})}\BibitemShut {NoStop}%
\bibitem [{\citenamefont {Bomb\'{\i}n}\ \emph {et~al.}(2019)\citenamefont
  {Bomb\'{\i}n}, \citenamefont {Comparin}, \citenamefont {Bertaina},
  \citenamefont {Mazzanti}, \citenamefont {Giorgini},\ and\ \citenamefont
  {Boronat}}]{Bombin-Boronat_PRA2019}%
  \BibitemOpen
  \bibfield  {author} {\bibinfo {author} {\bibfnamefont {R.}~\bibnamefont
  {Bomb\'{\i}n}}, \bibinfo {author} {\bibfnamefont {T.}~\bibnamefont
  {Comparin}}, \bibinfo {author} {\bibfnamefont {G.}~\bibnamefont {Bertaina}},
  \bibinfo {author} {\bibfnamefont {F.}~\bibnamefont {Mazzanti}}, \bibinfo
  {author} {\bibfnamefont {S.}~\bibnamefont {Giorgini}}, \ and\ \bibinfo
  {author} {\bibfnamefont {J.}~\bibnamefont {Boronat}},\ }\bibfield  {title}
  {\bibinfo {title} {\emph {Two-dimensional repulsive Fermi polarons with
  short- and long-range interactions}},\ }\href {\doibase
  10.1103/PhysRevA.100.023608} {\bibfield  {journal} {\bibinfo  {journal}
  {Phys. Rev. A}\ }\textbf {\bibinfo {volume} {100}},\ \bibinfo {pages}
  {023608} (\bibinfo {year} {2019})}\BibitemShut {NoStop}%
\end{thebibliography}

%

\end{document}